%% file: output.tex
\documentclass[amsmath,trackchanges,twocolumn,twocolappendix]{aastex702}

\input{newcommand}
\received{July 16, 2026}
\revised{October 20, 2026}

\submitjournal{ApJ}

\shorttitle{X-shooter observations on \id}
\shortauthors{Morishita et al.}

\begin{document}

\title{
Deep UV Spectroscopy of an Extremely Metal-Poor Galaxy at $z=5.725$
}

\correspondingauthor{Takahiro Morishita}
\author[0000-0002-8512-1404]{Takahiro Morishita}
\affiliation{Astronomical Institute, Graduate School of Science, Tohoku University, 6–3 Aoba, Sendai 980-8578, Japan}
\email{morishita@astr.tohoku.ac.jp}

\author[0000-0001-9935-6047]{Massimo Stiavelli}
\affiliation{Space Telescope Science Institute, 3700 San Martin Drive, Baltimore, MD 21218, USA}
\affiliation{The William H. Miller III, Dept. of Physics \& Astronomy, Johns Hopkins University, Baltimore, MD 21218, USA}
\email{mstiavel@stsci.edu}

\author[0009-0002-8965-1303]{Zhaoran Liu}
\affiliation{MIT Kavli Institute for Astrophysics and Space Research, 70 Vassar Street, Cambridge, MA 02139, USA}
\email{zrliu@mit.edu}

\author[0000-0003-1427-2456]{Matteo Messa}
\affiliation{INAF -- OAS, Osservatorio di Astrofisica e Scienza dello Spazio di Bologna, via Gobetti 93/3, I-40129 Bologna, Italy}
\email{matteo.messa@inaf.it}

\author[0000-0002-5057-135X]{Eros~Vanzella}
\affiliation{INAF -- OAS, Osservatorio di Astrofisica e Scienza dello Spazio di Bologna, via Gobetti 93/3, I-40129 Bologna, Italy}
\email{eros.vanzella@inaf.it}






\begin{abstract}
We present deep X-shooter follow-up spectroscopy of AMORE6, an exceptionally metal-poor galaxy at $z=5.725$. The new spectrum provides stringent upper limits on rest-frame ultraviolet metal lines, including N{\sc v}, C{\sc iv}, N{\sc iv}], N{\sc iii}], and O{\sc iii}]. The non-detection of C{\sc iv} breaks the density--metallicity degeneracy present in previous analyses based on the [O{\sc iii}]/H$\beta$ ratio, excluding the high-density, chemically enriched gas scenario and strengthening the evidence for an extremely metal-poor environment.
The X-shooter spectrum also reveals a narrow Ly$\alpha$ line with a velocity dispersion $\sigma_v=23.2\pm0.5$\,km\,s$^{-1}$ and a nearly symmetric profile centered at the systemic redshift measured from H$\beta$. Combining the Ly$\alpha$ line width with the measured source size, we infer {a dynamical mass $M_{\rm dyn}=(1$--$22)\times10^7\,M_\odot$, assuming that the line width traces gravitational motions.}
Spectral energy distribution modeling using Pop~III and Pop~II stellar population libraries yields stellar masses $(1$--$5)\times10^{5}\,M_\odot$. Although all models reproduce the observed blue UV continuum and \ha\ equivalent width (EW), none simultaneously reproduces the H$\beta$ luminosity and EW{, with the observed H$\beta$ luminosity exceeding the model predictions by up to $3.7\sigma$. Possible explanations include departures from standard recombination conditions or limitations in the adopted stellar population models.
We discuss possible interpretations of \id, including massive-star formation in an atomic-cooling halo and a gravitationally bound star cluster.}
\end{abstract}

\keywords{}

\section{Introduction}\label{sec:intro}
The formation of the first generation of stars—Population III (hereafter Pop~III)—marks a fundamental milestone in cosmic history. Formed from pristine, metal-free gas shortly after the Big Bang, these stars are thought to have initiated the first episodes of chemical enrichment in the intergalactic medium and seeded the formation of the earliest galaxies and black holes \citep[e.g.,][]{bromm03,wise2012}.

The advent of the \textit{James Webb Space Telescope} (JWST) has enabled the discovery and characterization of galaxies and star-forming regions within the first billion years of cosmic history. Remarkably, oxygen and other heavy elements have been detected in galaxies at redshifts $z\simgt10$ \citep[][]{curtis-lake23,roberts-borsani23,bunker23b,castellano24,napolitano25,harikane26,rodighiero26}, and now a few at $z\simgt14$ \citep{carniani24,schouws24,naidu25,donnan26}. These observations indicate that chemical enrichment can proceed rapidly, especially in relatively massive galaxies, consistent with the weak evolution of the mass--metallicity relation observed at $z>3$ \citep[e.g.,][]{laseter23,sanders24,morishita24b,scholte25}.

The lowest metallicities measured to date, typically inferred from strong-line diagnostics, are found in low-mass galaxies with stellar masses below $10^{7}\,M_\odot$ \citep[e.g.,][]{chemerynska24,hsiao25,trussler26}. These findings suggest that the search for galaxies hosting Pop~III-like stellar populations should focus on extremely low-mass systems, where the impact of chemical feedback may still be limited. Among them, several intriguing candidates have recently been identified through strong gravitational lensing, with stellar masses below $10^6\,M_\odot$ \citep[][also {\it Hebe} detected $\sim3$\,kpc away from the luminous galaxy GN-z11; \citealt{maiolino26,ubler26}]{vanzella23,nakajima25,morishita25,cai25,vanzella26}. 
In particular, many of these systems reside at relatively low redshifts,  providing a unique opportunity for detailed spectroscopic investigation through access to key Pop III diagnostic lines that would be difficult to observe at $z>10$. These galaxies exhibit exceptionally low values of $R3\,\equiv$\,\oiiir/\hb, implying oxygen abundances below $\approx1\%$ of the solar value. 
Furthermore, a recent follow-up spectroscopy of LAP1-B revealed strong \civ\ emission \citep{nakajima25}, suggesting the presence of very massive stars and revealing chemical abundance patterns consistent with enrichment by Pop~III supernovae.

Among these candidates, the strongly lensed galaxy AMORE6 at $z=5.725$ \citep{morishita25} currently provides one of the most stringent constraints on metallicity, with an upper limit of $12+\log({\rm O/H})<5.78$ (2$\sigma$), corresponding to only $\approx0.12\%$ of the solar oxygen abundance. Existing NIRCam wide-field slitless spectroscopy (WFSS), taken as part of ALT \citep{naidu24}, shows no detectable oxygen emission in either of the two lensed images, {\id-A and \id-B across the lensing critical curve, which are magnified by $\mu_{\rm A}=39.3_{-3.5}^{+3.7}$ and $\mu_{\rm B}=77.7_{-5.9}^{+8.4}$ respectively \citep[][see also \citealt{mahler18,richard21}]{bergamini23b}.} This upper limit on metallicity approaches a particularly interesting regime, as even a single pair-instability supernova (PISN) is expected to enrich the interstellar medium of a low-mass galaxy such as \id\ beyond this level \citep[e.g.,][]{bromm03}.

In addition to its extremely low metallicity, \id\ is characterized with a very low stellar mass ($\sim4\times10^5\,M_\odot$), strong Ly$\alpha$ emission detected with MUSE at the systemic redshift, a very high specific star-formation rate, and an exceptionally blue ultraviolet continuum slope of $\beta_{\rm UV}=-2.9\pm0.3$ \citep{morishita25}. Together, these properties suggest a young stellar population that may contain very massive ($>100\,M_\odot$) stars and are broadly consistent with a galaxy experiencing its first major episode of star formation \citep[e.g.,][]{schaerer02,bouwens10,katz25}.

Despite these remarkable characteristics, the current evidence falls short of a definitive identification of \id\ as a Pop~III galaxy. First, although the upper limit on $R3$ is among the most stringent reported for such candidates, it still corresponds to an oxygen abundance of approximately $>0.1\%$ solar. For comparison, most metal-poor low-mass stars in our Galaxy probe iron abundance of ${\rm [Fe/H]}<-4$ \citep[e.g.,][]{2002Natur.419..904C,2005Natur.434..871F,ji26}, though oxygen abundances, which would enable direct comparisons with high-redshift studies, are often not constrained. Similarly, damped Ly$\alpha$ absorbers (DLAs) observed along quasar sightlines have revealed gas-phase metallicities as low as ${\rm [Z/H]}\sim-4$--$-5$ \citep{tornatore07,fumagalli11}, an order of magnitude lower \citep[also,][]{2026arXiv261016509B}. Consequently, while the metallicity limit for \id\ is exceptionally low among other galaxies studied with JWST, it does not by itself establish the presence of a truly primordial stellar population.

Second, and more importantly, the analysis of \citet{morishita25} left open the possibility that the observed weakness of \oiii\ emission could be caused by an extremely high gas density. In such environments, collisional de-excitation can suppress \oiii\ emission even in chemically enriched gas, causing metallicity estimates based solely on the \oiii/\hb\ ratio to be biased toward artificially low values. Distinguishing between these scenarios requires deep rest-frame ultraviolet spectroscopy, as UV metal lines provide complementary constraints on both chemical enrichment and gas properties \citep[e.g.,][]{mingozzi22}.

In this paper, we investigate this alternative high-density scenario using newly acquired VLT/X-shooter spectroscopy. We also exploit the detected Ly$\alpha$ emission to constrain the dynamical and baryonic properties of \id, thereby assessing whether it satisfies the conditions expected for a bona fide Pop~III galaxy candidate. In addition, we revisit the stellar mass estimate using a more comprehensive suite of stellar population models that includes several extreme Pop~III scenarios. The inferred stellar mass depends sensitively on the adopted initial mass function (IMF), particularly because both its shape and upper-mass limit may differ substantially in metal-free environments \citep[e.g.,][]{2021MNRAS.508.4175C,2025MNRAS.539.2561C}. Theoretically, the first stars are expected to form initially in minihalos ($\sim10^{5-6}\,M_\odot$), where gas cools primarily through H$2$ molecules, giving rise to a top-heavy IMF with characteristic stellar masses of tens to hundreds of solar masses \citep[e.g.,][]{1999ApJ...527L...5B,2011ApJ...727..110C,hirano14}. If H$_2$ cooling is suppressed by a sufficiently strong Lyman--Werner (LW) radiation field, however, star formation can be delayed until the host halo reaches the atomic-cooling threshold ($T_{\rm vir}\gtrsim10^4$\,K, corresponding to halo masses of $\sim10^{7-8}\,M_\odot$, e.g., \citealt{2008MNRAS.387.1021G,StiavelliBook}). Under these conditions, the much higher gas accretion rates may lead to substantially different stellar mass distributions and, in extreme cases, the formation of very massive stars ($\gg100\,M_\odot$) \citep[e.g.,][]{2001ApJ...546..635O,bromm03,2024ApJ...970...14S,2026arXiv261003877K}. Observational constraints on the IMF would therefore provide valuable insight into the physical conditions of primordial star formation \citep[see also][]{2025arXiv251211790F} and the dominant cooling mechanism that initiated it.

This paper is organized as follows. In Section~\ref{sec:data}, we describe the observations and data reduction. In Section~\ref{sec:uv}, we present the spectroscopic analysis. In Section~\ref{sec:sedmodels}, we perform dedicated spectral energy distribution modeling and estimate the stellar mass. In Section~\ref{sec:disc}, we discuss the nature of \id\ and its implications for the search for primordial stellar populations. Throughout this paper, we adopt the AB magnitude system \citep{oke83,fukugita96} and a flat $\Lambda$CDM cosmology with $\Omega_{\rm m}=0.310$ and $H_0=67.7\,{\rm km\,s^{-1}\,Mpc^{-1}}$ \citep{planck20}.

\begin{figure*}
\centering
	\includegraphics[width=0.9\textwidth]{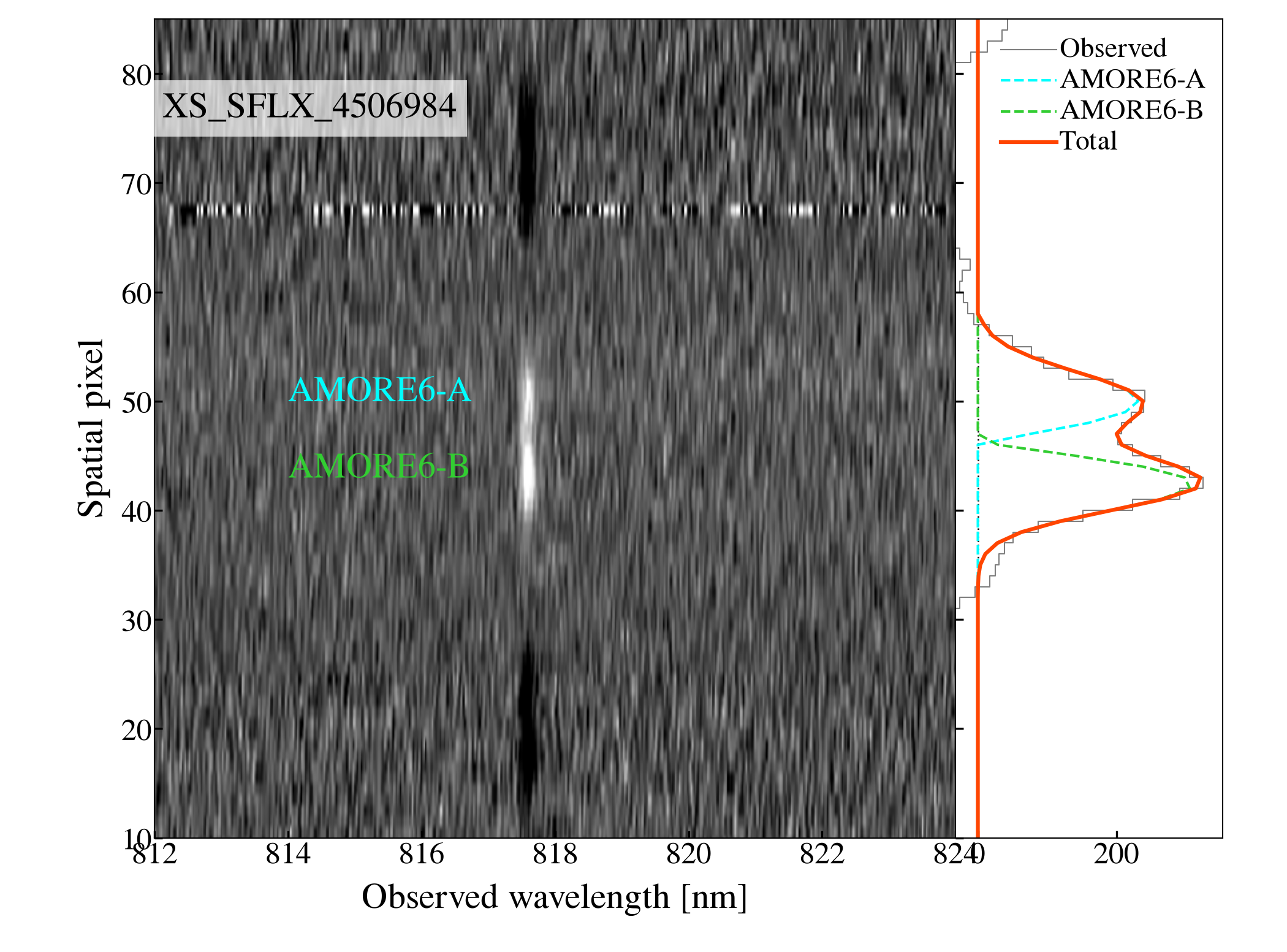}
	\caption{
    An example of a two-dimensional X-shooter VIS-arm spectrum near the \ly\ wavelength, corresponding to a single OB ($3600$\,sec). The approximate positions of \id-A and \id-B are indicated. In the right panel, the observed spatial profile along the cross-dispersion, used in optimal extraction, is shown (gray line). Gaussian models of the individual components and the total component are shown (cyan, green dashed lines, and red line).
    }
\label{fig:lya2d}
\end{figure*}

\begin{figure*}
\centering
	\includegraphics[width=0.49\textwidth]{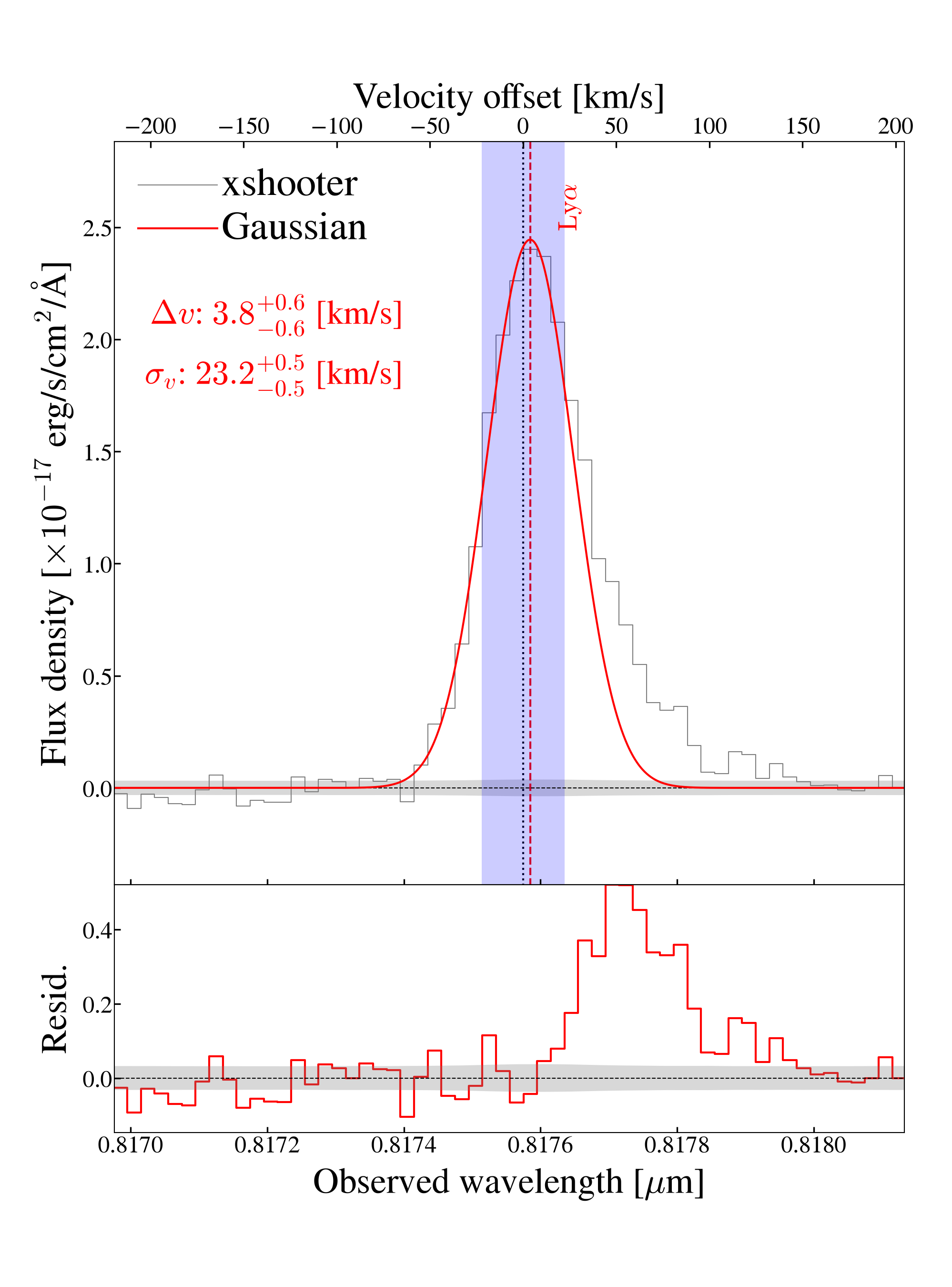}
	\includegraphics[width=0.49\textwidth]{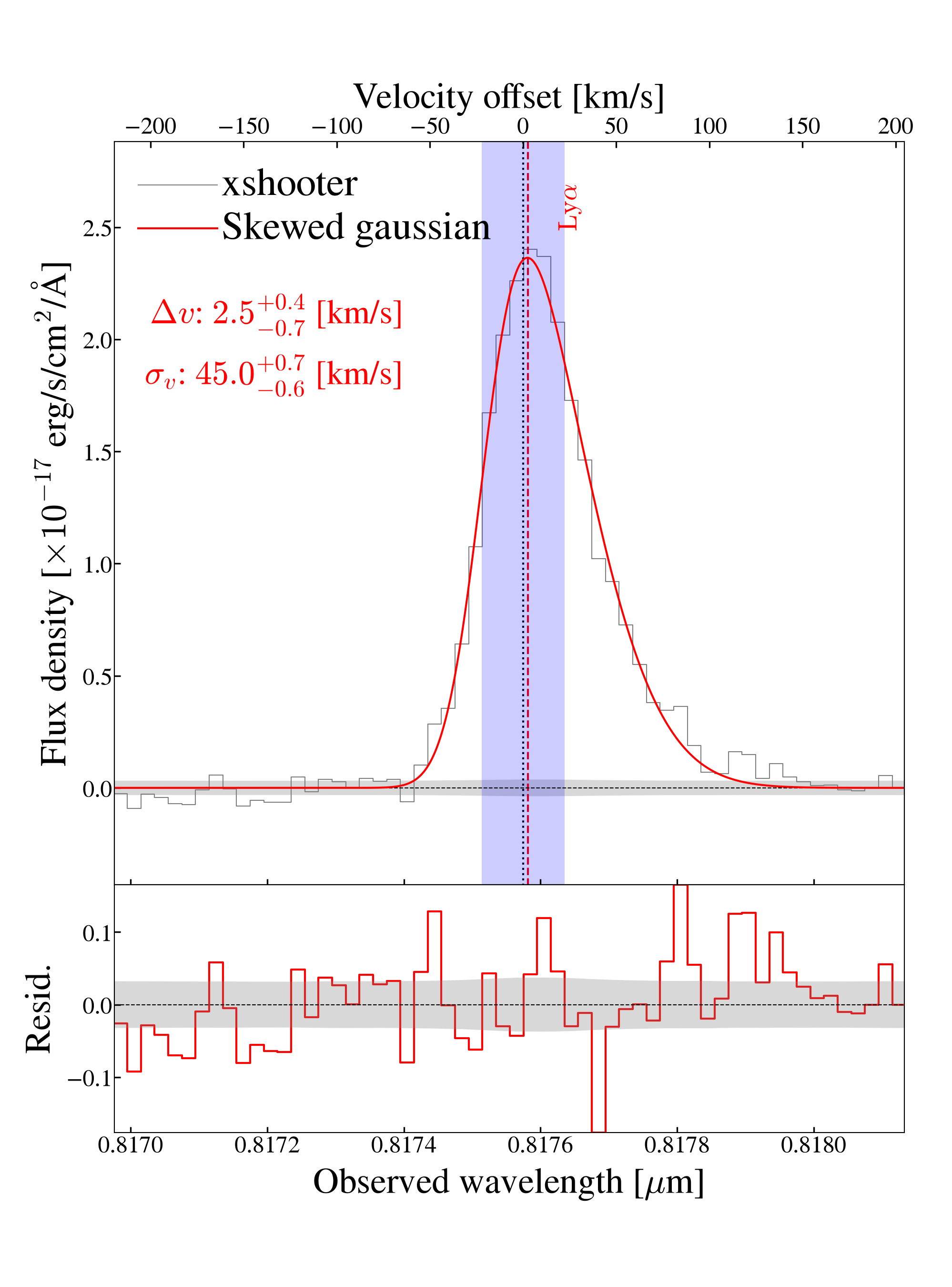}
	\caption{
    (Left): The fit to the one-dimensional \ly\ line (red solid line). The observed line here is the total flux (\id-A+B), derived via the optimal extraction. The model is fit only to the blue side of the observed line peak, resulting in enhanced residual emission on the red side. The upper horizontal axis shows the velocity offset relative to $z_{\rm H\beta}$. The vertical hatched region (blue) indicates the uncertainty range of $z_{\rm H\beta}$. The bottom panels show the residual between the data and model.
    (Right): Same as the left panel, but the model here is a skewed Gaussian. The model line center is indicated by a red dashed vertical line.
    }
\label{fig:lya}
\end{figure*}

%
\section{X-shooter observations}\label{sec:data}

X-shooter observations (PI: T. Morishita; DDT Program ID: 115.29G6.001) were carried out during August and September 2025. The slit width was set to $0.\!''9$.  In this paper, we focus on the visible (VIS) and near-infrared (NIR) arms, which cover \ly\ and other rest-frame UV emission lines. The adopted slit width corresponds to a spectral resolution of
$R \sim 8900$ in the VIS arm and $R \sim 5400$ in the NIR arm. The observations are summarized in Appendix~\ref{sec:log}. 

We use the ESO Science Data Products (Level 2) available from the archive. We note that {five OBs were excluded from the analysis because they did not pass the quality assessment. These include three OBs with poor seeing ($>0.\!''9$), resulting in increased slit losses. As the seeing in the retained OBs was comparable to or better than the slit width, the effective spectral resolution for the compact source may be higher than the nominal slit-limited resolution (see Sec.~\ref{ssec:lya}). The resulting on-source integration time is 7.7\,hrs.}

The \ly\ emission line is clearly detected in each OB with S/N $>15$. We therefore perform an optimal extraction of the one-dimensional spectrum for each OB using the cross-dispersion light profile of the \ly\ emission, measured within $\pm7.5$\,\AA\ of the line center. To extract the spectra of the two lensed images (\ida\ and \idb), whose separation is $\sim 1.\!''2$, we model the cross-dispersion profile in each OB with a two-component Gaussian function. We then extract the one-dimensional spectrum of each image using the corresponding Gaussian components (Fig.~\ref{fig:lya2d}).

The extracted one-dimensional spectra from all OBs are combined using inverse-variance weighting. The resulting spectrum is converted from air wavelengths to vacuum wavelengths and corrected for the barycentric radial velocity, ensuring consistency with the NIRCam-WFSS spectrum reported in \citet{morishita25}. We adopt the flux calibration provided by the ESO pipeline products.

In the following spectroscopic analysis in Sec.~\ref{sec:uv}, we adopt the flux and upper limits obtained from the total (\id-A and -B combined) spectrum, to maximize the signal-to-noise ratio. In the photometric analysis in Sec.~\ref{sec:sedmodels}, we compare the obtained SED parameters with the flux measurement of \id-B.

\begin{figure*}
\centering
	\includegraphics[width=0.245\textwidth]{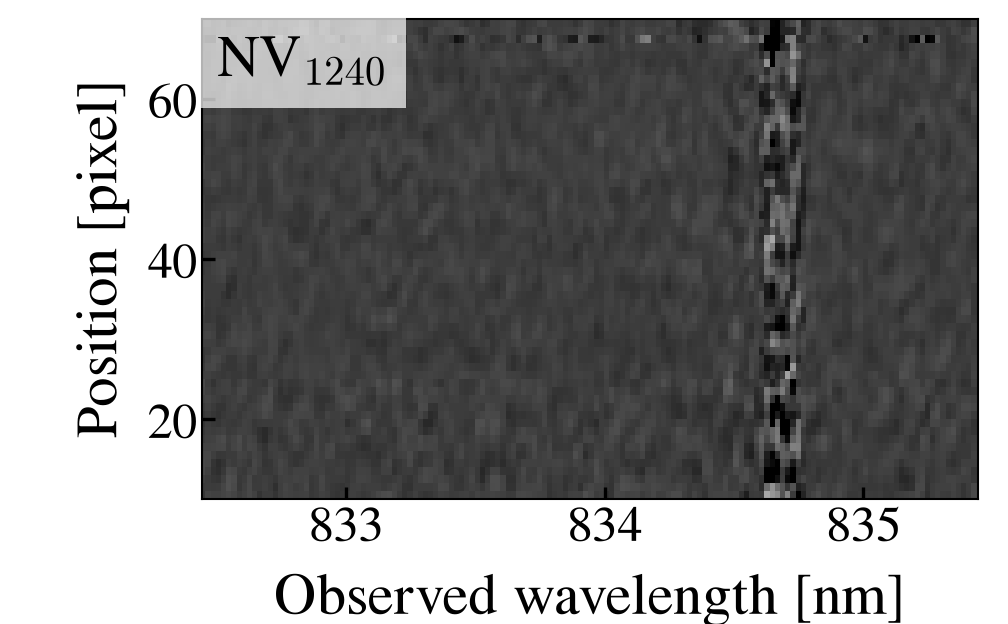}
	\includegraphics[width=0.245\textwidth]{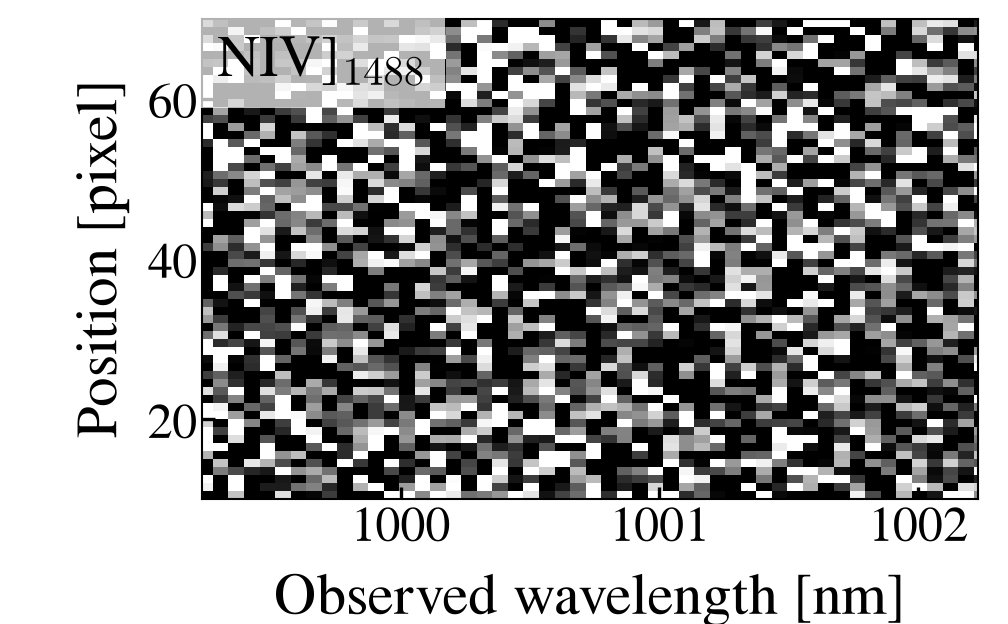}
	\includegraphics[width=0.245\textwidth]{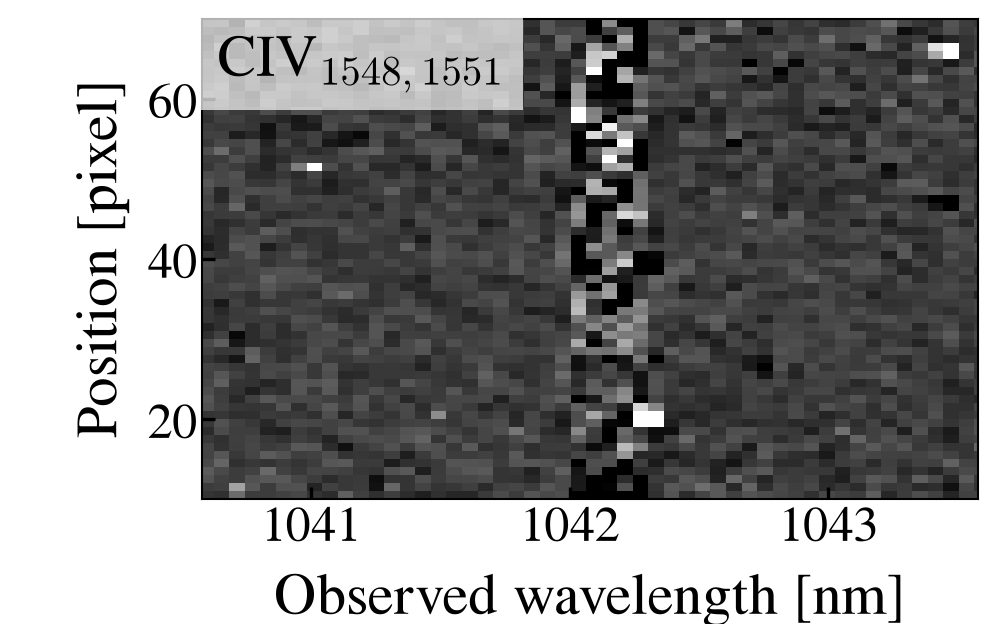}
	\includegraphics[width=0.245\textwidth]{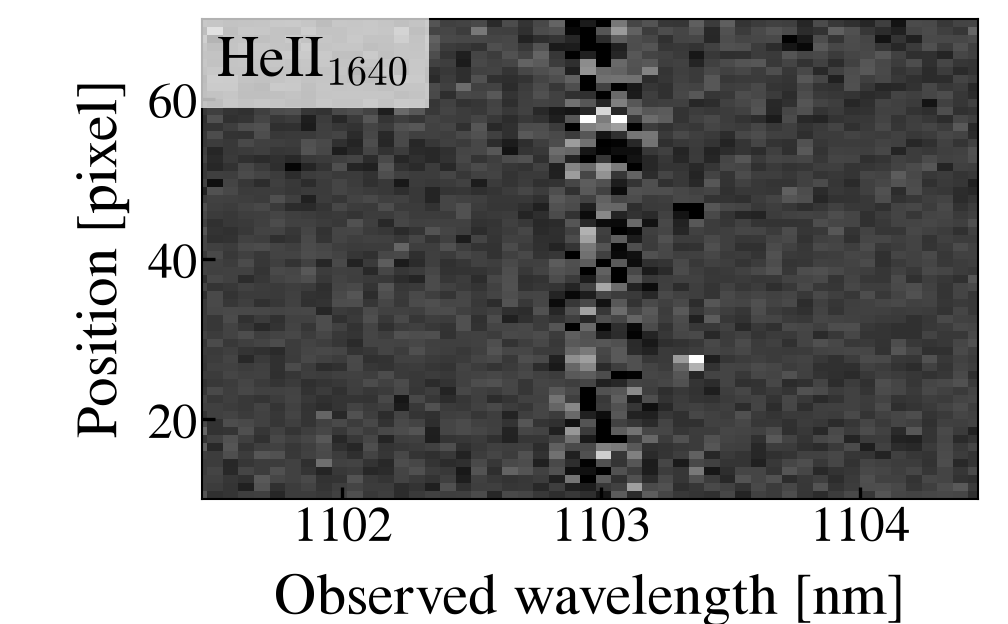}
	\includegraphics[width=0.245\textwidth]{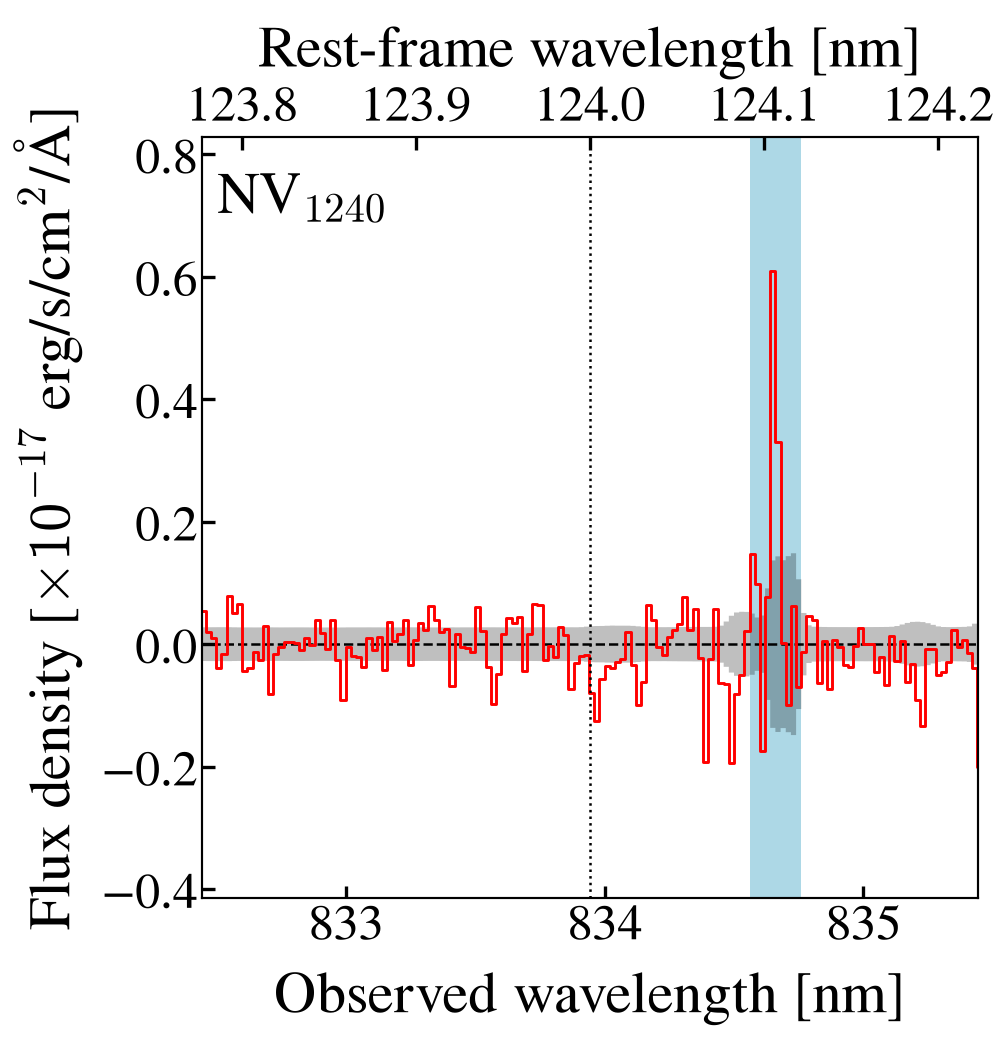}
	\includegraphics[width=0.245\textwidth]{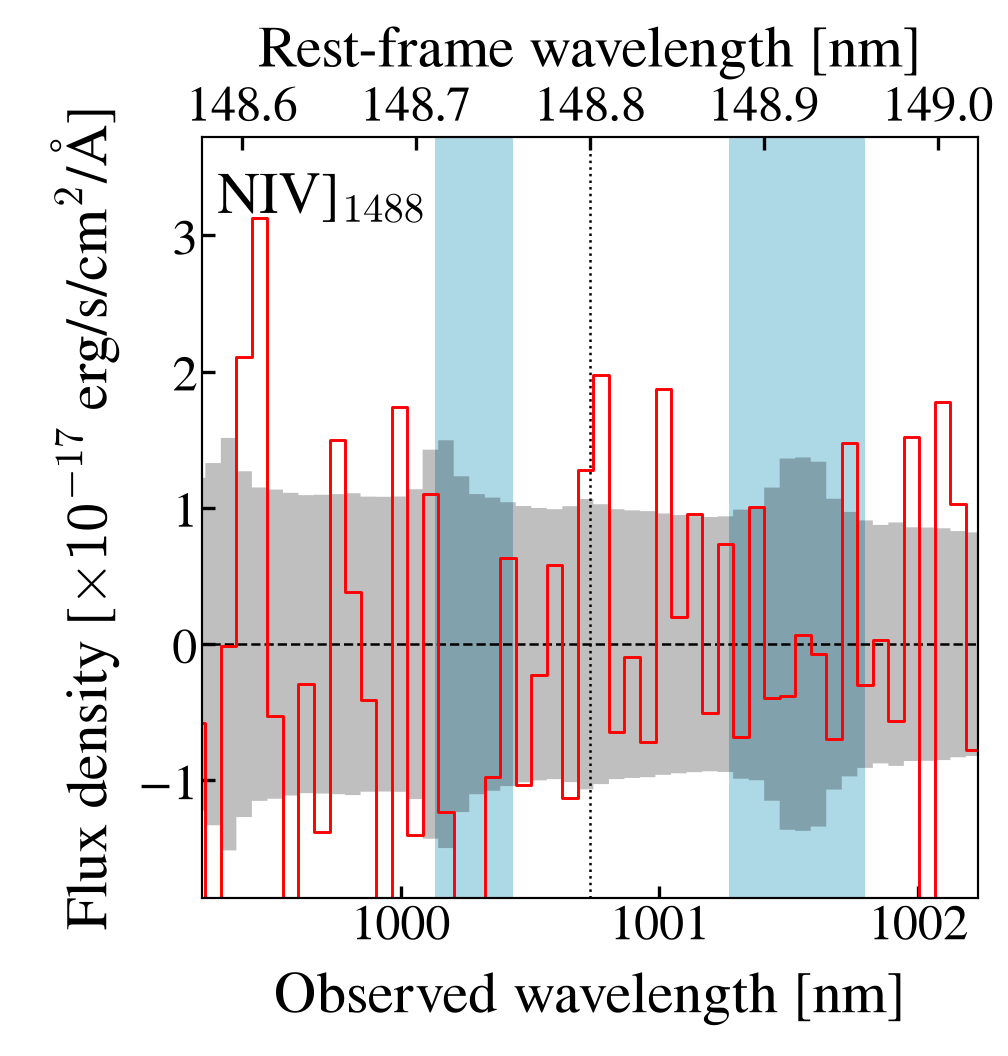}
	\includegraphics[width=0.245\textwidth]{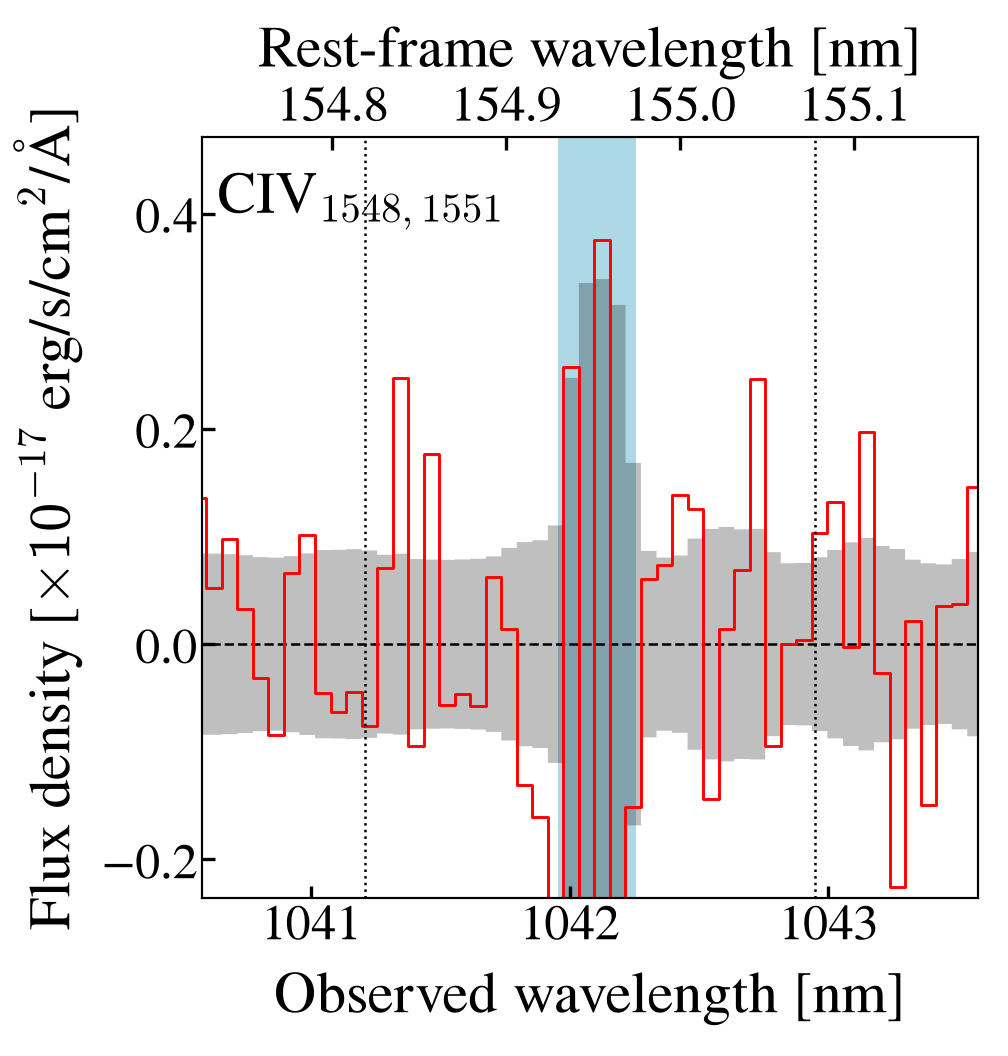}
	\includegraphics[width=0.245\textwidth]{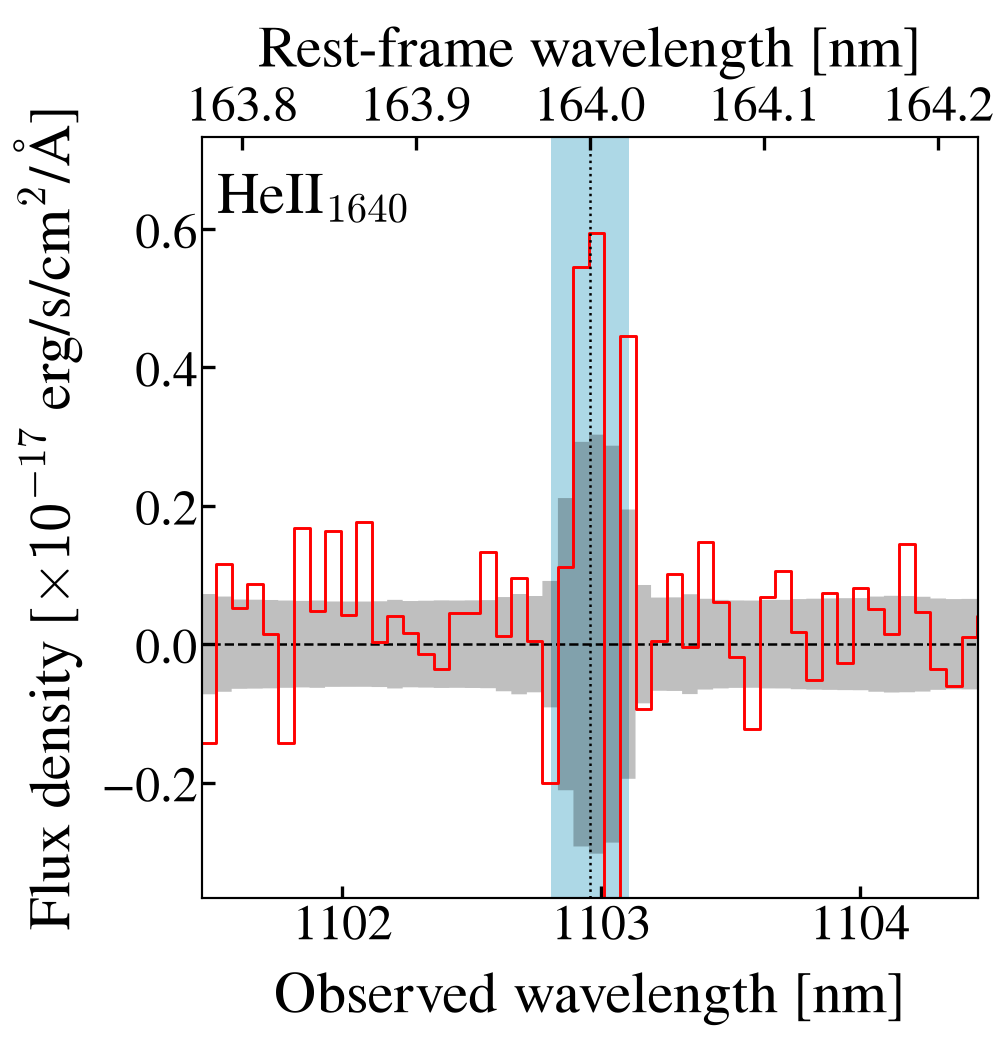}
	\includegraphics[width=0.246\textwidth]{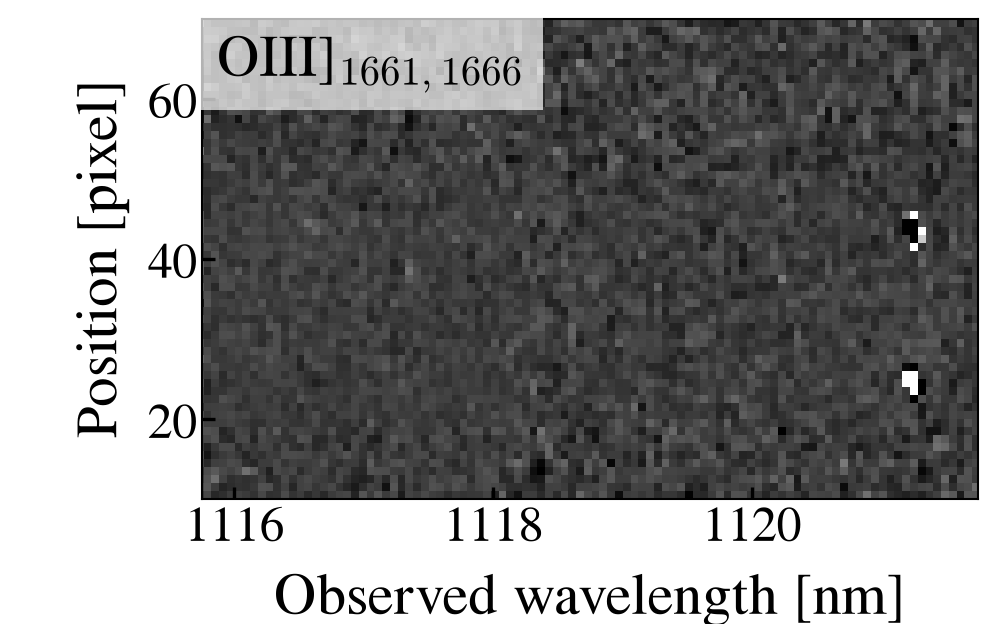}
	\includegraphics[width=0.246\textwidth]{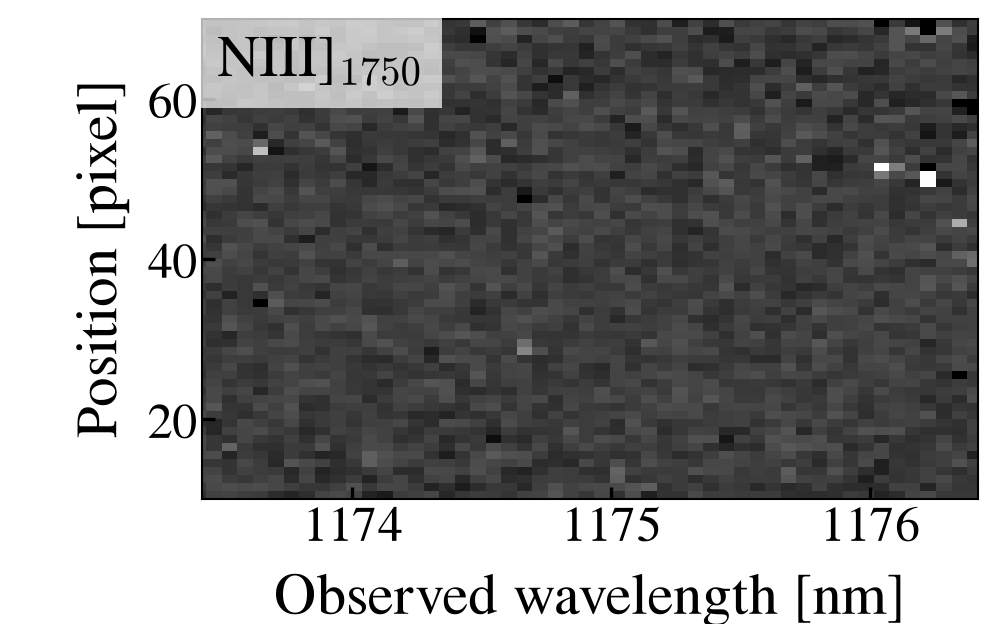}
	\includegraphics[width=0.246\textwidth]{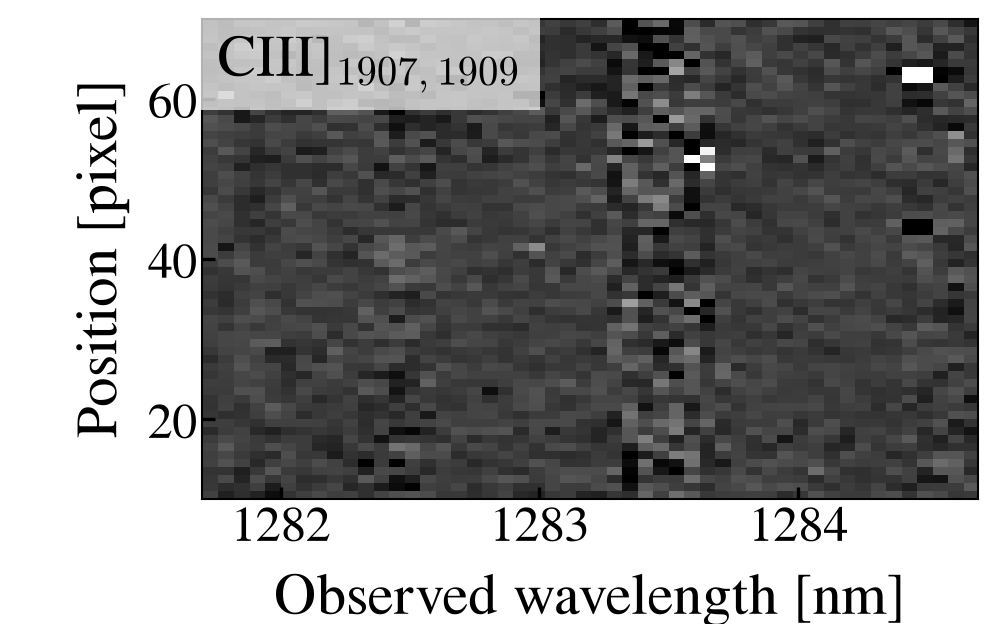}
	\includegraphics[width=0.246\textwidth]{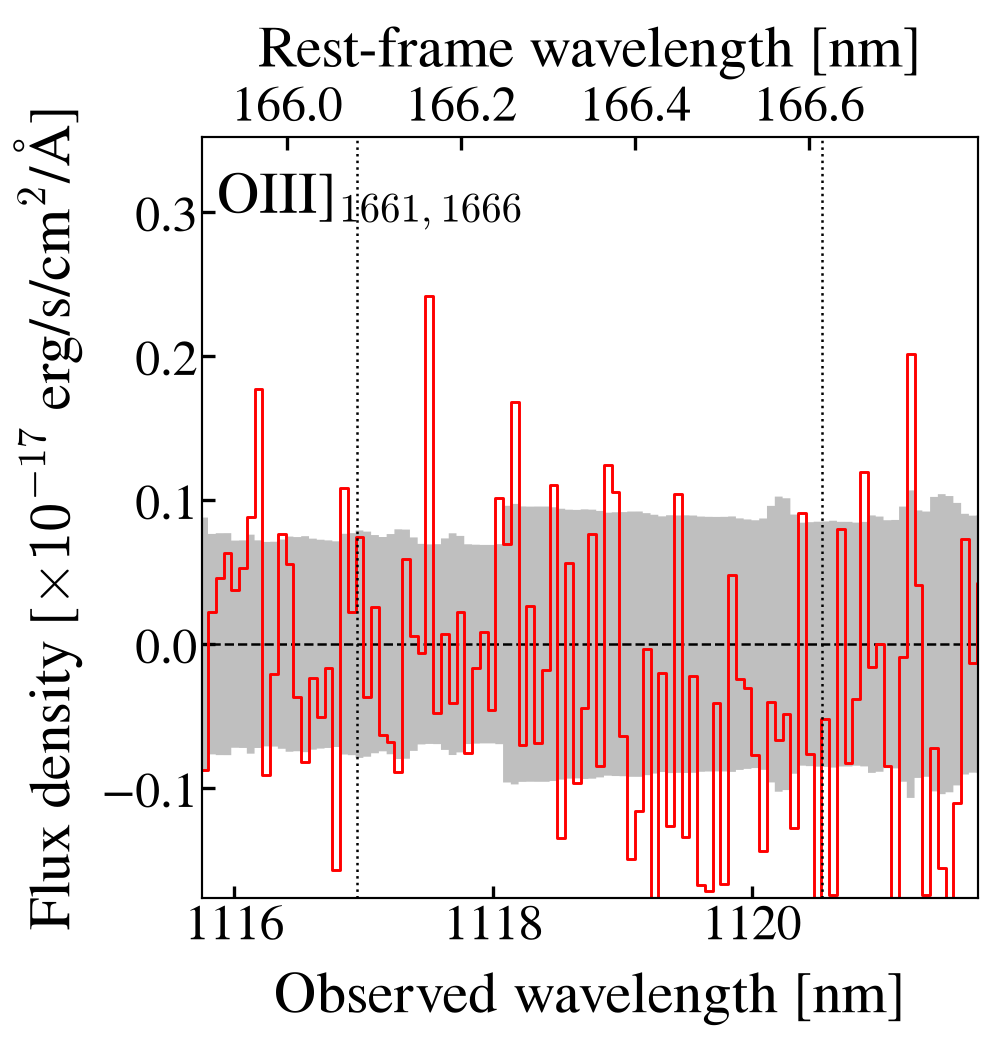}
	\includegraphics[width=0.246\textwidth]{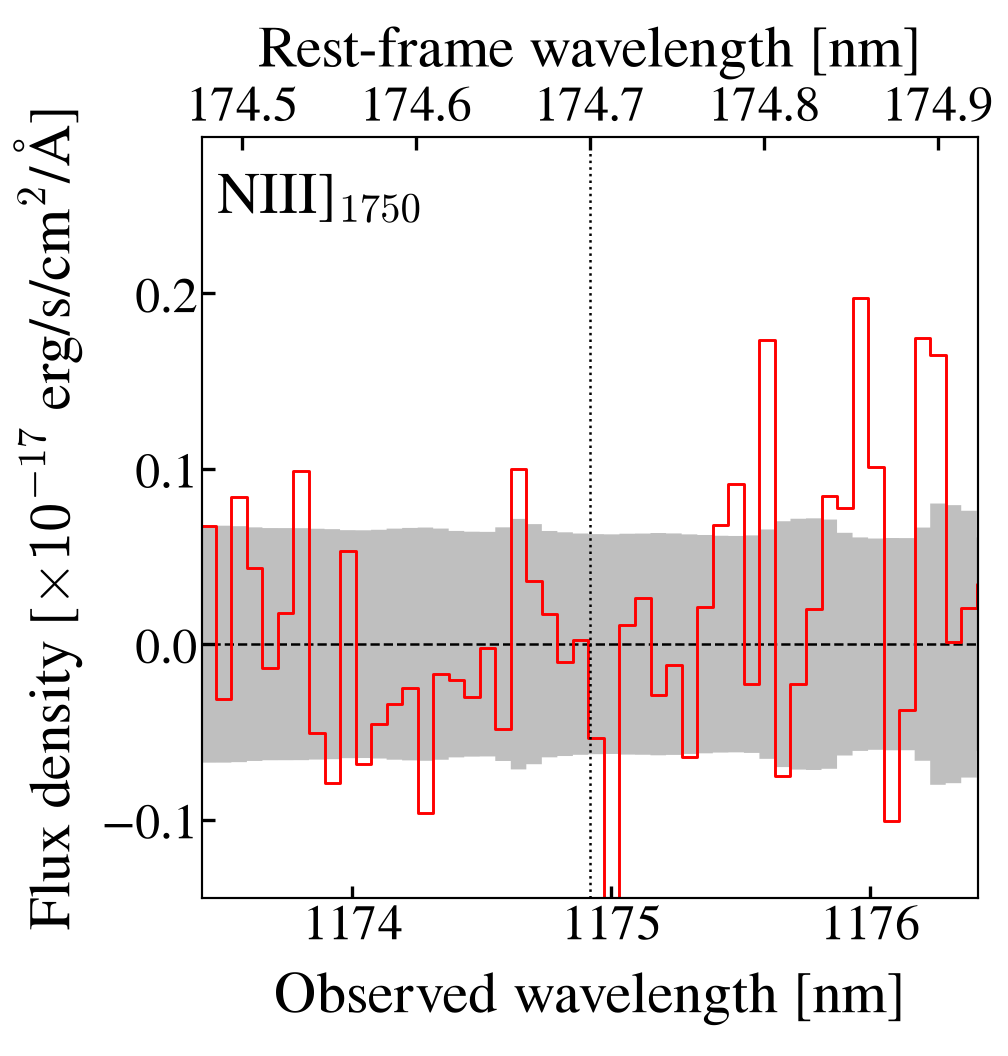}
	\includegraphics[width=0.246\textwidth]{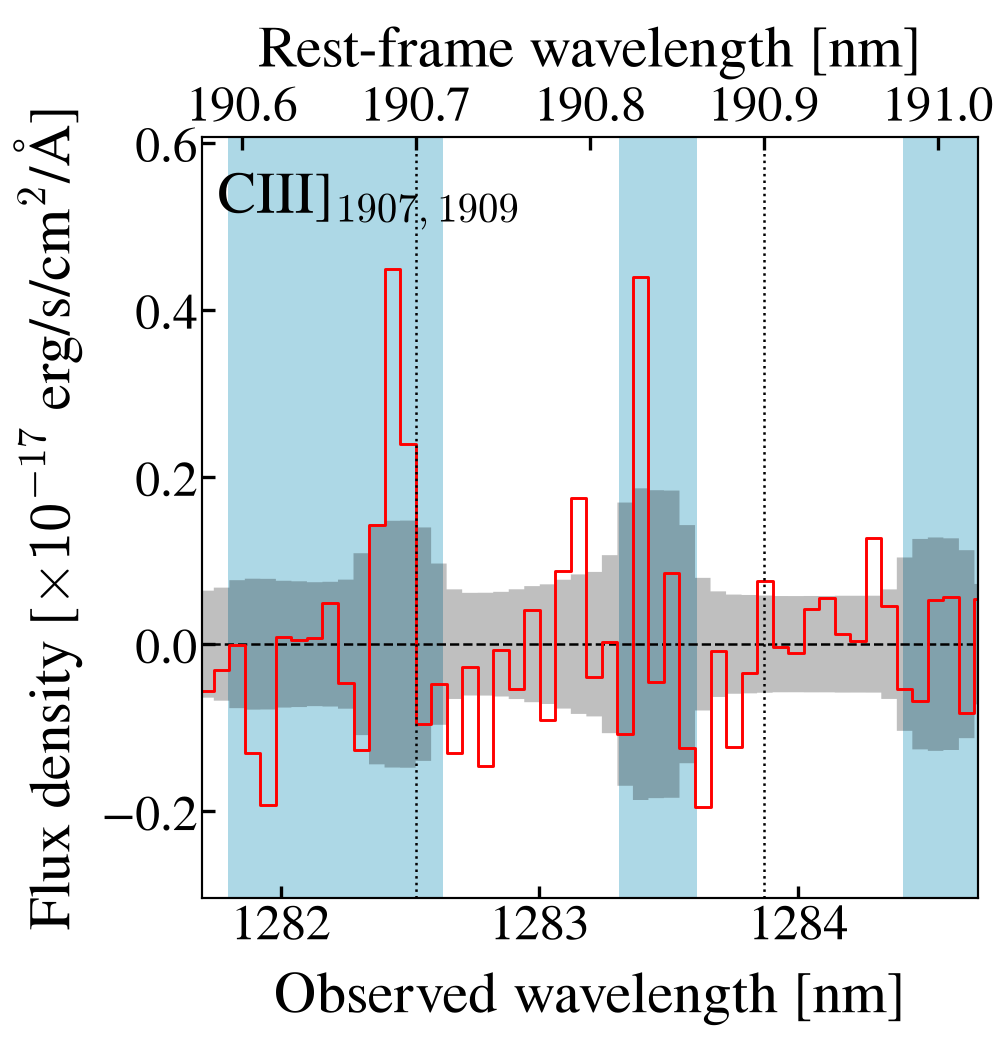}
	\caption{
    Close-up views of X-shooter {two-dimensional (top)} and one-dimensional spectra (bottom) for rest-frame UV lines. The expected wavelength of each UV line is indicated with a vertical dotted line. Spectra are shown with red solid lines, and uncertainties ($1\,\sigma$) with gray hatch. Vertical blue hatches indicate bright sky OH lines.
    }
\label{fig:uvlines}
\end{figure*}

\begin{deluxetable*}{lccccccc}
\tablecaption{X-shooter UV line fluxes and rest-frame equivalent widths of AMORE6\label{tab:uvline_limits}}
\tablehead{
\colhead{} &\colhead{} & \multicolumn{2}{c}{\id-A+B combined} & \multicolumn{2}{c}{\id-A} & \multicolumn{2}{c}{\id-B}\\
\cline{3-4} \cline{5-6} \cline{7-8}
\colhead{Line} &
\colhead{$\lambda_{\rm obs}$} &
\colhead{Line flux} &
\colhead{EW$_0$} &
\colhead{Line flux} &
\colhead{EW$_0$} &
\colhead{Line flux} &
\colhead{EW$_0$}  \\
\colhead{} &
\colhead{[\AA]} &
\colhead{[$10^{-17}$ erg s$^{-1}$ cm$^{-2}$]} &
\colhead{[\AA]} &
\colhead{[$10^{-17}$ erg s$^{-1}$ cm$^{-2}$]} &
\colhead{[\AA]} &
\colhead{[$10^{-17}$ erg s$^{-1}$ cm$^{-2}$]} &
\colhead{[\AA]}
}
\startdata
\ly\ & 8176 & $4.59 \pm 0.03$ & $186.9 \pm 1.3$ & $1.62 \pm 0.02$ & $196.7 \pm 2.6$ & $2.64 \pm 0.02$ & $161.7 \pm 1.5$ \\
NV$_{1240}$ & 8339 & $<0.11$ & $<9.4$ & $<0.07$ & $<18.2$ & $<0.08$ & $<10.2$ \\
NIV]$_{1488}$ & 10007 & $<4.44$ & $<298.9$ & $<3.83$ & $<767.1$ & $<2.98$ & $<302.2$ \\
CIV$_{1548}$ & 10412 & $<0.38$ & $<28.9$ & $<0.28$ & $<62.9$ & $<0.27$ & $<30.3$ \\
CIV$_{1551}$ & 10429 & $<0.37$ & $<27.9$ & $<0.26$ & $<59.0$ & $<0.26$ & $<29.6$ \\
HeII$_{1640}$ & 11030 & $<1.17$ & $<105.3$ & $<0.83$ & $<221.6$ & $<0.82$ & $<111.1$ \\
OIII]$_{1661}$ & 11170 & $<0.35$ & $<33.1$ & $<0.26$ & $<71.8$ & $<0.24$ & $<34.5$ \\
OIII]$_{1606}$ & 11205 & $<0.39$ & $<37.0$ & $<0.28$ & $<79.7$ & $<0.27$ & $<39.1$ \\
NIII]$_{1750}$ & 11749 & $<0.30$ & $<32.8$ & $<0.22$ & $<70.7$ & $<0.21$ & $<34.5$ \\
CIII]$_{1907}$ & 12825 & $<0.62$ & $<89.8$ & $<0.45$ & $<191.3$ & $<0.44$ & $<95.3$ \\
CIII]$_{1909}$ & 12839 & $<0.30$ & $<43.0$ & $<0.23$ & $<101.0$ & $<0.20$ & $<44.6$
\enddata
\tablecomments{Measurements are for the total spectrum of \id-A+-B combined, \id-A, and \id-B. The reported line fluxes and upper limits are not corrected for magnification. Quoted uncertainties are $1\,\sigma$. Upper limits are $3\,\sigma$. For flux upper limits, we assume a Gaussian line profile with $\sigma_v=23.2$\,km\,s$^{-1}$.
}
\end{deluxetable*}\label{tab:linefluxes}


\begin{figure*}
\centering
	\includegraphics[width=0.8\textwidth]{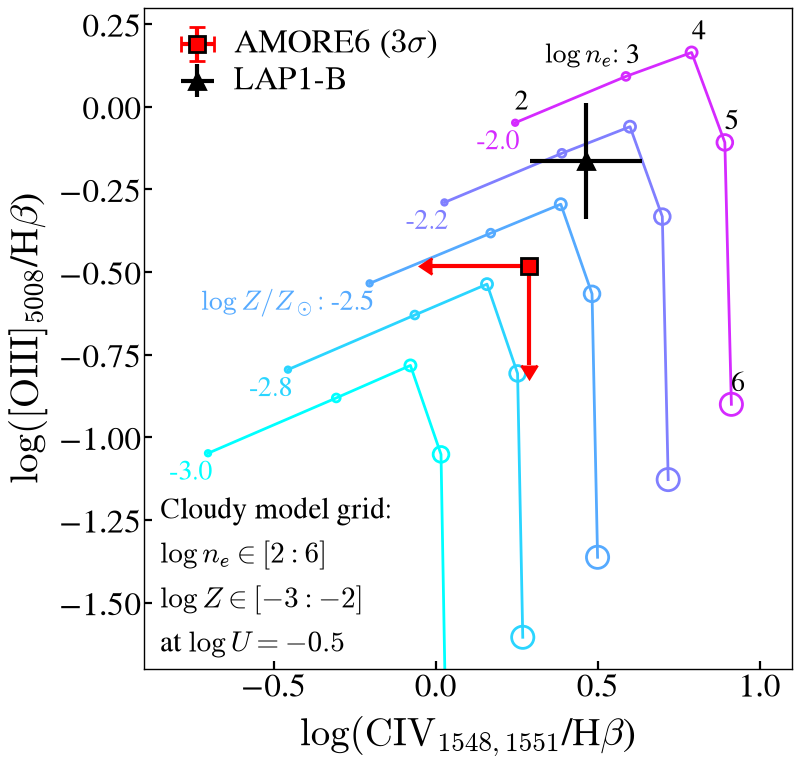}
	\caption{
    The line diagnostic of \id, in the \civ/\hb-\oiii/\hb\ space. $3\,\sigma$ upper limits are quoted. For comparison, we show \cl\ photoionization models for different physical parameters --- metallicity $\log Z_{\rm gas}/Z_\odot \in [-3:-2]$ (lines with different colors) and electron density $\log n_e/{\rm cm^{-3}} \in [2:6]$ (circles, size increasing with the density) calculated with $\log U=-0.5$. For comparison, the measurement for LAP1-B is shown \citep[black triangle,][]{nakajima25}. {Model tracks with different $\log U$, (C/O), and $\logZ$ are shown in Appendix~\ref{sec:logu}.}
    }
\label{fig:c4}
\end{figure*}

%
\section{Rest-frame UV line analysis}\label{sec:uv}

\subsection{Lyman-$\alpha$}\label{ssec:lya}

\id\ is a strong \ly\ emitter, originally discovered with MUSE \citep{mahler18,richard21}. As previously reported by \citet{messa26}, the higher resolving power of X-shooter ($R\sim8900$) reveals a remarkably narrow line profile with little excess flux on the red side (cf. $R\sim3000$ for MUSE). In addition, the line peak coincides with the systemic redshift measured from \hb. Based on these properties, \citet{messa26} argued that \id\ may exhibit a high escape fraction of \ly\ and LyC photons. Motivated by these unique spectral properties, we perform a detailed characterization of the Ly$\alpha$ profile below.


We characterize the line profile in two ways. First, we fit only the blue half of the line with a single Gaussian. By excluding the red side of the profile, which may be affected by processes such as resonant scattering in the surrounding neutral gas, we obtain an observed width $\sigma_{v}=23.2\pm0.5$\,km\,s$^{-1}$, before correcting for instrumental broadening (Fig.~\ref{fig:lya}). This measurement closely reflects the intrinsic gas dynamics within the star-forming regions of \id.

Second, we fit the line with a skewed Gaussian by introducing a skewness parameter $s$:
\begin{equation}
f(\lambda) = \exp\left(-\frac{(\lambda-\lambda_0)^2}{2\sigma_v^2}\right)
\left[
1 + \mathrm{erf}\left(
\frac{s(\lambda-\lambda_0)}
{\sqrt{2}\sigma_v}
\right)
\right],
\end{equation}
where $\mathrm{erf}$ is the error function. While the fit provides a reasonably good representation of the overall line shape, we observe a faint excess on the red side of the line. The line width is $\sigma_v=45.0_{-0.6}^{+0.7}$\,km\,s$^{-1}$.

Remarkably, the line peak is at $+3.8$\,km\,s$^{-1}$ and $+4.6$\,km\,s$^{-1}$ from the systemic redshift determined from \hb\ for the symmetric and asymmetric fits, respectively. Given the large uncertainty in the \hb\ line centroid, $\sim\pm25$\,km\,s$^{-1}$, arising from the slitless nature of the observations \citep{morishita25}, the measured offset is consistent with zero.

{We then correct the line width for the instrumental resolution through
\begin{equation}
\sigma_{v, int} = \sqrt{\sigma_{v}^2 - \sigma_{lsf}^2},
\end{equation}
where we adopt an instrumental line spread function $\sigma_{lsf}$. For the nominal resolving power of $R=8900$ at the wavelength of \ly, this corresponds to $14.3$\,km\,s$^{-1}$. Because the seeing in the retained OBs is comparable to or better than the $0.\!''9$ slit width, the effective instrumental broadening of the compact source may be smaller than the nominal slit-limited value. The deconvolved width therefore represents a conservative lower bound under this assumption. 
With this correction, we obtain $\sigma_{v, int}=\sigmasmint$\,km\,s$^{-1}$ and $\sigmaasmint$\,km\,s$^{-1}$ for the symmetric and asymmetric fits, respectively.
We note that this correction accounts only for instrumental broadening; thermal broadening and resonant scattering effects remain in the inferred line width (see Sec.~\ref{sec:atomic}).
}

To measure the line flux, we directly integrate the spectrum over the wavelength range 817.2--818.2\,nm. The width of the integration window corresponds to approximately $\pm200$\,km\,s$^{-1}$ from the line center. The line flux is reported in Table~\ref{tab:linefluxes}.

\subsection{Other UV lines}\label{ssec:uv}
The X-shooter wavelength coverage includes several rest-frame UV emission lines, including \nv\ in the VIS arm, and \oiiiuv, \niv, \niii, \heii, \civ, and \ciii\ in the NIR arm. Figure~\ref{fig:uvlines} shows zoomed-in views of the spectral regions surrounding these transitions. Among them, \heii\ and \ciii$_{1907}$ coincide with strong sky emission features and {show relatively large uncertainties;} these lines are therefore not useful for constraining the physical properties of \id. The remaining transitions are free from sky contamination; however, none are detected in our spectrum. We derive $3\sigma$ upper limits assuming the \ly\ line width obtained from the symmetric Gaussian fit. The measured upper limits are reported in Table~\ref{tab:linefluxes}.

The non-detection of metal emission lines in the deep X-shooter spectrum supports the previous claim that \id\ resides in a very low-metallicity environment. To quantify this constraint, we investigate UV line-to-\hb\ ratios. Among the accessible UV lines, CIV provides the strongest constraint because of its sensitivity to both metallicity and electron density. We show the \civ-to-\hb\ line ratio in Figure~\ref{fig:c4} and compare with photoionization model grids calculated using {\tt Cloudy} {(ver.25.00)} \cite{1998PASP..110..761F,chatzikos23} (see Sec.~\ref{sec:sed-pop3} for the detailed prescription). Here we adopt a nebular model grid with ionization parameter $\log U=-0.5$ irradiated by an instantaneous burst of metal-free stars, while exploring a range of electron densities and gas metallicities. 
The adopted ionization parameter is identical to that used by \citet{morishita25} in their determination of the oxygen abundance (also by \citealt{nakajima25} for LAP1-B), thereby providing a consistent framework for comparison with the \oiii-to-\hb\ line ratio. 

{Given the absence of metal lines, the ionization parameter of \id\ remains observationally undetermined. The inferred chemical abundances may therefore depend sensitively on the adopted ionization parameter. Recent studies of high redshift galaxies show variations in the ionization parameters. For example, \citet{2026arXiv260611345I} analyzed a large sample of galaxies at $1<z<10$ and found that lower ionization parameters, $\log U\sim-1.5$ to $-1$, were favored. However, we note $(i)$ their sample predominantly spans $7<\logm<9.5$ and the oxygen abundance range of $7<12+$\,\logoh\,$<8$, substantially higher stellar masses and oxygen abundances than \id, and $(ii)$ their new calibration between \logoh\ and $\log U$ shows a strong negative correlation, suggesting that $\log U$ could be even larger at lower \logoh. As such, we proceed with $\log U=-0.5$ in the following analysis, while noting the uncertainty associated with this assumption.}

The analysis of \citet{morishita25} left open the possibility that an extremely high gas density could suppress the \oiii\ emission, leading to an underestimate of the metallicity when inferred solely from the \oiii-to-\hb\ ratio. The new upper limit on the \civ-to-\hb\ ratio obtained here excludes this high-density scenario ($n_e > 10^5$\,cm$^{-3}$) for moderately enriched gas ($\log Z_{\rm gas}/Z_\odot > -2.5$), regardless of the choice of radiation sources assumed in this study. {Adopting a lower $\log U$ would relax the constraint to $\log Z_{\rm gas}/Z_\odot \sim -2$, but the conclusion remains qualitatively consistent. We present model grids spanning different ionization parameters, enhanced carbon-to-oxygen abundance ratios (C/O), and different ionizing-source metallicities in Appendix~\ref{sec:logu}.}

The upper limits from the other UV emission lines do not provide meaningful constraints on the nebular properties of \id. {For example, the upper limit of the \heii\ equivalent width from the X-shooter spectrum is $<110$\,\AA, whereas predicted values for Pop~III environments calculated with photo-ionization models (Sec.~\ref{sec:sedresults}) range at much lower values $<10$\,\AA.}

\input{table_sed}

\input{sedfitting}

\section{Discussion}\label{sec:disc}

\subsection{Dynamical properties from \ly}\label{ssec:mdyn}
One of the defining properties of \id\ is its low mass, comparable to those of local dwarf galaxies and massive globular clusters. Indeed, our X-shooter spectrum reveals an exceptionally narrow \ly\ line (Sec.~\ref{ssec:lya}). 
We estimate the dynamical mass of the system using this velocity information and compare it with the stellar mass inferred from the SED modeling.
Adopting the intrinsic line width $\sigma_{v,int}=\sigmasmint$\,km\,s$^{-1}$ (measured at the blue side of \ly) and the effective radius $R_e=36_{-9}^{+8}$\,pc in the source plane \citep{morishita25}, the dynamical mass can be estimated as
\begin{equation}
M_{\rm dyn}
=
K(n)\,K(q)\,
\frac{\sigma_{v,int}^2 R}{G}
=
\mvir,
\end{equation}
where $K(n)$ and $K(q)$ are virial correction factors that account for the measured S\'ersic index ($n=1.7$) and axis ratio ($q=0.6$), respectively \citep{vanderwel22}. 

{
The blue side of \ly, however, is subject to attenuation, and thus the measured line width may underestimate the width of the intrinsic emission profile. Adopting the line width $\sigma_{v,int}=\sigmaasmint$\,km\,s$^{-1}$ (measured using the full line shape of \ly) yields $M_{\rm dyn}=\mvirr$.

Another caveat is that the \ly\ line width may contain contributions from thermal motions and resonant scattering, and therefore does not necessarily provide a direct measurement of the gravitational velocity dispersion, while \citet{messa26} suggests an extremely low hydrogen column density, $\log({\rm N_{HI}})\simlt14$, from a detailed line profile analysis. In particular, the thermal velocity dispersion of hydrogen is $\sim14$\,km\,s$^{-1}$ at $T = 2.5 \times 10^4$\,K. Further subtracting this component yields $M_{\rm dyn}=\mvirrther$. As such, the dynamical mass of \id\ may lie between these estimates, although neither necessarily traces the gravitational velocity dispersion. 
In the following discussion, we adopt the median value, while the difference is included in the uncertainty, such that $M_{\rm dyn}=\mvirmed$. 
}

{Given that the dynamical mass is estimated within a small radius, it primarily reflects the baryonic mass, $M_{\rm baryon} \approx M_{\rm dyn}$. The contribution from dark matter within such a small radius is negligible. For example, assuming an NFW dark matter halo with $M_{200}=10^8$--$10^{10}\,M_\odot$ and $c_{200}=3$--8, the predicted dark matter mass enclosed within the effective radius of AMORE6 is approximately $2\times10^5$--$3\times10^6\,M_\odot$. This corresponds to approximately $0.3$--$5.5\,\%$ of the nominal \ly-based dynamical mass. The result suggests that the enclosed mass, if reliably inferred from gravitational motions, would be baryon-dominated.}

By adopting this baryonic mass, and the stellar mass inferred from the SED fitting, we estimate gas-to-baryon mass fraction:
\begin{equation}
\frac{M_{\rm gas}}{M_{\rm baryon}}
=
1 - \frac{M_{*}}{M_{\rm baryon}}
=
\fgas,
\end{equation}
indicating that the majority of the baryonic mass is still in the form of gas rather than stars.

Another important characteristic is the dynamical timescale. Using the effective radius and the observed velocity dispersion, we estimate
\begin{equation}
    t_{\rm dyn} \sim \frac{R_e}{\sigma_{v,int}} \approx 1~{\rm Myr},
\end{equation}
although we note that this estimate combines the stellar effective radius with the velocity dispersion measured from the ionized gas. Nevertheless, the inferred dynamical timescale is comparable to the estimated stellar population age of \id\ \citep{morishita25}. This suggests that \id\ is being observed at an exceptionally early stage of its evolution, with its dynamical properties still reflecting those established during its recent episode of star formation. Such a short dynamical age is also consistent with the interpretation that the current starburst represents the first major episode of star formation in the system.



\subsection{Counting stars}\label{sec:count}

One of the most important questions regarding \id\ is whether its stellar population is dominated by Pop~III stars. Our SED modeling, which explores a wide range of stellar population models including Pop~IIs, yields comparably good fits to the available photometric data. As a result, the current observations alone cannot conclusively determine the nature of the underlying stellar population. 

An alternative approach is to estimate the number of stars required to reproduce the observed luminosity of \id. While such estimates are necessarily model-dependent, they provide useful insight into the scale of the stellar population.



We can estimate the number of stars using the observed recombination-line luminosity, following the approach of \citet{visbal25}. The stellar mass required to reproduce the observed hydrogen recombination luminosity can be expressed as
\begin{equation}
M =
M_{\rm PopIII}\frac{
L_{\rm H\alpha}
}{
(1-f_{\rm esc})\,Q({\rm H})\,c_{\rm H\alpha}
},
\end{equation}
where $M_{\rm PopIII}$ is the mass of an individual Pop~III star, $f_{\rm esc}$ is the escape fraction of ionizing photons, $Q({\rm H})$ is the hydrogen-ionizing photon production rate of the star, and $c_{\rm H\alpha}=1.21\times10^{-12}$\,erg is the energy conversion factor for H$\alpha$ emission.

Using the lensing-corrected luminosity $L_{\rm H\beta}=2.8\times10^6\,L_\odot$, the Case B conversion factor H$\alpha$/H$\beta=2.87$, and assuming $f_{\rm esc}=0.1$, we obtain $N=50$ for zero-age main-sequence (ZAMS) stars of $500\,M_\odot$, adopting $Q({\rm H})=7.38\times10^{50}$\,s$^{-1}$ from \citet{schaerer02}. The required number decreases to $N=23$ if $M_{\rm PopIII}=1000\,M_\odot$ (with $Q({\rm H})=1.607\times10^{51}$\,s$^{-1}$) is adopted. On the other hand, the inferred number increases rapidly toward lower stellar masses: $N=92$ for $M_{\rm PopIII}=300\,M_\odot$ ($Q({\rm H})=4.029\times10^{50}$\,s$^{-1}$) and $N=1490$ for $M_{\rm PopIII}=40\,M_\odot$ ($Q({\rm H})=2.469\times10^{49}$\,s$^{-1}$).

The analysis above suggests that, even under the most optimistic assumptions, the observed luminosity of \id\ requires tens to thousands of massive Pop~III stars. We have seen that the key factor is the upper limit of the Pop~III stellar mass, which remains uncertain. Theoretical studies predict stars of $\simgt1000\,M_\odot$ \citep[e.g.,][]{hirano14}, and as high as $10^5\,M_\odot$ \citep[e.g.,][]{2010MNRAS.402.1249S} --- in such cases, the required number of stars considerably decreases to $N\approx$\,a few to a few tens. Nevertheless, it is still unclear whether such massive stars can form in large numbers within a single star-forming event. 

\subsection{On the nature of \id}\label{sec:nature}

\id\ is an unusually strong Ly$\alpha$ emitter with a narrow line profile and a high inferred Ly$\alpha$ escape fraction, ranging from $\lyesclow$ to $\approx\lyeschigh$, depending on the assumed stellar population model (Sec.~\ref{ssec:lya}). The extremely blue UV continuum slope, strong nebular emission, large equivalent widths, and other photometric and spectroscopic properties all favor a very young stellar population. Taken together, these observations are qualitatively consistent with a recent burst of star formation occurring within the past few Myr, potentially involving the nearly simultaneous formation of a large population of massive stars. {Below, we consider two possible scenarios to explain the nature of \id.}

\subsubsection{Is \id\ hosted in an atomic-cooling halo?}\label{sec:atomic}

{One possible explanation for the extraordinary properties of \id\ is a burst of Pop~III star formation in an atomic-cooling halo (Sec.~\ref{sec:intro}). At the redshift of \id, the critical halo mass for atomic cooling is $M_{\rm halo}\simeq7.7\times10^7\,M_\odot$ \citep{2024ApJ...962...49B}. Above this threshold, gas can cool through atomic hydrogen transitions even when molecular hydrogen cooling is strongly suppressed by LW radiation \citep{2008MNRAS.387.1021G,2024ApJ...962...49B}. The dynamical mass inferred from the \ly\ line width (Sec.~\ref{ssec:mdyn}) probes the mass enclosed within the compact emitting region, rather than the total virial mass of the host halo. Although the inferred enclosed mass is below the atomic-cooling threshold, the total halo mass may be substantially larger, making the atomic-cooling interpretation plausible.}

{This interpretation also offers a plausible explanation for a key question concerning \id: what could have delayed the onset of star formation until such a late cosmic epoch? Molecular hydrogen cooling can operate in lower-mass halos, but is susceptible to suppression by external LW radiation. Using the model of \citet{2023MNRAS.522..330I}, we estimate a mean LW background intensity of $J_{21}\sim17$ at $z=5.725$, obtained by modestly extrapolating their fitting relation calibrated over $6<z<23$. Here, $J_{21}$ denotes the specific intensity at the Lyman limit normalized to $10^{-21}\,{\rm erg\,s^{-1}\,cm^{-2}\,Hz^{-1}\,sr^{-1}}$. Adopting $J_{21}=17$ and the approximate molecular-cooling threshold from \citet{2001ApJ...548..509M},
\begin{equation}
    M_{\rm th} = \left(1.25\times10^5+
2.86\times10^6J_{21}^{0.47}\right)\,M_\odot,
\end{equation} 
yields a threshold halo mass of $\sim1.1\times10^7\,M_\odot$, below which H$_2$ cooling is suppressed by photodissociation. We caution that this estimate depends on the assumed radiation spectrum, gas chemistry, and treatment of molecular self-shielding. In particular, it does not necessarily imply that molecular cooling was suppressed until the atomic-cooling threshold was reached.}

LW feedback could be further enhanced by radiation from neighboring galaxies. In fact, \id\ is located in a relatively underdense region between two evolved galaxy overdensities at distances of $\sim3$\,pMpc \citep{morishita25a}. {Estimating the additional LW contribution from these galaxies is challenging because of uncertainties in their LW photon escape fractions, star formation histories, and three-dimensional separations from \id. Nevertheless, this additional irradiation may have further suppressed molecular hydrogen cooling in the progenitor halo, potentially delaying the onset of star formation until the halo became sufficiently massive for efficient atomic cooling. Such a scenario could provide a natural explanation for the exotic star-forming properties observed in \id.}

\subsubsection{{Is \id\ a gravitationally bound young star cluster?}}\label{ssec:cluster}

{The previous discussion assumes that \id\ is a galaxy residing in an individual dark matter halo. An alternative interpretation is that \id\ represents a gravitationally bound young star cluster rather than a galaxy undergoing star formation in a massive halo. A similar possibility has been discussed for compact, metal-poor systems. LAP1-B, with its extremely low mass $\simlt10^4\,M_\odot$, has been proposed as a candidate for such a cluster \citep{vanzella23}.}

\begin{figure*}
\centering
	\includegraphics[width=0.8\textwidth]{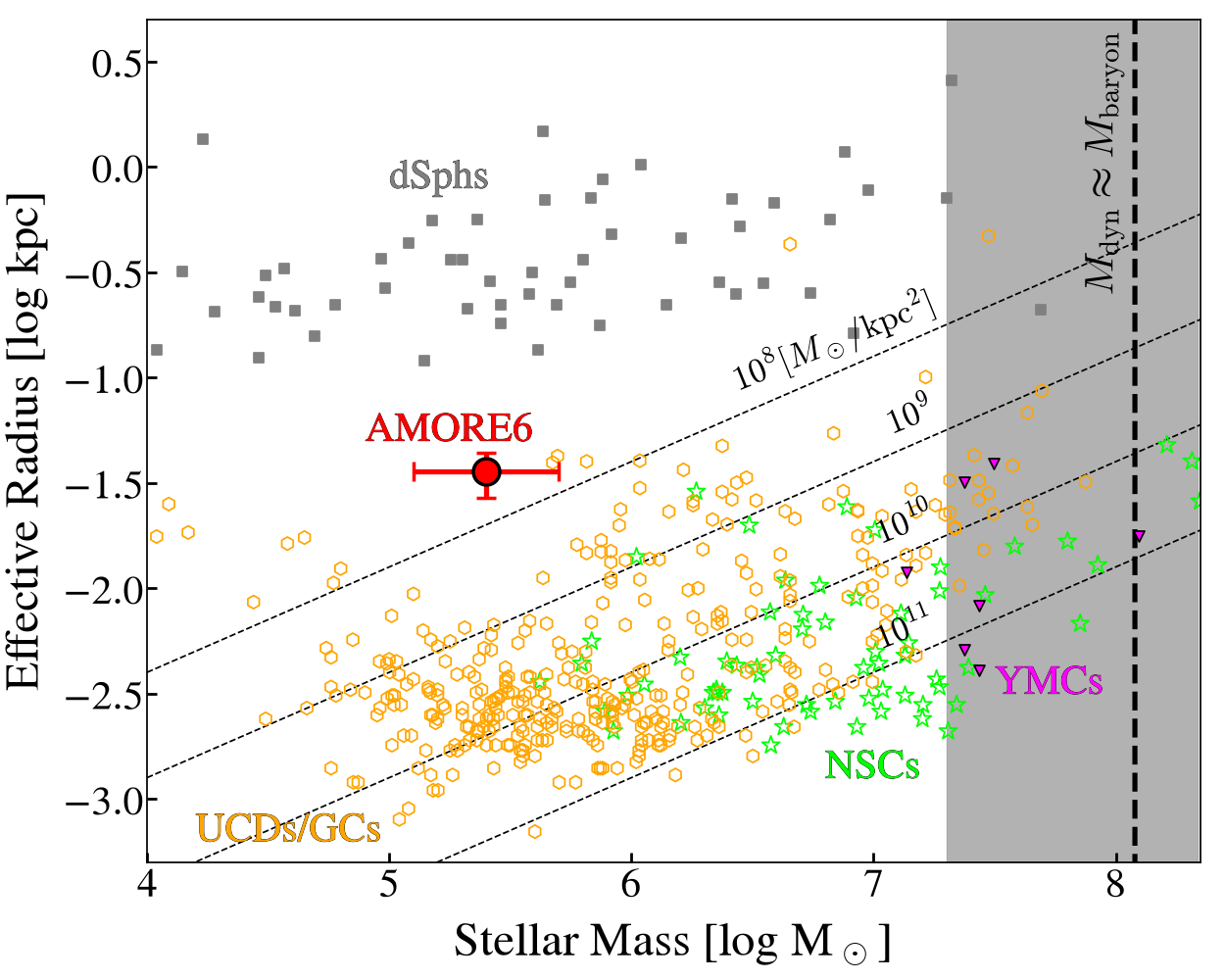}
	\caption{
    \id\ (red circle) is compared with various types of local objects, including dwarf spheroidal galaxies (dSphs, gray squares), nuclear star clusters (NSCs, green stars), ultra-compact dwarfs and globular clusters (UCDs/GCs, yellow hexagons), and young massive clusters (YMCs, magenta inverted triangles), from \cite{norris14}. The baryonic mass of \id, approximated from the dynamical mass measured within $R_e=36$\,pc (see Sec.~\ref{ssec:mdyn} for details), is also shown (vertical dashed line) with uncertainties (gray hatched regions).
    }
\label{fig:sizem}
\end{figure*}

{Figure~\ref{fig:sizem} compares \id\ with nearby stellar systems in the stellar mass--size plane, including dwarf spheroidal galaxies (dSphs), ultra-compact dwarfs (UCDs), and globular clusters (GCs). The small physical size of \id\ ($R_e\simeq36$\,pc) and its proximity to the region occupied by UCDs and massive star clusters make this interpretation worth considering.}

{However, several observational properties disfavor a star-cluster interpretation. First, inspection of the existing JWST/NIRCam imaging reveals no detectable extended host galaxy or nearby stellar component associated with either of the two highly magnified images, \id-A and \id-B \citep{morishita25}. Although a sufficiently faint underlying host cannot be excluded, the absence of detectable host emission in both images argues against a compact star-forming cluster embedded in a more extended galaxy.

Second, \id\ exhibits a smooth, apparently single-component light profile, without the multiple compact clumps observed in some other strongly lensed star-forming systems at similarly high magnifications. This morphology is consistent with an isolated, compact star-forming system. The effective radius of \id\ is also substantially larger than that of typical globular clusters, which generally have radii of only a few parsecs (Fig.~\ref{fig:sizem}).

Finally, the stellar mass inferred from SED modeling, $M_\ast\sim(1$--$5)\times10^5\,M_\odot$, is substantially smaller than the estimated dynamical mass. If the observed velocity dispersion primarily traces the self-gravity of a stellar cluster, this discrepancy would require an additional component or large non-virial contributions to the line width. We therefore consider the interpretation of \id\ as an isolated, extremely low-mass galaxy to be more plausible, while acknowledging that the current observations cannot definitively exclude a bound star cluster.}

\subsubsection{Implications for the future evolution}
We end by briefly discussing the potential future evolution. Given the unusual properties of AMORE6---its extremely low metallicity, compact size, dynamical mass estimate, and very young age---it is natural to ask what kind of present-day system such an object might ultimately become. In the stellar mass--size plane (Fig.~\ref{fig:sizem}), \id\ occupies a largely unpopulated region between dSphs and UCDs/GCs. Although its current stellar mass is comparable to that of dSphs, its effective radius is substantially smaller, suggesting that \id\ may represent a short-lived evolutionary phase that is rarely observed in the local Universe. Another notable distinction is that \id\ appears to be gas-rich, whereas these local systems are generally gas-poor or entirely devoid of gas. 

Interpreting the current position of AMORE6 in the stellar mass–size plane nevertheless requires caution. The observed half-light radius is dominated by the youngest massive stars, whose spatial distribution need not reflect that of the longer-lived stellar population. Furthermore, both stellar evolution and feedback-driven dynamical evolution may substantially modify the structure of the system over the next several tens of Myr. Assuming no external gas accretion, the maximum stellar mass that could be attained if all remaining gas were converted into stars is $\approx10^{7}$--$10^{8}\,M_\odot$. This value should be regarded as a strict upper limit, however, as realistic star formation efficiencies are likely to be much lower.

Stellar feedback is likely to play a key role in determining the subsequent evolution of \id. As discussed in Sec.~\ref{sec:count}, the system is likely dominated by very massive stars whose intense ionizing radiation, stellar winds, and eventual supernova explosions inject substantial amounts of energy into the surrounding gas. From the inferred virial mass and effective radius, we estimate an escape velocity of approximately $100~{\rm km\,s^{-1}}$, corresponding to a gas binding energy of order $2\times10^{53}$\,erg. Pair-instability supernovae or hypernovae can each release $\sim10^{52}$\,erg, while normal core-collapse supernovae contribute $\sim10^{51}$\,erg \citep[e.g.,][]{2002ApJ...567..532H}. Consequently, feedback from a relatively modest number of massive stars may be sufficient to expel a substantial fraction of the remaining gas, particularly if the explosions occur within the short duration of the starburst. Such rapid gas removal would not only suppress further star formation but could also dynamically expand the stellar system, shifting its position in the stellar mass--size plane. Therefore, even in the absence of external gas accretion or mergers, the final stellar mass and structural properties of \id\ remain highly uncertain.

\section{Summary}

In this paper, we have presented a comprehensive spectroscopic and photometric analysis of AMORE6, an extremely metal-poor galaxy candidate at $z=5.725$. Our main findings are summarized as follows:

\begin{itemize}

\item We obtained deep VLT/X-shooter spectroscopy covering Ly$\alpha$ and the major rest-frame ultraviolet metal emission lines. No UV metal-line emission is detected, placing stringent upper limits on \nv, \civ, \oiiiuv, \niv, \niii, and \ciii.

\item The new upper limit on the \civ-to-\hb\ ratio breaks the density--metallicity degeneracy present in previous analyses based on the [O\,{\sc iii}]/\hb\ ratio alone. The observations exclude the high-density enriched-gas scenario and strengthen the conclusion that AMORE6 resides in an extremely metal-poor environment with $\log(Z_{\rm gas}/Z_\odot)\lesssim-2$.

\item The high-resolution X-shooter spectrum reveals an unusually narrow Ly$\alpha$ line with an intrinsic velocity dispersion of $\sigma_v=23.2$\,km\,s$^{-1}$ and a peak consistent with the systemic redshift measured from \hb. Combining the velocity width with the measured effective radius yields a dynamical mass of $M_{\rm dyn}=(1$--$22)\times10^7\,M_\odot$, implying a gas-rich system with a gas fraction of $\fgas${, assuming that the dynamical mass represents the baryonic mass within the effective radius}.

\item We performed dedicated SED modeling using both Pop III and Pop II stellar population libraries. All acceptable models consistently indicate a remarkably small stellar mass of $(1$--$5)\times10^5\,M_\odot$ and a very young stellar population with an age of $\sim1$\,Myr.

\item Although all models successfully reproduce the observed blue UV continuum slope and are broadly consistent with the inferred H$\alpha$ equivalent width, none simultaneously reproduces the observed \hb\ line flux and equivalent width. This discrepancy suggests that the ionizing conditions in AMORE6 are not fully captured by current stellar population and photoionization models.

\item Independent estimates based on the hydrogen recombination luminosity consistently require a population of tens to thousands of massive stars, depending on the adopted upper mass limit of the Pop III mass distribution.

\item We discussed two potential scenarios for the observed properties of \id, massive star formation in an atomic-cooling halo and gravitationally bound star cluster. Together with its gas-rich nature, very young age, and strong Ly$\alpha$ emission, these properties are consistent with a recent burst of star formation, potentially involving a substantial population of metal-poor stars.

\end{itemize}

The evidence presented in this work does not yet establish AMORE6 as a bona fide Pop III galaxy. Nevertheless, its unique combination of extremely low metallicity, low stellar and dynamical masses, high gas fraction, remarkably young stellar population, and unusual nebular properties makes it a particularly valuable laboratory currently known for studying massive star formation under near-primordial conditions. In particular, the inability of existing stellar population and photoionization models to simultaneously reproduce the observed blue UV continuum and the H$\beta$ emission highlights the need for improved theoretical models incorporating more realistic massive-star physics and ionizing environments. Future observations with JWST/NIRSpec, the next generation of 30-m class telescopes, and ultimately the \textit{Nancy Grace Roman Space Telescope} will provide the sensitivity and statistical samples needed to determine whether AMORE6 represents a rare example of massive metal-free star formation or the first member of a previously unexplored population of low-mass galaxies near the atomic-cooling threshold.

\vspace{0.5em}
\section*{Acknowledgements}
This study is based on observations collected at the European Southern Observatory under ESO programme 
DDT 115.29G6.001. We thank the anonymous referee for providing constructive comments. Some/all of the data presented in this paper were obtained from the Mikulski Archive for Space Telescopes (MAST) at the Space Telescope Science Institute. 
We acknowledge support for this work under NASA grant 80NSSC22K1294. This work was supported by JSPS KAKENHI Grant Number 25K24918.
We would like to thank Pietro Bergamini, Marco Chiaberge, Claudio Grillo, Piero Rosati, Tommaso Treu, Harley Katz, Alessandra Venditti, Akio Inoue, and Masashi Chiba for useful discussions. 

{
{\it Software:} 
Astropy \citep{astropy13,astropy18,astropy22}, 
\cl\ (ver.25.00) \citep{1998PASP..110..761F,chatzikos23},
gsf (ver.2.0.2) \citep{morishita19},
numpy \citep{numpy}.
}

\appendix

\section{X-shooter observation log}\label{sec:log}
We report the log of our VLT/X-shooter observations in Table~\ref{tab:xshooter_log}.

\begin{figure*}
\centering
	\includegraphics[width=0.48\textwidth]{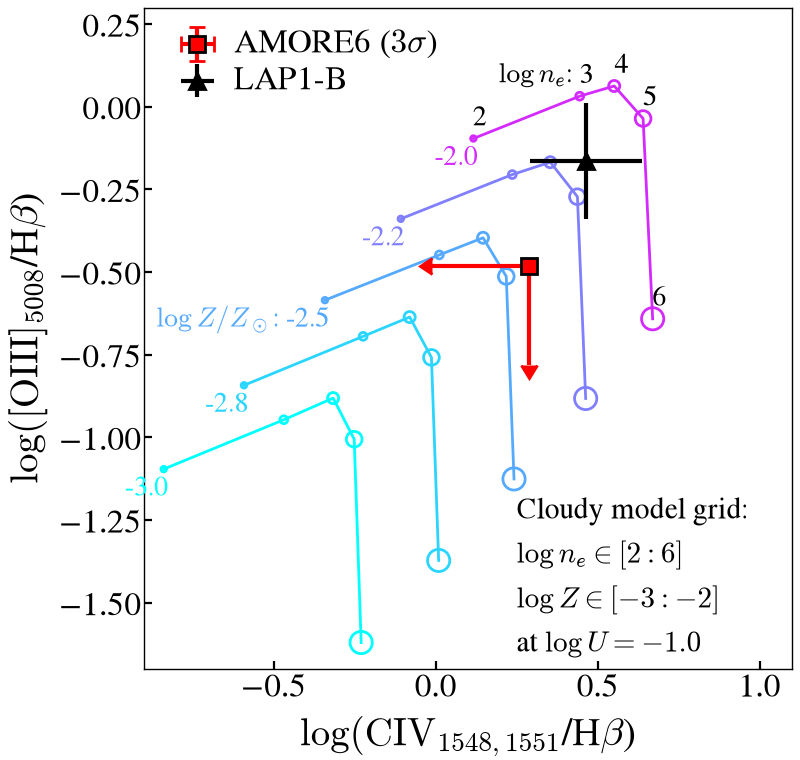}
	\includegraphics[width=0.48\textwidth]{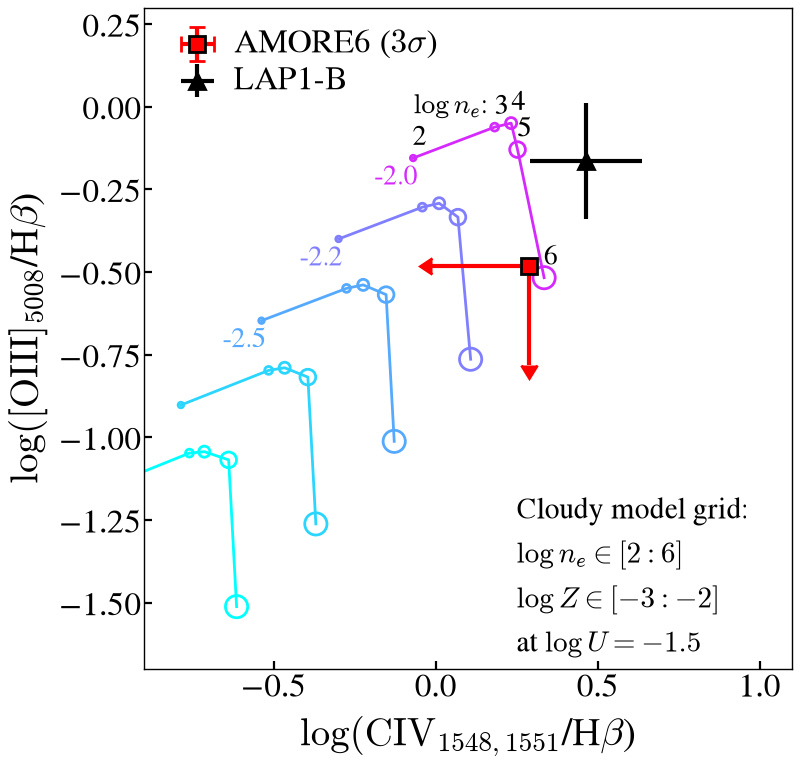}
	\includegraphics[width=0.48\textwidth]{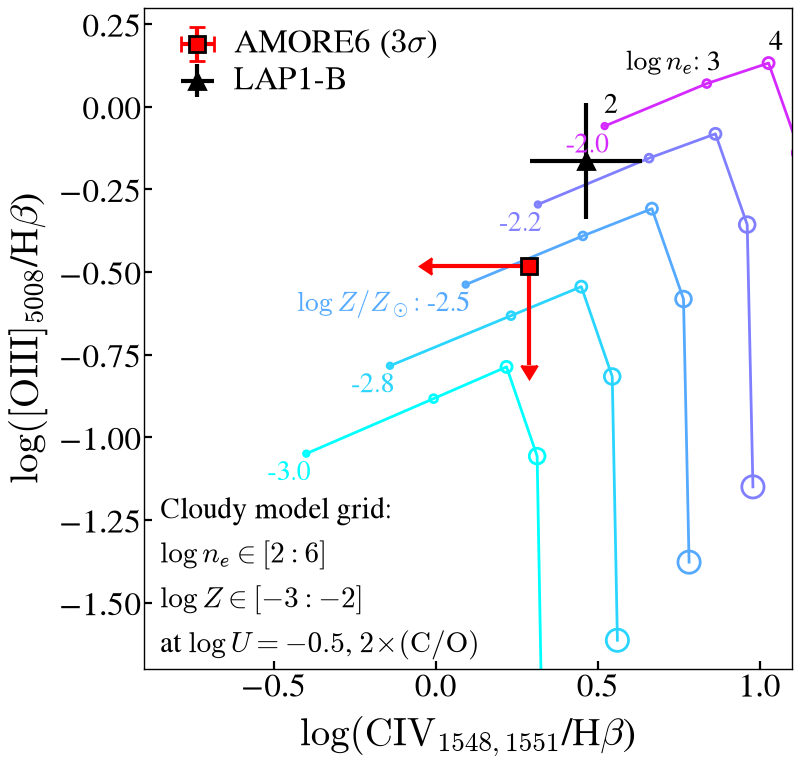}
	\includegraphics[width=0.48\textwidth]{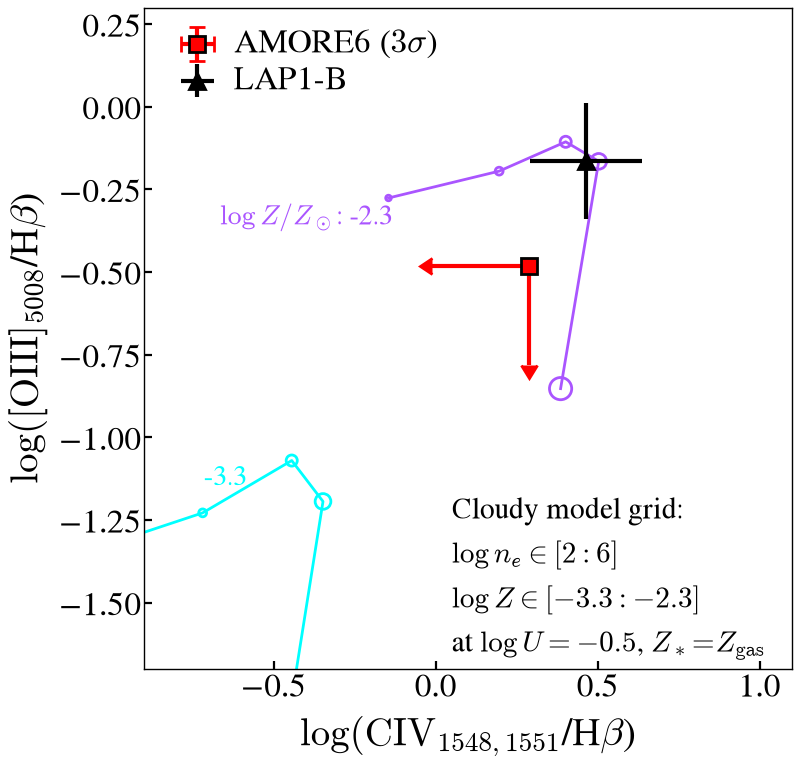}
	\caption{
    {Same as Fig.~\ref{fig:c4}, but with different setups for the photo-ionization model calculation. (Top): ionization parameter set to $\log U=-1$ (left) and $\log U = -1.5$ (right). The model tracks systematically shift toward the lower left at decreasing $\log U$, while the shape of the tracks remain similar. (Bottom left): the same ionization parameter as Fig.~\ref{fig:c4} but with an enhanced carbon-to-oxygen ratio to $2\times$ the solar ratio. The tracks shift horizontally toward right. (Bottom right): model tracks are calculated for Pop-II sources with {\tt BPASS}, where the ionizing source metallicity is matched to that of gas, $Z_{\rm gas} = Z_*$. Due to the limited availability of {\tt BPASS} models for metallicity at $\logZ<-2$, only two tracks are shown.}
    }
\label{fig:logu}
\end{figure*}

{
\section{Photoionization models and electron density inference}\label{sec:logu}
In Fig.~\ref{fig:logu}, we compare photo-ionization models with difference assumptions, namely lower ionization parameters ($\log U=-1, -1.5$), enhanced carbon-to-oxygen ratio to 2 times the solar ratio, and matched metallicity between the gas cloud and the ionizing source. 

Model tracks of lower ionization parameters systematically shift toward lower left at decreasing $\log U$, while the shape of the tracks remain similar. This suggests an increased metallicity upper limit for a given \oiii-to-\hb\ ratio. While the inference of the ionization parameter is not feasible for \id, we caution in Sec.~\ref{ssec:uv} that it remains high for the low stellar mass of \id. Even with the highest value expected from the interpolation of previous work $\log U\simgt -1.5$  \citep{2026arXiv260611345I}, the metallicity upper limit is $\log Z_{\rm gas}/Z_\odot \simlt -2$.

For the (C/O) enhanced model, the track shifts horizontally toward right. In this case, the constraint from the \oiii-to-\hb\ ratio remains similar, whereas the \civ-to-\hb\ ratio gives a stronger constraint, disfavoring the high-density gas scenario even more strongly.

Lastly, we also show the model tracks calculated with Pop-II ionizing sources. In these calculations, we set the gas metallicity the same as the ionizing source metallicity. Due to the limited availability of different metallicity templates by {\tt BPASS}, there are only two tracks available at the range of $\logZ < -2$. Since our models are calculated at a given $\log U$, the model tracks predict similar metallicity inference as for those in Fig.~\ref{fig:c4}.
}

\input{xshooter}


\bibliography{output}{}
\bibliographystyle{aasjournal}



\end{document}

%% file: newcommand.tex
\definecolor{fuchsia}{rgb}{0.54, 0.17, 0.89}
\definecolor{azure}{rgb}{0.0, 0.5, 1.0}
\definecolor{pgreen}{rgb}{0.12, 0.3, 0.17}
\definecolor{alizarin}{rgb}{0.82, 0.1, 0.26}

\newcommand{\oiii}{[\textrm{O}~\textsc{iii}]}
\newcommand{\oiiir}{$[\textrm{O}~\textsc{iii}]_{\lambda5007}$}

\newcommand{\oiiiuv}{\textrm{O}~\textsc{iii}]}

\newcommand{\niii}{[\textrm{N}~\textsc{iii}]}
\newcommand{\niv}{\textrm{N}~\textsc{iv}]}
\newcommand{\nv}{\textrm{N}~\textsc{v}}

\newcommand{\ciii}{\textrm{C}~\textsc{iii}]}
\newcommand{\civ}{\textrm{C}~\textsc{iv}}

\newcommand{\heii}{\textrm{He}~\textsc{ii}}

\newcommand{\simgt}{\,\rlap{\lower 3.5 pt \hbox{$\mathchar \sim$}} \raise
1pt \hbox {$>$}\,}
\newcommand{\simlt}{\,\rlap{\lower 3.5 pt \hbox{$\mathchar \sim$}} \raise
1pt \hbox {$<$}\,}
\newcommand{\Msun}{M_{\odot}}

\newcommand{\logm}{\log M_*/\Msun}
\newcommand{\logZ}{\log Z_*/Z_\odot}

\newcommand{\logoh}{$\log {\rm (O/H)}$}

\newcommand{\ly}{${\rm Ly\alpha}$}
\newcommand{\ha}{${\rm H\alpha}$}
\newcommand{\hb}{${\rm H\beta}$}

\newcommand{\Z}{{\rm [Z/H]}}

\newcommand{\gsf}{{\tt gsf}}
\newcommand{\ygdr}{{\tt Yggdrasil}}
\newcommand{\cl}{{\tt Cloudy}}
\newcommand{\bp}{{\tt BPASS}}

\newcommand{\id}{AMORE6} 
\newcommand{\ida}{\id-A}
\newcommand{\idb}{\id-B}

\newcommand{\sigmasmint}{18.3}
\newcommand{\sigmaasmint}{42.7}

\newcommand{\mvir}{1.9_{-0.4}^{+0.5}\times10^{7}\,M_\odot} 
\newcommand{\fgas}{0.97\pm0.03}

\newcommand{\mvirr}{1.7_{-0.4}^{+0.5}\times10^{8}\,M_\odot} 
\newcommand{\mvirrther}{9.3_{-2.1}^{+2.6}\times10^{7}\,M_\odot} 
\newcommand{\mvirmed}{1.2\pm1.0\times10^{8}\,M_\odot} 

\newcommand{\lyesclow}{77\,\%} 
\newcommand{\lyeschigh}{100\,\%} 


%% file: table_sed.tex
\begin{deluxetable*}{lcccccccc}
\tablecaption{Summary of SED fitting setups.\label{tab:sedmodels}}
\tablewidth{0pt}
\tablehead{
\colhead{Model} &
\colhead{Library} &
\colhead{IMF} &
\colhead{$Z_\ast/Z_\odot$} &
\colhead{$Z_{\rm gas}/Z_\odot$} &
\colhead{Age} &
\colhead{SFH} &
\colhead{$\log U$} &
\colhead{$\log n_{\rm H}/{\rm cm^{-3}}$}
}
\startdata
(a) Pop-III pristine & \ygdr & Pop-III.1 & 0 & 0 & 1\,Myr & Inst. burst & $-0.5$ & 3 \\
(b) Pop-III pristine & \ygdr & Pop-III.2 & 0 & 0 & 1\,Myr & Inst. burst & $-0.5$ & 3 \\
(c) Pop-III pristine & \ygdr & Kroupa & 0 & 0 & 1\,Myr & Inst. burst & $-0.5$ & 3 \\
(d) Pop-III enriched & \ygdr & Pop-III.1 & 0 & $10^{-2.5}$ & 1\,Myr & Inst. burst & $-0.5$ & 3 \\
(e) Pop-III enriched & \ygdr & Pop-III.2 & 0 & $10^{-2.5}$ & 1\,Myr & Inst. burst & $-0.5$ & 3 \\
(f) Pop-III enriched & \ygdr & Kroupa & 0 & $10^{-2.5}$ & 1\,Myr & Inst. burst & $-0.5$ & 3 \\
(g) Pop-II & \bp & Chab300 & $10^{-2.5}$ & $10^{-2.5}$ & 1\,Myr & Inst. burst & $-0.5$ & 3 \\
(h) Pop-II & \bp & 135\_300 & $10^{-2.5}$ & $10^{-2.5}$ & 1\,Myr & Inst. burst & $-0.5$ & 3 \\
(i) Pop-II & \bp & 135\_100 & $10^{-2.5}$ & $10^{-2.5}$ & 1\,Myr & Inst. burst & $-0.5$ & 3 \\
\enddata

\end{deluxetable*}

%% file: sedfitting.tex
\section{SED modeling}\label{sec:sedmodels}

\subsection{SED modeling setup}\label{sec:sedsetup}
We aim to obtain the stellar mass of \id\ via spectral energy distribution (SED) fitting to the observed photometric fluxes using SED fitting code {\tt gsf} (ver.2.0.2) \citep{morishita19}. For a young system such as AMORE6, stellar population models spanning extreme physical conditions are essential, as ordinary stellar radiation sources may not sufficiently reproduce the observed blue UV slope, as well as the nebular properties.

We consider three classes of models: 1.~Pop-III stars in a pristine gas cloud, 2.~Pop-III in an enriched cloud, and 3.~Pop-II in an enriched cloud. The first class assumes a very early stage following the formation of the first generation of stars, by utilizing the \ygdr\ stellar library \citep{zackrisson11}. 
The second class assumes a situation where some first-generation stars have undergone supernova explosions and enriched the surrounding gas, similar to the case of LAP1-B \citep{vanzella23,nakajima25} where emission lines from heavy elements are detected. 
For the third class, we assume stars with nonzero metallicity from \bp\ \citep{eldridge17} in an enriched gas cloud. For each class, we also explore different IMFs to the degree available for each library.

Nebular emission models are calculated using \cl, using the ionizing spectrum specified for each model. We set the geometry to ``sphere expanding" with an initial radius of 10\,pc and iterate the calculation to convergence in the parameter grid of metallicity, electron density, and ionization parameter. In the Pop-II models, we set the metallicity of gas equal to the stellar metallicity. 

We adopt the NIRCam+HST photometry reported in \citet{morishita25}. Our approach combines stellar population templates with calculated nebular emission. 
The relative contribution of the nebular component to the stellar continuum is treated as a free parameter in the fitting procedure, following the default implementation in \gsf. This effectively allows different ionizing-photon escape fractions without explicitly parameterizing the gas covering fraction. We fix redshift to $z=5.725$ in the fitting.
{We note that {\tt gsf} does not include detailed \ly\ radiative-transfer modeling (ISM/IGM absorption and resonant scattering causing line broadening), whereas \id\ exhibits a prominent \ly\ emission, likely detectable in the photometry. For this reason, we exclude the F090W data from fitting. While the F814W filter overlaps with the \ly\ wavelength, excluding the data point does not affect the final result, due to its relatively large flux uncertainty.}

For all models, we adopt an instantaneous-burst star-formation history with an age of $1$\,Myr. While the second and third model classes can in principle accommodate older stellar populations, our tests showed that neither introducing additional age parameters nor adopting different ages improved the quality of the fit. Similarly, \id\ is found to exhibit negligible dust attenuation, with the best-fitting solutions consistently yielding $A_V \approx 0$.
We therefore fix the stellar population age to $1$\,Myr and the dust attenuation to $A_V=0$ for all models. This choice reduces the number of free parameters and ensures that all model families are fit with an identical parameter set, enabling a direct comparison of the resulting goodness-of-fit and inferred physical properties. The details of each model setup are described in the following subsections, while a summary is provided in Table~\ref{tab:sedmodels}. Throughout the SED modeling, we fix the ionization parameter and electron density to $\log U=-0.5$ and $n_e=10^3\,{\rm cm}^{-3}$, respectively.

\subsubsection{Pop-III in a pristine cloud}\label{sec:sed-prist}

We first consider the case where both the stellar population and the surrounding gas cloud are pristine. In this setup, we combine the \ygdr\ stellar population library with nebular emission models assuming zero metallicity gas. The nebular emission is calculated with \cl\ using \ygdr\ spectra as the ionizing source. The gas-phase metallicity is fixed to $Z_{\rm gas}=0$, corresponding to a chemically pristine environment. This setup represents the idealized case of genuine first-generation star formation without previous chemical enrichment.

We adopt three IMFs available in \ygdr, namely Pop-III.1, Pop-III.2, and Pop-III.Kroupa. The Pop-III.1 IMF corresponds to an extremely top-heavy IMF with a power-law IMF ($dN/dM \propto M^{-\alpha}$) of slope $\alpha = 2.35$ over the stellar mass range $50$--$500\,M_\odot$. The Pop-III.2 IMF adopts a less extreme log-normal form with a characteristic stellar mass of $10\,M_\odot$, as was presented in \citet{raiter10} (also \citealt{tumlinson06}). The Pop-III.Kroupa model adopts the \citet{kroupa01} IMF over the mass range of $0.1$--$100\,M_\odot$. These three (very distinct) IMFs bracket plausible theoretical expectations for primordial star formation and also allow us to compare the resulting stellar mass of the system. In all cases, we assume an instantaneous burst and set the stellar population and cloud ages to $1$\,Myr.

\subsubsection{Pop-III in an enriched cloud}\label{sec:sed-pop3}

We next consider a scenario in which the stellar population itself remains Pop-III-like, while the surrounding gas cloud has already experienced mild chemical enrichment. In this setup, we combine the \ygdr\ stellar population library with nebular emission models assuming non-zero gas metallicity. This setup allows us to investigate whether the observed SED can be reproduced by primordial stellar populations embedded in a slightly enriched interstellar medium.

Our analysis of the X-shooter spectrum suggests a metallicity $\simlt 10^{-2.5}~Z_\odot$. Motivated by this constraint, we here adopt $\log (Z_{\rm gas}/Z_\odot) = -2.5$ for the nebular calculation. Changing this assumption toward even lower metallicities does not affect our conclusions, since the resulting nebular spectra remain similarly weak in metal-line emission.

\subsubsection{Pop-II}\label{sec:sed-pop2}
Finally, we consider Pop-II stellar populations with non-zero stellar and gas metallicities. In this setup, we combine the binary stellar population models from \bp\ with nebular emission models of non-zero metallicity. The Pop-II models provide an important comparison set to evaluate whether the observed SED properties of \id\ can be explained by ordinary metal-poor stellar populations without invoking primordial massive stars.

For the stellar component, we adopt three IMFs available in \bp: ``Chab300'', ``135\_300'', and ``135\_100''. The former adopts an exponential turnover below $1\,M_\odot$, similar to the \citet{chabrier03} IMF, while following a power-law slope of $\alpha=2.3$ up to $300\,M_\odot$ at higher masses. The latter two adopt a broken power-law IMF with a slope of $\alpha=1.3$ over the same low-mass range $0.1$--$0.5\,M_\odot$ and $\alpha=2.35$ over the high-mass range $0.5$--$300\,M_\odot$ and $0.5$--$100\,M_\odot$, respectively. In these models, we fix the stellar metallicity to $\logZ=-2.5$, the stellar population age to $1$\,Myr, and dust attenuation to zero, for the same reason as above. We calculate the nebular emission with \cl\ using the same gas-phase metallicity.

\begin{figure*}
\centering
	\includegraphics[width=0.8\textwidth]{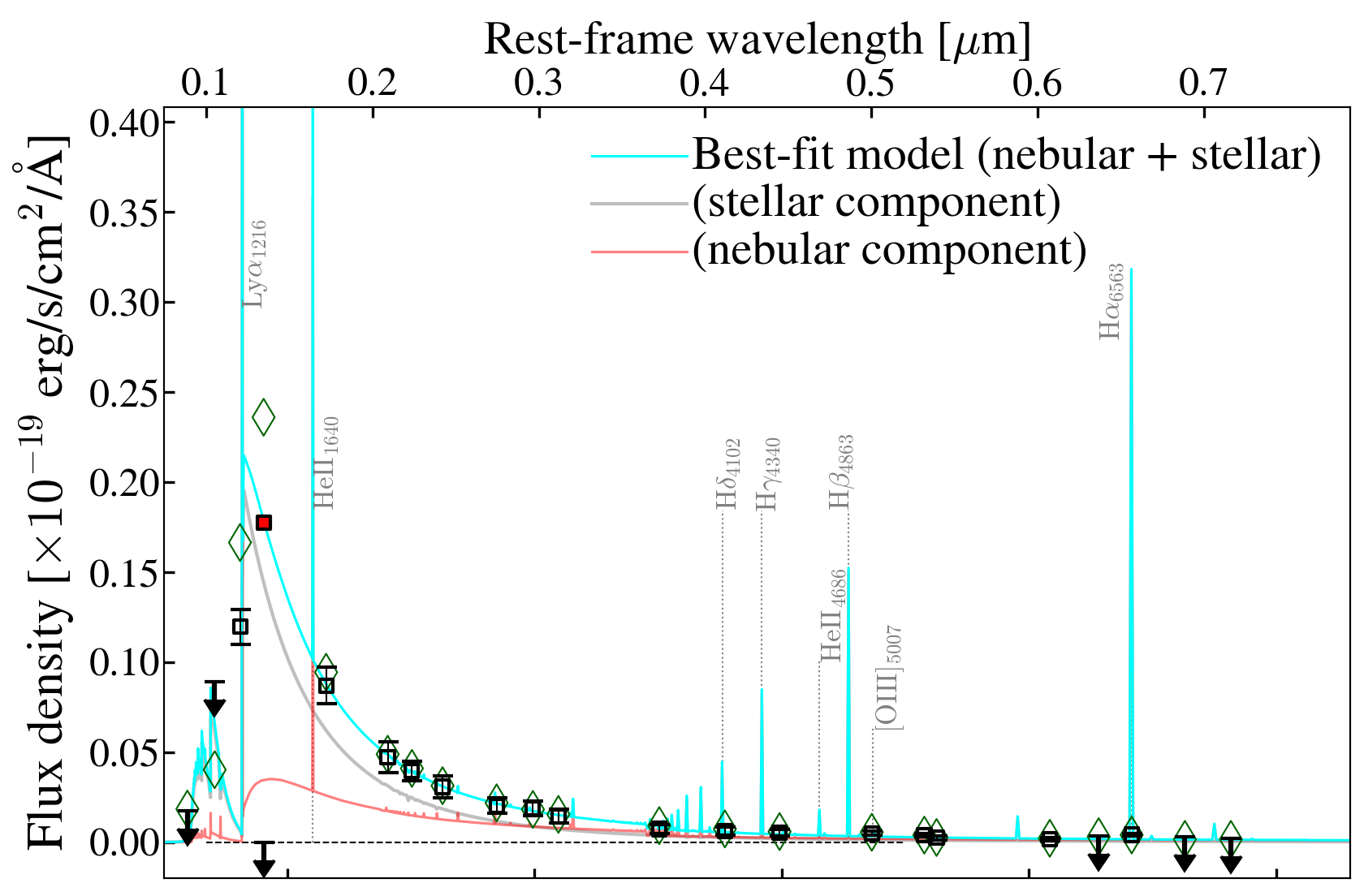}
	\includegraphics[width=0.8\textwidth]{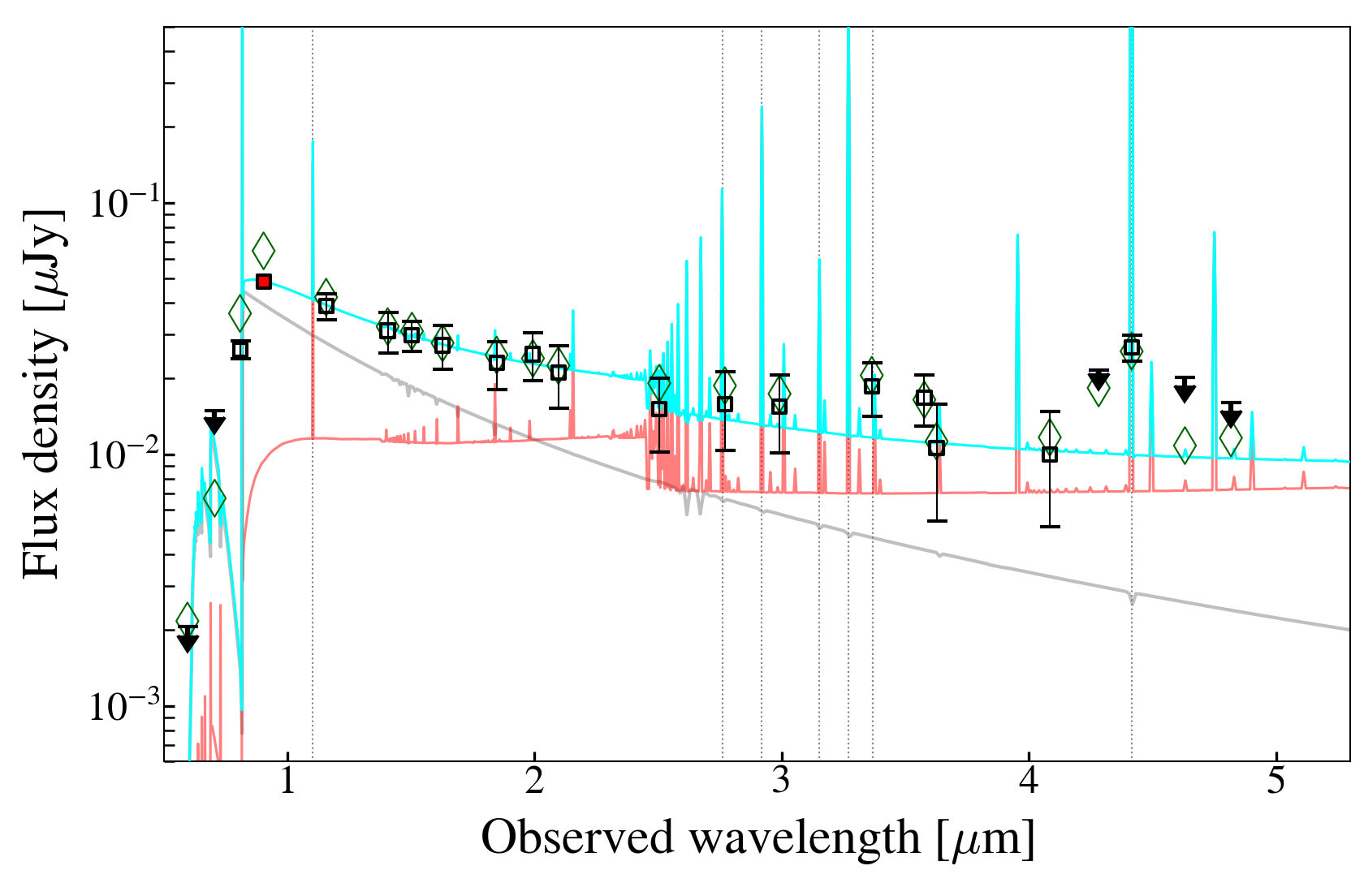}
	\caption{
    (Top):
    Spectral energy distribution of \id.
    The spectral energy distribution is modeled using the photometry of \idb, in the units of $10^{-19}\,{\rm erg\,s^{-1}\,cm^{-2}\,\AA^{-1}}$. Observed photometric fluxes are shown with 1\,$\sigma$ uncertainties/upper limits (squares/arrows). The best-fitting full model spectrum (nebular+stellar; cyan line), stellar component (gray line), and nebular component (red line) are shown. Pseudo photometric fluxes for the best-fitting model are shown (green diamonds). The NIRCam F090W data point (red square) was excluded from the fitting (Sec.~\ref{sec:sedsetup}). 
    (Bottom): Same as the top panel but in the units of $\mu$Jy. 
    }
\label{fig:sed}
\end{figure*}

%
\subsection{Results of SED modeling}\label{sec:sedresults}

Table~\ref{tab:sed_results} summarizes the main parameters derived from our SED modeling. We focus on several key observables that can be compared with independent measurements where available, including the stellar mass, UV spectral slope ($\beta_{\rm UV}$), hydrogen emission-line equivalent widths, and line fluxes.

First, all models successfully reproduce the extremely blue UV continuum slope, yielding $\beta_{\rm UV}\sim-3$ to $-2.9$. For comparison, the UV slope measured directly from the photometric data is $\beta_{\rm UV}=-2.9\pm0.3$. The agreement indicates that both the Pop-III and metal-poor Pop-II models naturally produce the observed blue continuum without requiring dust attenuation.

While there is no direct spectroscopic measurement of \ha, \id\ exhibits a large flux excess at the wavelength of \ha. Using the flux excess in F444W relative to the continuum estimated from F410M, \citet{morishita25} measured $\mathrm{EW_{H\alpha}}=1900^{+3200}_{-1100}$\,\AA. All models considered here are broadly consistent with this estimate within the large uncertainty, while we note that the largest equivalent widths are produced by the pristine Pop-III models, reaching $\mathrm{EW_{H\alpha}}\sim2300$\,\AA. 

In contrast, none of the models successfully reproduces the {\hb\ integrated line flux, $f_{\rm H\beta}=2.9\pm0.5\times10^{-20}\,{\rm erg\,s^{-1}\,cm^{-2}}$, originally observed in the NIRCam/WFSS spectrum at the position of \id-B and then corrected for magnification $\mu_{\rm B}$ \citep{morishita25}}. The closest prediction is obtained from the pristine models (a-c), yielding a flux of $1\times10^{-20}\,{\rm erg s^{-1} cm^{-2}}$ (magnification corrected), yet still deviating from the observation {by $3.7\,\sigma$.}
Although larger \hb\ fluxes can be obtained by increasing the relative contribution of the nebular component, doing so simultaneously reddens the UV continuum because of the intrinsically red nebular continuum (Figure~\ref{fig:sed}). Consequently, the observed combination of a very blue UV slope and a large \hb\ equivalent width is difficult to reproduce within the current model framework. Indeed, the nebular templates we prepared reach a maximum \hb\ equivalent width of $\sim700$\,\AA, already lower than the value inferred from the NIRCam/WFSS spectrum, $900\pm150$\,\AA. Changing the adopted SFH or gas properties ($\log U$ and $n_{\rm H}$) does not resolve the tension. While the current \hb\ measurement should be verified with higher-S/N spectroscopy, a persistent discrepancy would suggest the need for a different ionizing environment, such as a harder ionizing spectrum from the inclusion of rotation and Pop-III binaries \citep[e.g.,][]{2022A&A...666A.199S,2026MNRAS.547ag386W}, or peculiar gas geometry capable of producing stronger nebular emission while preserving the blue UV continuum \citep[also][]{2009MNRAS.399...37J}.
{Similarly, recent spectroscopy showed flat Balmer decrements in several $z\sim6$ systems \citep[e.g.,][]{cameron23,topping24}. \citet{2024ApJ...974..180Y} proposed that density-bounded nebulae that are optically thin to higher-order Lyman-series photons could explain enhanced \hb.}

In addition to the physical parameters, Table~\ref{tab:sed_results} reports the goodness-of-fit metrics, $\chi^2$. All models provide statistically acceptable fits, with $\chi^2\sim0.8$. We note that the number of free parameters is the same across all models, thus the difference in e.g., the Bayesian Information Criterion (BIC) is very small. Adding more age bins of $>1$\,Myr to enriched models does not improve the fit, simply increasing BIC thus statistically disfavored. While subtle, the differences in $\chi^2$ are primarily driven by the nebular emission properties, especially around the flux excess of \ha\ and \hb, rather than the UV slope alone, as all models here reproduce the observed blue UV slope.
We remind that the fitting procedure excludes the spectroscopic data. 

Despite the different assumptions regarding the stellar population and nebular emission, all models consistently indicate that \id\ is a low-mass galaxy, within $M_\ast\sim$\,1--5\,$\times10^5\,M_\odot$. 
Given the overall excellent fitting quality for all the models and the matched parameter space, the difference mainly arises from the adopted IMFs, especially the presence of very massive stars ($>100\,M_\odot$) and low mass stars. Indeed, the Kroupa IMF, which has an extended mass range toward a lower mass, results in the most massive stellar mass estimate, whereas the models with $>300\,M_\odot$ predict the lowest mass estimate.

By comparing the intrinsic \ly\ strength from fitting models, we can estimate the escape fraction of \ly. {This fraction ranges from \lyesclow\ (Pop-III models) to $\approx100\,\%$ (Pop-II models), which is considerably high among high-$z$ sources reported in the literature} \citep[e.g.,][see also \citealt{messa26} for a detailed discussion]{2026arXiv260304517G}.

\input{table_sedresults_all}

%% file: table_sedresults_all.tex
\begin{deluxetable*}{lcccccccccccc}
\tablecaption{SED fitting results of \id-B for different stellar population models.\label{tab:sed_results}}
\tablewidth{0pt}
\tablehead{
\colhead{Model} & \colhead{$\chi^2$} & \colhead{$\log(M_\star/M_\odot)$} & \colhead{$M_{\rm UV}$} & \colhead{$\beta_{\rm UV}$} & \colhead{$f_{\rm Ly\alpha}$} & \colhead{$f_{\rm He\,\textsc{ii}}$} & \colhead{$f_{\rm H\beta}$} & \colhead{EW$_{\rm 0,Ly\alpha}$} & \colhead{EW$_{\rm 0, He\,\textsc{ii}}$} & \colhead{EW$_{\rm 0,H\beta}$} & \colhead{EW$_{\rm 0,H\alpha}$} \\
\colhead{} & \colhead{} & \colhead{} & \colhead{[mag]} & \colhead{} & \colhead{[$10^{-19}$]} & \colhead{[$10^{-20}$]} & \colhead{[$10^{-20}$]} & \colhead{[\AA]} & \colhead{[\AA]} & \colhead{[\AA]} & \colhead{[\AA]}
}
\startdata
(a) & 0.75 & $5.1^{+0.1}_{-0.1}$ & $-14.7^{+0.1}_{-0.1}$ & $-3.0^{+0.1}_{-0.1}$ & $4.3^{+0.5}_{-0.5}$ & $0.9^{+0.1}_{-0.1}$ & $1.0^{+0.1}_{-0.1}$ & $223^{+56}_{-40}$ & $10^{+2}_{-2}$ & $359^{+37}_{-33}$ & $2250^{+150}_{-150}$ \\
(b) & 0.76 & $5.4^{+0.1}_{-0.1}$ & $-14.7^{+0.1}_{-0.1}$ & $-3.0^{+0.1}_{-0.1}$ & $4.2^{+0.6}_{-0.6}$ & $0.5^{+0.1}_{-0.1}$ & $1.0^{+0.1}_{-0.1}$ & $221^{+54}_{-50}$ & $6^{+1}_{-1}$ & $361^{+38}_{-44}$ & $2250^{+160}_{-200}$ \\
(c) & 0.77 & $5.7^{+0.1}_{-0.1}$ & $-14.7^{+0.1}_{-0.1}$ & $-3.0^{+0.1}_{-0.1}$ & $4.0^{+0.6}_{-0.5}$ & $0.3^{+0.1}_{-0.1}$ & $1.0^{+0.1}_{-0.1}$ & $216^{+64}_{-38}$ & $4^{+1}_{-1}$ & $325^{+47}_{-34}$ & $2060^{+210}_{-170}$ \\
(d) & 0.77 & $5.1^{+0.1}_{-0.1}$ & $-14.8^{+0.1}_{-0.1}$ & $-3.0^{+0.1}_{-0.1}$ & $3.8^{+0.4}_{-0.4}$ & $0.9^{+0.1}_{-0.1}$ & $1.0^{+0.1}_{-0.1}$ & $195^{+36}_{-37}$ & $10^{+1}_{-2}$ & $337^{+27}_{-33}$ & $2090^{+110}_{-140}$ \\
(e) & 0.78 & $5.4^{+0.1}_{-0.1}$ & $-14.8^{+0.1}_{-0.1}$ & $-3.0^{+0.1}_{-0.1}$ & $3.6^{+0.5}_{-0.4}$ & $0.5^{+0.1}_{-0.1}$ & $0.9^{+0.1}_{-0.1}$ & $186^{+44}_{-36}$ & $5^{+1}_{-1}$ & $329^{+35}_{-34}$ & $2060^{+150}_{-150}$ \\
(f) & 0.79 & $5.7^{+0.1}_{-0.1}$ & $-14.7^{+0.1}_{-0.1}$ & $-3.0^{+0.1}_{-0.1}$ & $3.3^{+0.5}_{-0.4}$ & $0.3^{+0.1}_{-0.1}$ & $0.9^{+0.1}_{-0.1}$ & $175^{+49}_{-34}$ & $3^{+1}_{-1}$ & $286^{+41}_{-34}$ & $1830^{+190}_{-170}$ \\
(g) & 0.82 & $5.2^{+0.1}_{-0.1}$ & $-14.7^{+0.1}_{-0.1}$ & $-2.9^{+0.1}_{-0.1}$ & $3.0^{+0.4}_{-0.4}$ & $<0.1$ & $0.9^{+0.1}_{-0.1}$ & $148^{+34}_{-33}$ & $<1$ & $285^{+37}_{-43}$ & $1780^{+170}_{-210}$ \\
(h) & 0.82 & $5.3^{+0.1}_{-0.1}$ & $-14.7^{+0.1}_{-0.1}$ & $-2.9^{+0.1}_{-0.1}$ & $3.0^{+0.4}_{-0.4}$ & $<0.1$ & $0.9^{+0.1}_{-0.1}$ & $149^{+36}_{-32}$ & $<1$ & $284^{+40}_{-41}$ & $1780^{+180}_{-200}$ \\
(i) & 0.82 & $5.5^{+0.1}_{-0.1}$ & $-14.6^{+0.1}_{-0.1}$ & $-2.9^{+0.1}_{-0.1}$ & $3.0^{+0.4}_{-0.5}$ & $<0.1$ & $0.9^{+0.1}_{-0.1}$ & $150^{+30}_{-33}$ & $<1$ & $279^{+33}_{-43}$ & $1750^{+160}_{-220}$ \\
\hline
Observed & & & & $-2.9\pm0.3$ & $3.4\pm0.3$ & $<10.6$ & $2.9\pm0.5$ & $161.7 \pm 1.5$ & $<111.1$ & $940^{+250}_{-210}$ & $1900^{+3200}_{-1100}$\\
\enddata
\tablenotetext{}{Quantities are corrected for magnification. Quoted uncertainties correspond to the 16th and 84th percentiles of the posterior distributions. 
Equivalent widths are rest-frame values.
{\bf 
The observed \ly\ flux was corrected for magnification by multiplying $1/\mu_{\rm B}$ to the X-shooter \ly\ flux of \id-B (Table~\ref{tab:linefluxes}). 
The reported uncertainty includes the one in the magnification factor.
The observed \hb\ flux was corrected for magnification by multiplying $1/\mu_{\rm B}$ to the reported value of \id-B in \citet{morishita25}.}
}
\end{deluxetable*}

%% file: xshooter.tex
\begin{deluxetable*}{lcccccc}
\tablecaption{Observational log of the X-shooter data, retrieved from the ESO archive.\label{tab:xshooter_log}}
\tablewidth{0pt}
\tablehead{
\colhead{Dataset ID} &
\colhead{Date Obs} &
\colhead{MJD-OBS} &
\colhead{Exptime} &
\colhead{Airmass} &
\colhead{Seeing} &
\colhead{Arm}
}
\startdata
XS\_SFLX\_4506966 & 2025-08-14 08:10 & 60901.3410 & 3600 & 1.021 & 0.56 & UVB/VIS/NIR \\
%
%
%
%
XS\_SFLX\_4506975 & 2025-08-24 06:04 & 60911.2531 & 3600 & 1.018 & 0.76 & UVB/VIS/NIR \\
XS\_SFLX\_4506978 & 2025-08-24 08:11 & 60911.3412 & 3600 & 1.062 & 0.82 & UVB/VIS/NIR \\
XS\_SFLX\_4506981 & 2025-08-29 07:16 & 60916.3033 & 3600 & 1.024 & 0.80 & UVB/VIS/NIR \\
XS\_SFLX\_4506984 & 2025-08-29 08:24 & 60916.3503 & 3600 & 1.120 & 0.89 & UVB/VIS/NIR \\
XS\_SFLX\_4507001 & 2025-09-24 04:23 & 60942.1827 & 3600 & 1.008 & 0.64 & UVB/VIS/NIR \\
XS\_SFLX\_4506987 & 2025-09-24 05:30 & 60942.2294 & 3600 & 1.021 & 0.54 & UVB/VIS/NIR \\
XS\_SFLX\_4507083 & 2025-09-25 02:06 & 60943.0877 & 2400 & 1.214 & 0.40 & UVB/VIS/NIR \\
\enddata
\tablenotetext{}{Each row corresponds to a unique reduced X-shooter spectrum available in the ESO archive. The UVB, VIS, and NIR products associated with the same observation are grouped into a single entry. The total integration time represented by the reduced spectra is 27600\,s (7.7\,hrs).}
\end{deluxetable*}\label{tab:xshoot}

%% file: output.bib
@ARTICLE{2026arXiv261003877K,
       author = {{Kiyuna}, Masaki and {Chon}, Sunmyon and {Omukai}, Kazuyuki and {Schneider}, Raffaella},
        title = "{Internal Radiative Feedback Triggers Global Collapse and Direct-Collapse Black Hole Formation in Metal-free Atomic-Cooling Haloes}",
      journal = {arXiv e-prints},
         year = 2026,
        month = oct,
          eid = {arXiv:2610.03877},
        pages = {arXiv:2610.03877},
archivePrefix = {arXiv},
       eprint = {2610.03877},
 primaryClass = {astro-ph.GA},
       adsurl = {https://ui.adsabs.harvard.edu/abs/2026arXiv261003877K}
}

@ARTICLE{2026arXiv261016509B,
       author = {{Bosman}, Sarah E.~I. and {Davies}, Frederick B. and {Welsh}, Louise and {Farina}, Emanuele P. and {Gaikwad}, Prakash and {Spina}, Benedetta and {Maity}, Barun and {Cooke}, Ryan and {Protu{\v{s}}ov{\'a}}, Klaudia and {D'Odorico}, Valentina and {Ba{\~n}ados}, Eduardo and {Bolton}, James S. and {Davies}, Rebecca L. and {Fan}, Xiaohui and {Haehnelt}, Martin G. and {Hennawi}, Joseph F. and {Hill}, Tahlia J. and {Hwang}, Seyeon and {Ir{\v{s}}i{\v{c}}}, Vid and {Keating}, Laura and {Kulkarni}, Girish and {Lai}, Samuel and {Ryan-Weber}, Emma V. and {Seyberlich}, Lasse and {Wolfson}, Molly},
        title = "{Nearly pristine gas in a galaxy 1 billion years after the Big Bang}",
      journal = {arXiv e-prints},
         year = 2026,
        month = oct,
          eid = {arXiv:2610.16509},
        pages = {arXiv:2610.16509},
          doi = {10.48550/arXiv.2610.16509},
archivePrefix = {arXiv},
       eprint = {2610.16509},
 primaryClass = {astro-ph.GA},
       adsurl = {https://ui.adsabs.harvard.edu/abs/2026arXiv261016509B}
}

@ARTICLE{2001ApJ...548..509M,
       author = {{Machacek}, Marie E. and {Bryan}, Greg L. and {Abel}, Tom},
        title = "{Simulations of Pregalactic Structure Formation with Radiative Feedback}",
      journal = {\apj},
         year = 2001,
        month = feb,
       volume = {548},
       number = {2},
        pages = {509-521},
          doi = {10.1086/319014},
archivePrefix = {arXiv},
       eprint = {astro-ph/0007198},
 primaryClass = {astro-ph},
       adsurl = {https://ui.adsabs.harvard.edu/abs/2001ApJ...548..509M}
}

@ARTICLE{1998PASP..110..761F,
       author = {{Ferland}, G.~J. and {Korista}, K.~T. and {Verner}, D.~A. and {Ferguson}, J.~W. and {Kingdon}, J.~B. and {Verner}, E.~M.},
        title = "{CLOUDY 90: Numerical Simulation of Plasmas and Their Spectra}",
      journal = {\pasp},
         year = 1998,
        month = jul,
       volume = {110},
       number = {749},
        pages = {761-778},
          doi = {10.1086/316190},
       adsurl = {https://ui.adsabs.harvard.edu/abs/1998PASP..110..761F}
}

@ARTICLE{chatzikos23,
       author = {{Chatzikos}, M. and {Bianchi}, S. and {Camilloni}, F. and {Chakraborty}, P. and {Gunasekera}, C.~M. and {Guzm{\'a}n}, F. and {Milby}, J.~S. and {Sarkar}, A. and {Shaw}, G. and {van Hoof}, P.~A.~M. and {Ferland}, G.~J.},
        title = "{The 2023 Release of Cloudy}",
      journal = {\rmxaa},
         year = 2023,
        month = oct,
       volume = {59},
        pages = {327-343},
          doi = {10.22201/ia.01851101p.2023.59.02.12},
archivePrefix = {arXiv},
       eprint = {2308.06396},
 primaryClass = {astro-ph.GA},
       adsurl = {https://ui.adsabs.harvard.edu/abs/2023RMxAA..59..327C}
}

@ARTICLE{2024ApJ...974..180Y,
       author = {{Yanagisawa}, Hiroto and {Ouchi}, Masami and {Nakajima}, Kimihiko and {Yajima}, Hidenobu and {Umeda}, Hiroya and {Baba}, Shunsuke and {Nakagawa}, Takao and {Nakane}, Minami and {Matsumoto}, Akinori and {Ono}, Yoshiaki and {Harikane}, Yuichi and {Isobe}, Yuki and {Xu}, Yi and {Zhang}, Yechi},
        title = "{Balmer Decrement Anomalies in Galaxies at z {\ensuremath{\sim}} 6 Found by JWST Observations: Density-bounded Nebulae or Excited H I Clouds?}",
      journal = {\apj},
         year = 2024,
        month = oct,
       volume = {974},
       number = {2},
          eid = {180},
        pages = {180},
          doi = {10.3847/1538-4357/ad7097},
archivePrefix = {arXiv},
       eprint = {2403.20118},
 primaryClass = {astro-ph.GA},
       adsurl = {https://ui.adsabs.harvard.edu/abs/2024ApJ...974..180Y}
}

@ARTICLE{2026arXiv260611345I,
       author = {{Isobe}, Yuki and {Curti}, Mirko and {Maiolino}, Roberto and {Duan}, Qiao and {McClymont}, William and {Pusk{\'a}s}, D{\'a}vid and {D'Eugenio}, Francesco and {Rinaldi}, Pierluigi and {Trussler}, James A.~A. and {Scholtz}, Jan and {Looser}, Tobias J. and {Nelson}, Erica and {Ji}, Xihan and {Langeroodi}, Danial and {Tacchella}, Sandro and {Jones}, Gareth C. and {Juod{\v{z}}balis}, Ignas and {Pascalau}, Robert G. and {Hsiao}, Tiger Yu-Yang and {{\"U}bler}, Hannah and {Baker}, William M. and {Bunker}, Andrew J. and {Carniani}, Stefano and {Charlot}, St{\'e}phane and {Curtis-Lake}, Emma and {Geris}, Sophia and {Koller}, Maria and {Lyu}, Jianwei and {Robertson}, Brant and {Williams}, Christina C. and {Wu}, Zihao},
        title = "{JADES: the mass-metallicity relation at $z=1-10$. New calibrations, extremely metal-poor galaxies, and chemical diversity}",
      journal = {arXiv e-prints},
         year = 2026,
        month = jun,
          eid = {arXiv:2606.11345},
        pages = {arXiv:2606.11345},
          doi = {10.48550/arXiv.2606.11345},
archivePrefix = {arXiv},
       eprint = {2606.11345},
 primaryClass = {astro-ph.GA},
       adsurl = {https://ui.adsabs.harvard.edu/abs/2026arXiv260611345I}
}

@ARTICLE{2025arXiv251211790F,
       author = {{Fujimoto}, Seiji and {Asada}, Yoshihisa and {Naidu}, Rohan P. and {Chisholm}, John and {Atek}, Hakim and {Brammer}, Gabriel and {Berg}, Danielle A. and {Schaerer}, Daniel and {Kokorev}, Vasily and {Furtak}, Lukas J. and {Richard}, Johan and {Venditti}, Alessandra and {Bromm}, Volker and {Adamo}, Angela and {Claeyssens}, Adelaide and {Dessauges-Zavadsky}, Miroslava and {Fei}, Qinyue and {Hsiao}, Tiger Yu-Yang and {Korber}, Damien and {Munoz}, Julian B. and {Pan}, Richard and {Saldana-Lopez}, Alberto},
        title = "{GLIMPSE-D: An Exotic Balmer-Jump Object at z=6.20? Revisiting Photometric Selection and the Cosmic Abundance of Pop III Galaxies}",
      journal = {arXiv e-prints},
         year = 2025,
        month = dec,
          eid = {arXiv:2512.11790},
        pages = {arXiv:2512.11790},
          doi = {10.48550/arXiv.2512.11790},
archivePrefix = {arXiv},
       eprint = {2512.11790},
 primaryClass = {astro-ph.GA},
       adsurl = {https://ui.adsabs.harvard.edu/abs/2025arXiv251211790F}
}

@ARTICLE{2001ApJ...546..635O,
       author = {{Omukai}, Kazuyuki},
        title = "{Primordial Star Formation under Far-Ultraviolet Radiation}",
      journal = {\apj},
         year = 2001,
        month = jan,
       volume = {546},
       number = {2},
        pages = {635-651},
          doi = {10.1086/318296},
archivePrefix = {arXiv},
       eprint = {astro-ph/0011446},
 primaryClass = {astro-ph},
       adsurl = {https://ui.adsabs.harvard.edu/abs/2001ApJ...546..635O}
}

@ARTICLE{2024ApJ...970...14S,
       author = {{Sugimura}, Kazuyuki and {Ricotti}, Massimo and {Park}, Jongwon and {Garcia}, Fred Angelo Batan and {Yajima}, Hidenobu},
        title = "{Violent Starbursts and Quiescence Induced by Far-ultraviolet Radiation Feedback in Metal-poor Galaxies at High Redshift}",
      journal = {\apj},
         year = 2024,
        month = jul,
       volume = {970},
       number = {1},
          eid = {14},
        pages = {14},
          doi = {10.3847/1538-4357/ad499a},
archivePrefix = {arXiv},
       eprint = {2403.04824},
 primaryClass = {astro-ph.GA},
       adsurl = {https://ui.adsabs.harvard.edu/abs/2024ApJ...970...14S}
}

@ARTICLE{2011ApJ...727..110C,
       author = {{Clark}, Paul C. and {Glover}, Simon C.~O. and {Klessen}, Ralf S. and {Bromm}, Volker},
        title = "{Gravitational Fragmentation in Turbulent Primordial Gas and the Initial Mass Function of Population III Stars}",
      journal = {\apj},
         year = 2011,
        month = feb,
       volume = {727},
       number = {2},
          eid = {110},
        pages = {110},
          doi = {10.1088/0004-637X/727/2/110},
archivePrefix = {arXiv},
       eprint = {1006.1508},
 primaryClass = {astro-ph.GA},
       adsurl = {https://ui.adsabs.harvard.edu/abs/2011ApJ...727..110C}
}

@ARTICLE{1999ApJ...527L...5B,
       author = {{Bromm}, Volker and {Coppi}, Paolo S. and {Larson}, Richard B.},
        title = "{Forming the First Stars in the Universe: The Fragmentation of Primordial Gas}",
      journal = {\apjl},
         year = 1999,
        month = dec,
       volume = {527},
       number = {1},
        pages = {L5-L8},
          doi = {10.1086/312385},
archivePrefix = {arXiv},
       eprint = {astro-ph/9910224},
 primaryClass = {astro-ph},
       adsurl = {https://ui.adsabs.harvard.edu/abs/1999ApJ...527L...5B}
}

@ARTICLE{2024ApJ...962...49B,
       author = {{Bovill}, Mia Sauda and {Stiavelli}, Massimo and {Wiggins}, Alessa Ibrahim and {Ricotti}, Massimo and {Trenti}, Michele},
        title = "{Kindling the First Stars. I. Dependence of Detectability of the First Stars with JWST on the Population III Stellar Masses}",
      journal = {\apj},
         year = 2024,
        month = feb,
       volume = {962},
       number = {1},
          eid = {49},
        pages = {49},
          doi = {10.3847/1538-4357/ad148a},
archivePrefix = {arXiv},
       eprint = {2210.10190},
 primaryClass = {astro-ph.GA},
       adsurl = {https://ui.adsabs.harvard.edu/abs/2024ApJ...962...49B}
}

@ARTICLE{2023MNRAS.522..330I,
       author = {{Incatasciato}, Andrea and {Khochfar}, Sadegh and {O{\~n}orbe}, Jose},
        title = "{Modelling the cosmological Lyman-Werner background radiation field in the early Universe}",
      journal = {\mnras},
         year = 2023,
        month = jun,
       volume = {522},
       number = {1},
        pages = {330-349},
          doi = {10.1093/mnras/stad1008},
archivePrefix = {arXiv},
       eprint = {2301.08242},
 primaryClass = {astro-ph.GA},
       adsurl = {https://ui.adsabs.harvard.edu/abs/2023MNRAS.522..330I}
}

@ARTICLE{2002ApJ...567..532H,
       author = {{Heger}, A. and {Woosley}, S.~E.},
        title = "{The Nucleosynthetic Signature of Population III}",
      journal = {\apj},
         year = 2002,
        month = mar,
       volume = {567},
       number = {1},
        pages = {532-543},
          doi = {10.1086/338487},
archivePrefix = {arXiv},
       eprint = {astro-ph/0107037},
 primaryClass = {astro-ph},
       adsurl = {https://ui.adsabs.harvard.edu/abs/2002ApJ...567..532H}
}

@ARTICLE{2026arXiv260304517G,
       author = {{Goovaerts}, Ilias and {Rafelski}, Marc and {Beckett}, Alexander and {Oyarz{\`u}n}, Grecco and {Citro}, Annalisa and {Hasan}, Farhanul and {Nedkova}, Kalina V and {Hawcroft}, Calum and {Koekemoer}, Anton M and {Revalski}, Mitchell and {Hayes}, Matthew J and {Scarlata}, Claudia and {Lucas}, Ray A and {Grogin}, Norman A and {Stark}, David V and {Suin}, Paolo and {Pirzkal}, Nor},
        title = "{MXDFz4.4: A LyC emitter 250Myr after the epoch of reionization and a first test of Ly-alpha morphology as a tracer of LyC escape at high redshift}",
      journal = {arXiv e-prints},
         year = 2026,
        month = mar,
          eid = {arXiv:2603.04517},
        pages = {arXiv:2603.04517},
          doi = {10.48550/arXiv.2603.04517},
archivePrefix = {arXiv},
       eprint = {2603.04517},
 primaryClass = {astro-ph.GA},
       adsurl = {https://ui.adsabs.harvard.edu/abs/2026arXiv260304517G}
}

@ARTICLE{2008MNRAS.387.1021G,
       author = {{Greif}, Thomas H. and {Johnson}, Jarrett L. and {Klessen}, Ralf S. and {Bromm}, Volker},
        title = "{The first galaxies: assembly, cooling and the onset of turbulence}",
      journal = {\mnras},
         year = 2008,
        month = jul,
       volume = {387},
       number = {3},
        pages = {1021-1036},
          doi = {10.1111/j.1365-2966.2008.13326.x},
archivePrefix = {arXiv},
       eprint = {0803.2237},
 primaryClass = {astro-ph},
       adsurl = {https://ui.adsabs.harvard.edu/abs/2008MNRAS.387.1021G}
}

@ARTICLE{2005Natur.434..871F,
       author = {{Frebel}, Anna and {Aoki}, Wako and {Christlieb}, Norbert and {Ando}, Hiroyasu and {Asplund}, Martin and {Barklem}, Paul S. and {Beers}, Timothy C. and {Eriksson}, Kjell and {Fechner}, Cora and {Fujimoto}, Masayuki Y. and {Honda}, Satoshi and {Kajino}, Toshitaka and {Minezaki}, Takeo and {Nomoto}, Ken'ichi and {Norris}, John E. and {Ryan}, Sean G. and {Takada-Hidai}, Masahide and {Tsangarides}, Stelios and {Yoshii}, Yuzuru},
        title = "{Nucleosynthetic signatures of the first stars}",
      journal = {\nat},
         year = 2005,
        month = apr,
       volume = {434},
       number = {7035},
        pages = {871-873},
          doi = {10.1038/nature03455},
archivePrefix = {arXiv},
       eprint = {astro-ph/0503021},
 primaryClass = {astro-ph},
       adsurl = {https://ui.adsabs.harvard.edu/abs/2005Natur.434..871F}
}

@ARTICLE{2002Natur.419..904C,
       author = {{Christlieb}, N. and {Bessell}, M.~S. and {Beers}, T.~C. and {Gustafsson}, B. and {Korn}, A. and {Barklem}, P.~S. and {Karlsson}, T. and {Mizuno-Wiedner}, M. and {Rossi}, S.},
        title = "{A stellar relic from the early Milky Way}",
      journal = {\nat},
         year = 2002,
        month = oct,
       volume = {419},
       number = {6910},
        pages = {904-906},
          doi = {10.1038/nature01142},
archivePrefix = {arXiv},
       eprint = {astro-ph/0211274},
 primaryClass = {astro-ph},
       adsurl = {https://ui.adsabs.harvard.edu/abs/2002Natur.419..904C}
}

@ARTICLE{2009MNRAS.399...37J,
       author = {{Johnson}, Jarrett L. and {Greif}, Thomas H. and {Bromm}, Volker and {Klessen}, Ralf S. and {Ippolito}, Joseph},
        title = "{The first galaxies: signatures of the initial starburst}",
      journal = {\mnras},
         year = 2009,
        month = oct,
       volume = {399},
       number = {1},
        pages = {37-47},
          doi = {10.1111/j.1365-2966.2009.15158.x},
archivePrefix = {arXiv},
       eprint = {0902.3263},
 primaryClass = {astro-ph.CO},
       adsurl = {https://ui.adsabs.harvard.edu/abs/2009MNRAS.399...37J}
}

@ARTICLE{2025MNRAS.539.2561C,
       author = {{Chon}, Sunmyon and {Omukai}, Kazuyuki},
        title = "{Formation of supermassive stars and dense star clusters in metal-poor clouds exposed to strong FUV radiation}",
      journal = {\mnras},
         year = 2025,
        month = may,
       volume = {539},
       number = {3},
        pages = {2561-2582},
          doi = {10.1093/mnras/staf598},
archivePrefix = {arXiv},
       eprint = {2412.14900},
 primaryClass = {astro-ph.GA},
       adsurl = {https://ui.adsabs.harvard.edu/abs/2025MNRAS.539.2561C}
}

@ARTICLE{2021MNRAS.508.4175C,
       author = {{Chon}, Sunmyon and {Omukai}, Kazuyuki and {Schneider}, Raffaella},
        title = "{Transition of the initial mass function in the metal-poor environments}",
      journal = {\mnras},
         year = 2021,
        month = dec,
       volume = {508},
       number = {3},
        pages = {4175-4192},
          doi = {10.1093/mnras/stab2497},
archivePrefix = {arXiv},
       eprint = {2103.04997},
 primaryClass = {astro-ph.GA},
       adsurl = {https://ui.adsabs.harvard.edu/abs/2021MNRAS.508.4175C}
}

@ARTICLE{2022A&A...666A.199S,
       author = {{Sibony}, Y. and {Liu}, B. and {Simmonds}, C. and {Meynet}, G. and {Bromm}, V.},
        title = "{Impact of Population III homogeneous stellar evolution on early cosmic reionisation}",
      journal = {\aap},
         year = 2022,
        month = oct,
       volume = {666},
          eid = {A199},
        pages = {A199},
          doi = {10.1051/0004-6361/202244146},
archivePrefix = {arXiv},
       eprint = {2205.15125},
 primaryClass = {astro-ph.SR},
       adsurl = {https://ui.adsabs.harvard.edu/abs/2022A&A...666A.199S}
}

@ARTICLE{2026MNRAS.547ag386W,
       author = {{Wasserman}, Joel and {Zackrisson}, Erik and {Dhandha}, Jiten and {Fialkov}, Anastasia and {Noble}, Leon and {Majumdar}, Suman},
        title = "{Ultraviolet photon production rates of the first stars: Impact on the He II {\ensuremath{\lambda}} 1640 {\r{A}} emission line from primordial star clusters and the 21-cm signal from cosmic dawn}",
      journal = {\mnras},
         year = 2026,
        month = apr,
       volume = {547},
       number = {3},
          eid = {stag386},
        pages = {stag386},
          doi = {10.1093/mnras/stag386},
archivePrefix = {arXiv},
       eprint = {2507.21764},
 primaryClass = {astro-ph.GA},
       adsurl = {https://ui.adsabs.harvard.edu/abs/2026MNRAS.547ag386W}
}

@ARTICLE{2010MNRAS.402.1249S,
       author = {{Shang}, Cien and {Bryan}, Greg L. and {Haiman}, Z.},
        title = "{Supermassive black hole formation by direct collapse: keeping protogalactic gas H$_{2}$ free in dark matter haloes with virial temperatures T$_{vir}$ > rsim {}10$^{4}$ K}",
      journal = {\mnras},
         year = 2010,
        month = feb,
       volume = {402},
       number = {2},
        pages = {1249-1262},
          doi = {10.1111/j.1365-2966.2009.15960.x},
archivePrefix = {arXiv},
       eprint = {0906.4773},
 primaryClass = {astro-ph.CO},
       adsurl = {https://ui.adsabs.harvard.edu/abs/2010MNRAS.402.1249S}
}

@ARTICLE{mahler18,
       author = {{Mahler}, G. and {Richard}, J. and {Cl{\'e}ment}, B. and {Lagattuta}, D. and {Schmidt}, K. and {Patr{\'\i}cio}, V. and {Soucail}, G. and {Bacon}, R. and {Pello}, R. and {Bouwens}, R. and {Maseda}, M. and {Martinez}, J. and {Carollo}, M. and {Inami}, H. and {Leclercq}, F. and {Wisotzki}, L.},
        title = "{Strong-lensing analysis of A2744 with MUSE and Hubble Frontier Fields images}",
      journal = {\mnras},
         year = 2018,
        month = jan,
       volume = {473},
       number = {1},
        pages = {663-692},
          doi = {10.1093/mnras/stx1971},
archivePrefix = {arXiv},
       eprint = {1702.06962},
 primaryClass = {astro-ph.GA},
       adsurl = {https://ui.adsabs.harvard.edu/abs/2018MNRAS.473..663M}
}

@ARTICLE{tornatore07,
       author = {{Tornatore}, Luca and {Ferrara}, Andrea and {Schneider}, Raffaella},
        title = "{Population III stars: hidden or disappeared?}",
      journal = {\mnras},
         year = 2007,
        month = dec,
       volume = {382},
       number = {3},
        pages = {945-950},
          doi = {10.1111/j.1365-2966.2007.12215.x},
archivePrefix = {arXiv},
       eprint = {0707.1433},
 primaryClass = {astro-ph},
       adsurl = {https://ui.adsabs.harvard.edu/abs/2007MNRAS.382..945T}
}

@ARTICLE{cai25,
       author = {{Cai}, Sijia and {Li}, Mingyu and {Cai}, Zheng and {Wu}, Yunjing and {Yu}, Fujiang and {Dickinson}, Mark and {Sun}, Fengwu and {Fan}, Xiaohui and {Wang}, Ben and {Cullen}, Fergus and {Bian}, Fuyan and {Lin}, Xiaojing and {Zou}, Jiaqi},
        title = "{A Metal-free Galaxy at z = 3.19? Evidence of Late Population III Star Formation at Cosmic Noon}",
      journal = {\apjl},
         year = 2025,
        month = nov,
       volume = {993},
       number = {2},
          eid = {L52},
        pages = {L52},
          doi = {10.3847/2041-8213/ae1608},
archivePrefix = {arXiv},
       eprint = {2507.17820},
 primaryClass = {astro-ph.GA},
       adsurl = {https://ui.adsabs.harvard.edu/abs/2025ApJ...993L..52C}
}

@ARTICLE{ubler26,
       author = {{{\"U}bler}, Hannah and {Maiolino}, Roberto and {P{\'e}rez-Gonz{\'a}lez}, Pablo G. and {Isobe}, Yuki and {Jones}, Gareth C. and {Kumari}, Nimisha and {Charlot}, St{\'e}phane and {Rusta}, Elka and {Salvadori}, Stefania and {Nakajima}, Kimihiko and {Perna}, Michele and {Arribas}, Santiago and {Bunker}, Andrew J. and {Carniani}, Stefano and {D'Eugenio}, Francesco and {Rodr{\'\i}guez Del Pino}, Bruno and {Bertola}, Elena and {B{\"o}ker}, Torsten and {Chevallard}, Jacopo and {Circosta}, Chiara and {Cresci}, Giovanni and {Curti}, Mirko and {Curtis-Lake}, Emma and {Eisenstein}, Daniel J. and {Hainline}, Kevin and {Johnson}, Benjamin D. and {Parlanti}, Eleonora and {Rinaldi}, Pierluigi and {Robertson}, Brant and {Scholtz}, Jan and {Tacchella}, Sandro and {Venturi}, Giacomo and {Witstok}, Joris and {Zamora}, Sandra},
        title = "{GA-NIFS \& JADES: Confirmation of pristine gas near GN-z11}",
      journal = {arXiv e-prints},
         year = 2026,
        month = mar,
          eid = {arXiv:2603.20360},
        pages = {arXiv:2603.20360},
          doi = {10.48550/arXiv.2603.20360},
archivePrefix = {arXiv},
       eprint = {2603.20360},
 primaryClass = {astro-ph.GA},
       adsurl = {https://ui.adsabs.harvard.edu/abs/2026arXiv260320360U}
}

@ARTICLE{maiolino26,
       author = {{Maiolino}, Roberto and {{\"U}bler}, Hannah and {Perna}, Michele and {Witstok}, Joris and {Jones}, Gareth C. and {Perez-Gonzalez}, Pablo G. and {Nakajima}, Kimihiko and {Rusta}, Elka and {Salvadori}, Stefania and {Tacchella}, Sandro and {Madau}, Piero and {Trussler}, James A.~A. and {D'Eugenio}, Francesco and {Ji}, Xihan and {Scholtz}, Jan and {Carniani}, Stefano and {Isobe}, Yuki and {Katz}, Harley and {Arribas}, Santiago and {Baker}, William M. and {B{\"o}ker}, Torsten and {Bromm}, Volker and {Bunker}, Andrew J. and {Charlot}, Stephane and {Chevallard}, Jacopo and {Curti}, Mirko and {Curtis-Lake}, Emma and {Eisenstein}, Daniel and {Egami}, Eiichi and {Ferrara}, Andrea and {Graziani}, Luca and {Hainline}, Kevin and {Helton}, Jakob M. and {Ivey}, Lucy and {Jonson}, Benjamin and {Koller}, Maria and {Kumari}, Nimisha and {Marconi}, Alessandro and {Mazzolari}, Giovanni and {Laporte}, Nicolas and {Parlanti}, Eleonora and {Pascalau}, Robert and {Pentericci}, Laura and {Rinaldi}, Pierluigi and {Robertson}, Brant and {Rodr{\'\i}guez Del Pino}, Bruno and {Schneider}, Raffaella and {Venditti}, Alessandra and {Venturi}, Giacomo and {Willmer}, Christopher N.~A. and {Witten}, Callum and {Zamora}, Sandra},
        title = "{The search for Population III: Confirmation of a HeII emitter with no metal lines at z=10.6}",
      journal = {arXiv e-prints},
         year = 2026,
        month = mar,
          eid = {arXiv:2603.20362},
        pages = {arXiv:2603.20362},
          doi = {10.48550/arXiv.2603.20362},
archivePrefix = {arXiv},
       eprint = {2603.20362},
 primaryClass = {astro-ph.GA},
       adsurl = {https://ui.adsabs.harvard.edu/abs/2026arXiv260320362M}
}

@ARTICLE{scholte25,
       author = {{Scholte}, D. and {Cullen}, F. and {Carnall}, A.~C. and {Arellano-C{\'o}rdova}, K.~Z. and {Stanton}, T.~M. and {Barrufet}, L. and {Begley}, R. and {Bondestam}, C. and {Donnan}, C.~T. and {Dunlop}, J.~S. and {Leung}, H.-H. and {McLeod}, D.~J. and {McLure}, R.~J. and {Moustakas}, J.~M. and {Pollock}, C.~L. and {Shapley}, A.~E. and {Stevenson}, S. and {Zou}, H.},
        title = "{The JWST EXCELS survey: probing strong-line diagnostics and the chemical evolution of galaxies over cosmic time using T$_{e}$-metallicities}",
      journal = {\mnras},
         year = 2025,
        month = jun,
       volume = {540},
       number = {2},
        pages = {1800-1826},
          doi = {10.1093/mnras/staf834},
archivePrefix = {arXiv},
       eprint = {2502.10499},
 primaryClass = {astro-ph.GA},
       adsurl = {https://ui.adsabs.harvard.edu/abs/2025MNRAS.540.1800S}
}

@ARTICLE{napolitano25,
       author = {{Napolitano}, L. and {Castellano}, M. and {Pentericci}, L. and {Arrabal Haro}, P. and {Fontana}, A. and {Treu}, T. and {Bergamini}, P. and {Calabr{\`o}}, A. and {Mascia}, S. and {Morishita}, T. and {Roberts-Borsani}, G. and {Santini}, P. and {Vanzella}, E. and {Vulcani}, B. and {Zakharova}, D. and {Bakx}, T. and {Dickinson}, M. and {Grillo}, C. and {Leethochawalit}, N. and {Llerena}, M. and {Merlin}, E. and {Paris}, D. and {Rojas-Ruiz}, S. and {Rosati}, P. and {Wang}, X. and {Yoon}, I. and {Zavala}, J.},
        title = "{Seven wonders of Cosmic Dawn: JWST confirms a high abundance of galaxies and AGN at z ≃ 9─11 in the GLASS field}",
      journal = {\aap},
         year = 2025,
        month = jan,
       volume = {693},
          eid = {A50},
        pages = {A50},
          doi = {10.1051/0004-6361/202452090},
archivePrefix = {arXiv},
       eprint = {2410.10967},
 primaryClass = {astro-ph.GA},
       adsurl = {https://ui.adsabs.harvard.edu/abs/2025A&A...693A..50N}
}

@ARTICLE{roberts-borsani23,
       author = {{Roberts-Borsani}, Guido and {Treu}, Tommaso and {Chen}, Wenlei and {Morishita}, Takahiro and {Vanzella}, Eros and {Zitrin}, Adi and {Bergamini}, Pietro and {Castellano}, Marco and {Fontana}, Adriano and {Glazebrook}, Karl and {Grillo}, Claudio and {Kelly}, Patrick L. and {Merlin}, Emiliano and {Nanayakkara}, Themiya and {Paris}, Diego and {Rosati}, Piero and {Yang}, Lilan and {Acebron}, Ana and {Bonchi}, Andrea and {Boyett}, Kit and {Brada{\v{c}}}, Maru{\v{s}}a and {Brammer}, Gabriel and {Broadhurst}, Tom and {Calabr{\'o}}, Antonello and {Diego}, Jose M. and {Dressler}, Alan and {Furtak}, Lukas J. and {Filippenko}, Alexei V. and {Henry}, Alaina and {Koekemoer}, Anton M. and {Leethochawalit}, Nicha and {Malkan}, Matthew A. and {Mason}, Charlotte and {Mercurio}, Amata and {Metha}, Benjamin and {Pentericci}, Laura and {Pierel}, Justin and {Rieck}, Steven and {Roy}, Namrata and {Santini}, Paola and {Strait}, Victoria and {Strausbaugh}, Robert and {Trenti}, Michele and {Vulcani}, Benedetta and {Wang}, Lifan and {Wang}, Xin and {Windhorst}, Rogier A.},
        title = "{The nature of an ultra-faint galaxy in the cosmic dark ages seen with JWST}",
      journal = {\nat},
         year = 2023,
        month = jun,
       volume = {618},
       number = {7965},
        pages = {480-483},
          doi = {10.1038/s41586-023-05994-w},
archivePrefix = {arXiv},
       eprint = {2210.15639},
 primaryClass = {astro-ph.GA},
       adsurl = {https://ui.adsabs.harvard.edu/abs/2023Natur.618..480R}
}

@ARTICLE{rodighiero26,
       author = {{Rodighiero}, Giulia and {Ferrara}, Andrea and {Catone}, Michele and {Napolitano}, Lorenzo and {Cassata}, Paolo and {Gandolfi}, Giovanni and {Merlin}, Emiliano and {Grazian}, Andrea and {Renzini}, Alvio and {Bisigello}, Laura and {Castellano}, Marco and {P{\'e}rez-Gonz{\'a}lez}, Pablo G. and {P{\'e}rez-D{\'\i}az}, Borja and {Iani}, Edoardo and {Gruppioni}, Carlotta and {Finkelstein}, Steven L. and {Koekemoer}, Anton M. and {Bianchetti}, Alessandro and {Sinigaglia}, Francesco},
        title = "{EGS-z11-R0: a red, dust-rich galaxy at Cosmic Dawn}",
      journal = {arXiv e-prints},
         year = 2026,
        month = mar,
          eid = {arXiv:2603.15841},
        pages = {arXiv:2603.15841},
          doi = {10.48550/arXiv.2603.15841},
archivePrefix = {arXiv},
       eprint = {2603.15841},
 primaryClass = {astro-ph.GA},
       adsurl = {https://ui.adsabs.harvard.edu/abs/2026arXiv260315841R}
}

@ARTICLE{bouwens10,
       author = {{Bouwens}, R.~J. and {Illingworth}, G.~D. and {Oesch}, P.~A. and {Trenti}, M. and {Stiavelli}, M. and {Carollo}, C.~M. and {Franx}, M. and {van Dokkum}, P.~G. and {Labb{\'e}}, I. and {Magee}, D.},
        title = "{Very Blue UV-Continuum Slope {\ensuremath{\beta}} of Low Luminosity z \raisebox{-0.5ex}\textasciitilde 7 Galaxies from WFC3/IR: Evidence for Extremely Low Metallicities?}",
      journal = {\apjl},
         year = 2010,
        month = jan,
       volume = {708},
       number = {2},
        pages = {L69-L73},
          doi = {10.1088/2041-8205/708/2/L69},
archivePrefix = {arXiv},
       eprint = {0910.0001},
 primaryClass = {astro-ph.CO},
       adsurl = {https://ui.adsabs.harvard.edu/abs/2010ApJ...708L..69B}
}

@ARTICLE{katz25,
       author = {{Katz}, Harley and {Cameron}, Alex J. and {Saxena}, Aayush and {Barrufet}, Laia and {Choustikov}, Nichloas and {Cleri}, Nikko J. and {de Graff}, Anna and {Ellis}, Richard S. and {Fosbury}, Robert A.~E. and {Heintz}, Kasper E. and {Maseda}, Michael and {Matthee}, Jorryt and {McConachie}, Ian and {Oesch}, Pascal A.},
        title = "{21 Balmer Jump Street: The Nebular Continuum at High Redshift and Implications for the Bright Galaxy Problem, UV Continuum Slopes, and Early Stellar Populations}",
      journal = {The Open Journal of Astrophysics},
         year = 2025,
        month = jul,
       volume = {8},
          eid = {104},
        pages = {104},
          doi = {10.33232/001c.142570},
archivePrefix = {arXiv},
       eprint = {2408.03189},
 primaryClass = {astro-ph.GA},
       adsurl = {https://ui.adsabs.harvard.edu/abs/2025OJAp....8E.104K}
}

@ARTICLE{mingozzi22,
       author = {{Mingozzi}, Matilde and {James}, Bethan L. and {Arellano-C{\'o}rdova}, Karla Z. and {Berg}, Danielle A. and {Senchyna}, Peter and {Chisholm}, John and {Brinchmann}, Jarle and {Aloisi}, Alessandra and {Amor{\'\i}n}, Ricardo O. and {Charlot}, St{\'e}phane and {Feltre}, Anna and {Hayes}, Matthew and {Heckman}, Timothy and {Henry}, Alaina and {Hernandez}, Svea and {Kumari}, Nimisha and {Leitherer}, Claus and {Llerena}, Mario and {Martin}, Crystal L. and {Nanayakkara}, Themiya and {Ravindranath}, Swara and {Skillman}, Evan D. and {Sugahara}, Yuma and {Wofford}, Aida and {Xu}, Xinfeng},
        title = "{CLASSY IV. Exploring UV Diagnostics of the Interstellar Medium in Local High-z Analogs at the Dawn of the JWST Era}",
      journal = {\apj},
         year = 2022,
        month = nov,
       volume = {939},
       number = {2},
          eid = {110},
        pages = {110},
          doi = {10.3847/1538-4357/ac952c},
archivePrefix = {arXiv},
       eprint = {2209.09047},
 primaryClass = {astro-ph.GA},
       adsurl = {https://ui.adsabs.harvard.edu/abs/2022ApJ...939..110M}
}

@ARTICLE{morishita25,
       author = {{Morishita}, Takahiro and {Liu}, Zhaoran and {Stiavelli}, Massimo and {Treu}, Tommaso and {Bergamini}, Pietro and {Zhang}, Yechi},
        title = "{Pristine Massive Star Formation Caught at the Break of Cosmic Dawn}",
      journal = {arXiv e-prints},
         year = 2025,
        month = jul,
          eid = {arXiv:2507.10521},
        pages = {arXiv:2507.10521},
          doi = {10.48550/arXiv.2507.10521},
archivePrefix = {arXiv},
       eprint = {2507.10521},
 primaryClass = {astro-ph.CO},
       adsurl = {https://ui.adsabs.harvard.edu/abs/2025arXiv250710521M}
}

@ARTICLE{vanzella26,
       author = {{Vanzella}, E. and {Messa}, M. and {Zanella}, A. and {Bolamperti}, A. and {Castellano}, M. and {Loiacono}, F. and {Bergamini}, P. and {Roberts Borsani}, G. and {Adamo}, A. and {Fontana}, A. and {Treu}, T. and {Calura}, F. and {Grillo}, C. and {Lombardi}, M. and {Rosati}, P. and {Gilli}, R. and {Meneghetti}, M.},
        title = "{A pristine, star-forming complex at z = 4.19}",
      journal = {\aap},
         year = 2026,
        month = jan,
       volume = {705},
          eid = {L12},
        pages = {L12},
          doi = {10.1051/0004-6361/202557153},
archivePrefix = {arXiv},
       eprint = {2509.07073},
 primaryClass = {astro-ph.GA},
       adsurl = {https://ui.adsabs.harvard.edu/abs/2026A&A...705L..12V}
}

@ARTICLE{trussler26,
       author = {{Trussler}, James A.~A. and {Eisenstein}, Daniel J. and {Bunker}, Andrew J. and {Cameron}, Alex J. and {Carniani}, Stefano and {Charlot}, St{\'e}phane and {Chevallard}, Jacopo and {Conselice}, Christopher J. and {Curti}, Mirko and {Curtis-Lake}, Emma and {D'Eugenio}, Francesco and {Egami}, Eiichi and {Hainline}, Kevin and {Hausen}, Ryan and {Helton}, Jakob M. and {Hsiao}, Tiger Yu-Yang and {Ji}, Zhiyuan and {Johnson}, Benjamin D. and {Looser}, Tobias J. and {Maiolino}, Roberto and {Pusk{\'a}s}, D{\'a}vid and {Rinaldi}, Pierluigi and {Robertson}, Brant and {Sun}, Fengwu and {Tacchella}, Sandro and {{\"U}bler}, Hannah and {Williams}, Christina C. and {Willmer}, Christopher N.~A. and {Witstok}, Joris and {Wu}, Zihao},
        title = "{Out of oxygen: Extremely metal-poor galaxy candidates identified at $2.5 < z < 6.5$ with deep JADES medium-band imaging}",
      journal = {arXiv e-prints},
         year = 2026,
        month = mar,
          eid = {arXiv:2603.15761},
        pages = {arXiv:2603.15761},
          doi = {10.48550/arXiv.2603.15761},
archivePrefix = {arXiv},
       eprint = {2603.15761},
 primaryClass = {astro-ph.GA},
       adsurl = {https://ui.adsabs.harvard.edu/abs/2026arXiv260315761T}
}

@ARTICLE{donnan26,
       author = {{Donnan}, Callum T. and {McLeod}, Derek J. and {McLure}, Ross J. and {Dunlop}, James S. and {Cullen}, Fergus and {Dickinson}, Mark and {Arrabal Haro}, Pablo and {Taylor}, Anthony J. and {Bondestam}, Cecilia and {Liu}, Feng-Yuan and {Arellano-C{\'o}rdova}, Karla Z. and {Barrufet}, Laia and {Begley}, Ryan and {Carnall}, Adam C. and {Golawska}, Hanna and {Leung}, Ho-Hin and {Scholte}, Dirk and {Stanton}, Thomas M.},
        title = "{Spectroscopic Confirmation of a Large and Luminous Galaxy with Weak Emission Lines at z = 13.53}",
      journal = {\apj},
         year = 2026,
        month = may,
       volume = {1002},
       number = {2},
          eid = {134},
        pages = {134},
          doi = {10.3847/1538-4357/ae5c05},
archivePrefix = {arXiv},
       eprint = {2601.11515},
 primaryClass = {astro-ph.GA},
       adsurl = {https://ui.adsabs.harvard.edu/abs/2026ApJ..1002..134D}
}

@ARTICLE{harikane26,
       author = {{Harikane}, Yuichi and {Perez-Gonzalez}, Pablo G. and {Alvarez-Marquez}, Javier and {Ouchi}, Masami and {Nakazato}, Yurina and {Ono}, Yoshiaki and {Nakajima}, Kimihiko and {Umeda}, Hiroya and {Isobe}, Yuki and {Xu}, Yi and {Zhang}, Yechi},
        title = "{A UV-Luminous Galaxy at z=11 with Surprisingly Weak Star Formation Activity}",
      journal = {arXiv e-prints},
         year = 2026,
        month = jan,
          eid = {arXiv:2601.21833},
        pages = {arXiv:2601.21833},
          doi = {10.48550/arXiv.2601.21833},
archivePrefix = {arXiv},
       eprint = {2601.21833},
 primaryClass = {astro-ph.GA},
       adsurl = {https://ui.adsabs.harvard.edu/abs/2026arXiv260121833H}
}

@ARTICLE{naidu25,
       author = {{Naidu}, Rohan P. and {Oesch}, Pascal A. and {Brammer}, Gabriel and {Weibel}, Andrea and {Li}, Yijia and {Matthee}, Jorryt and {Chisholm}, John and {Pollock}, Clara L. and {Heintz}, Kasper E. and {Johnson}, Benjamin D. and {Shen}, Xuejian and {Hviding}, Raphael E. and {Leja}, Joel and {Tacchella}, Sandro and {Ganguly}, Arpita and {Witten}, Callum and {Atek}, Hakim and {Belli}, Sirio and {Bose}, Sownak and {Bouwens}, Rychard and {Dayal}, Pratika and {Decarli}, Roberto and {de Graaff}, Anna and {Fudamoto}, Yoshinobu and {Giovinazzo}, Emma and {Greene}, Jenny E. and {Illingworth}, Garth and {Inoue}, Akio K. and {Kane}, Sarah G. and {Labbe}, Ivo and {Leonova}, Ecaterina and {Marques-Chaves}, Rui and {Meyer}, Romain A. and {Nelson}, Erica J. and {Roberts-Borsani}, Guido and {Schaerer}, Daniel and {Simcoe}, Robert A. and {Stefanon}, Mauro and {Sugahara}, Yuma and {Toft}, Sune and {van der Wel}, Arjen and {van Dokkum}, Pieter and {Walter}, Fabian and {Watson}, Darach and {Weaver}, John R. and {Whitaker}, Katherine E.},
        title = "{A Cosmic Miracle: A Remarkably Luminous Galaxy at $z_{\rm{spec}}=14.44$ Confirmed with JWST}",
      journal = {arXiv e-prints},
         year = 2025,
        month = may,
          eid = {arXiv:2505.11263},
        pages = {arXiv:2505.11263},
          doi = {10.48550/arXiv.2505.11263},
archivePrefix = {arXiv},
       eprint = {2505.11263},
 primaryClass = {astro-ph.GA},
       adsurl = {https://ui.adsabs.harvard.edu/abs/2025arXiv250511263N}
}

@ARTICLE{schouws24,
       author = {{Schouws}, Sander and {Bouwens}, Rychard J. and {Ormerod}, Katherine and {Smit}, Renske and {Algera}, Hiddo and {Sommovigo}, Laura and {Hodge}, Jacqueline and {Ferrara}, Andrea and {Oesch}, Pascal A. and {Rowland}, Lucie E. and {van Leeuwen}, Ivana and {Stefanon}, Mauro and {Herard-Demanche}, Thomas and {Fudamoto}, Yoshinobu and {R{\"o}ttgering}, Huub and {van der Werf}, Paul},
        title = "{Detection of [OIII]88$\mu$m in JADES-GS-z14-0 at z=14.1793}",
      journal = {arXiv e-prints},
         year = 2024,
        month = sep,
          eid = {arXiv:2409.20549},
        pages = {arXiv:2409.20549},
          doi = {10.48550/arXiv.2409.20549},
archivePrefix = {arXiv},
       eprint = {2409.20549},
 primaryClass = {astro-ph.GA},
       adsurl = {https://ui.adsabs.harvard.edu/abs/2024arXiv240920549S}
}

@ARTICLE{carniani24,
       author = {{Carniani}, Stefano and {Hainline}, Kevin and {D'Eugenio}, Francesco and {Eisenstein}, Daniel J. and {Jakobsen}, Peter and {Witstok}, Joris and {Johnson}, Benjamin D. and {Chevallard}, Jacopo and {Maiolino}, Roberto and {Helton}, Jakob M. and {Willott}, Chris and {Robertson}, Brant and {Alberts}, Stacey and {Arribas}, Santiago and {Baker}, William M. and {Bhatawdekar}, Rachana and {Boyett}, Kristan and {Bunker}, Andrew J. and {Cameron}, Alex J. and {Cargile}, Phillip A. and {Charlot}, St{\'e}phane and {Curti}, Mirko and {Curtis-Lake}, Emma and {Egami}, Eiichi and {Giardino}, Giovanna and {Isaak}, Kate and {Ji}, Zhiyuan and {Jones}, Gareth C. and {Kumari}, Nimisha and {Maseda}, Michael V. and {Parlanti}, Eleonora and {P{\'e}rez-Gonz{\'a}lez}, Pablo G. and {Rawle}, Tim and {Rieke}, George and {Rieke}, Marcia and {Del Pino}, Bruno Rodr{\'\i}guez and {Saxena}, Aayush and {Scholtz}, Jan and {Smit}, Renske and {Sun}, Fengwu and {Tacchella}, Sandro and {{\"U}bler}, Hannah and {Venturi}, Giacomo and {Williams}, Christina C. and {Willmer}, Christopher N.~A.},
        title = "{Spectroscopic confirmation of two luminous galaxies at a redshift of 14}",
      journal = {\nat},
         year = 2024,
        month = sep,
       volume = {633},
       number = {8029},
        pages = {318-322},
          doi = {10.1038/s41586-024-07860-9},
archivePrefix = {arXiv},
       eprint = {2405.18485},
 primaryClass = {astro-ph.GA},
       adsurl = {https://ui.adsabs.harvard.edu/abs/2024Natur.633..318C}
}

@ARTICLE{castellano24,
       author = {{Castellano}, Marco and {Napolitano}, Lorenzo and {Fontana}, Adriano and {Roberts-Borsani}, Guido and {Treu}, Tommaso and {Vanzella}, Eros and {Zavala}, Jorge A. and {Arrabal Haro}, Pablo and {Calabr{\`o}}, Antonello and {Llerena}, Mario and {Mascia}, Sara and {Merlin}, Emiliano and {Paris}, Diego and {Pentericci}, Laura and {Santini}, Paola and {Bakx}, Tom J.~L.~C. and {Bergamini}, Pietro and {Cupani}, Guido and {Dickinson}, Mark and {Filippenko}, Alexei V. and {Glazebrook}, Karl and {Grillo}, Claudio and {Kelly}, Patrick L. and {Malkan}, Matthew A. and {Mason}, Charlotte A. and {Morishita}, Takahiro and {Nanayakkara}, Themiya and {Rosati}, Piero and {Sani}, Eleonora and {Wang}, Xin and {Yoon}, Ilsang},
        title = "{JWST NIRSpec Spectroscopy of the Remarkable Bright Galaxy GHZ2/GLASS-z12 at Redshift 12.34}",
      journal = {\apj},
         year = 2024,
        month = sep,
       volume = {972},
       number = {2},
          eid = {143},
        pages = {143},
          doi = {10.3847/1538-4357/ad5f88},
archivePrefix = {arXiv},
       eprint = {2403.10238},
 primaryClass = {astro-ph.GA},
       adsurl = {https://ui.adsabs.harvard.edu/abs/2024ApJ...972..143C}
}

@ARTICLE{bromm03,
       author = {{Bromm}, Volker and {Loeb}, Abraham},
        title = "{The formation of the first low-mass stars from gas with low carbon and oxygen abundances}",
      journal = {\nat},
         year = 2003,
        month = oct,
       volume = {425},
       number = {6960},
        pages = {812-814},
          doi = {10.1038/nature02071},
archivePrefix = {arXiv},
       eprint = {astro-ph/0310622},
 primaryClass = {astro-ph},
       adsurl = {https://ui.adsabs.harvard.edu/abs/2003Natur.425..812B}
}

@ARTICLE{Wise2012,
       author = {{Wise}, John H. and {Turk}, Matthew J. and {Norman}, Michael L. and {Abel}, Tom},
        title = "{The Birth of a Galaxy: Primordial Metal Enrichment and Stellar Populations}",
      journal = {\apj},
         year = 2012,
        month = jan,
       volume = {745},
       number = {1},
          eid = {50},
        pages = {50},
          doi = {10.1088/0004-637X/745/1/50},
archivePrefix = {arXiv},
       eprint = {1011.2632},
 primaryClass = {astro-ph.CO},
       adsurl = {https://ui.adsabs.harvard.edu/abs/2012ApJ...745...50W}
}

@ARTICLE{ji26,
       author = {{Ji}, Alexander P. and {Chandra}, Vedant and {Mejias-Torres}, Selenna and {Zhang}, Zhongyuan and {Eitner}, Philipp and {Schlaufman}, Kevin C. and {Andales}, Hillary Diane and {Do}, Ha and {Orrantia}, Natalie M. and {Tudmilla}, Rithika and {Thibodeaux}, Pierre N. and {Stassun}, Keivan G. and {Howell}, Madeline and {Tayar}, Jamie and {Bergemann}, Maria and {Casey}, Andrew R. and {Johnson}, Jennifer A. and {Carlberg}, Joleen K. and {Cerny}, William and {Fern{\'a}ndez-Trincado}, Jos{\'e} G. and {Hawkins}, Keith and {Kollmeier}, Juna A. and {Laporte}, Chervin F.~P. and {Limberg}, Guilherme and {Matsuno}, Tadafumi and {M{\'e}sz{\'a}ros}, Szabolcs and {Morrison}, Sean and {Nidever}, David L. and {Stringfellow}, Guy S. and {Schneider}, Donald P. and {Thai}, Riley},
        title = "{A nearly pristine star from the Large Magellanic Cloud}",
      journal = {Nature Astronomy},
         year = 2026,
        month = apr,
          doi = {10.1038/s41550-026-02816-7},
archivePrefix = {arXiv},
       eprint = {2509.21643},
 primaryClass = {astro-ph.SR},
       adsurl = {https://ui.adsabs.harvard.edu/abs/2026NatAs.tmp...73J}
}

@ARTICLE{fumagalli11,
       author = {{Fumagalli}, Michele and {O'Meara}, John M. and {Prochaska}, J. Xavier},
        title = "{Detection of Pristine Gas Two Billion Years After the Big Bang}",
      journal = {Science},
         year = 2011,
        month = dec,
       volume = {334},
       number = {6060},
        pages = {1245},
          doi = {10.1126/science.1213581},
archivePrefix = {arXiv},
       eprint = {1111.2334},
 primaryClass = {astro-ph.CO},
       adsurl = {https://ui.adsabs.harvard.edu/abs/2011Sci...334.1245F}
}

@ARTICLE{planck20,
       author = {{Planck Collaboration} and {Aghanim}, N. and {Akrami}, Y. and {Ashdown}, M. and {Aumont}, J. and {Baccigalupi}, C. and {Ballardini}, M. and {Banday}, A.~J. and {Barreiro}, R.~B. and {Bartolo}, N. and {Basak}, S. and {Battye}, R. and {Benabed}, K. and {Bernard}, J.-P. and {Bersanelli}, M. and {Bielewicz}, P. and {Bock}, J.~J. and {Bond}, J.~R. and {Borrill}, J. and {Bouchet}, F.~R. and {Boulanger}, F. and {Bucher}, M. and {Burigana}, C. and {Butler}, R.~C. and {Calabrese}, E. and {Cardoso}, J.-F. and {Carron}, J. and {Challinor}, A. and {Chiang}, H.~C. and {Chluba}, J. and {Colombo}, L.~P.~L. and {Combet}, C. and {Contreras}, D. and {Crill}, B.~P. and {Cuttaia}, F. and {de Bernardis}, P. and {de Zotti}, G. and {Delabrouille}, J. and {Delouis}, J.-M. and {Di Valentino}, E. and {Diego}, J.~M. and {Dor{\'e}}, O. and {Douspis}, M. and {Ducout}, A. and {Dupac}, X. and {Dusini}, S. and {Efstathiou}, G. and {Elsner}, F. and {En{\ss}lin}, T.~A. and {Eriksen}, H.~K. and {Fantaye}, Y. and {Farhang}, M. and {Fergusson}, J. and {Fernandez-Cobos}, R. and {Finelli}, F. and {Forastieri}, F. and {Frailis}, M. and {Fraisse}, A.~A. and {Franceschi}, E. and {Frolov}, A. and {Galeotta}, S. and {Galli}, S. and {Ganga}, K. and {G{\'e}nova-Santos}, R.~T. and {Gerbino}, M. and {Ghosh}, T. and {Gonz{\'a}lez-Nuevo}, J. and {G{\'o}rski}, K.~M. and {Gratton}, S. and {Gruppuso}, A. and {Gudmundsson}, J.~E. and {Hamann}, J. and {Handley}, W. and {Hansen}, F.~K. and {Herranz}, D. and {Hildebrandt}, S.~R. and {Hivon}, E. and {Huang}, Z. and {Jaffe}, A.~H. and {Jones}, W.~C. and {Karakci}, A. and {Keih{\"a}nen}, E. and {Keskitalo}, R. and {Kiiveri}, K. and {Kim}, J. and {Kisner}, T.~S. and {Knox}, L. and {Krachmalnicoff}, N. and {Kunz}, M. and {Kurki-Suonio}, H. and {Lagache}, G. and {Lamarre}, J.-M. and {Lasenby}, A. and {Lattanzi}, M. and {Lawrence}, C.~R. and {Le Jeune}, M. and {Lemos}, P. and {Lesgourgues}, J. and {Levrier}, F. and {Lewis}, A. and {Liguori}, M. and {Lilje}, P.~B. and {Lilley}, M. and {Lindholm}, V. and {L{\'o}pez-Caniego}, M. and {Lubin}, P.~M. and {Ma}, Y.-Z. and {Mac{\'\i}as-P{\'e}rez}, J.~F. and {Maggio}, G. and {Maino}, D. and {Mandolesi}, N. and {Mangilli}, A. and {Marcos-Caballero}, A. and {Maris}, M. and {Martin}, P.~G. and {Martinelli}, M. and {Mart{\'\i}nez-Gonz{\'a}lez}, E. and {Matarrese}, S. and {Mauri}, N. and {McEwen}, J.~D. and {Meinhold}, P.~R. and {Melchiorri}, A. and {Mennella}, A. and {Migliaccio}, M. and {Millea}, M. and {Mitra}, S. and {Miville-Desch{\^e}nes}, M.-A. and {Molinari}, D. and {Montier}, L. and {Morgante}, G. and {Moss}, A. and {Natoli}, P. and {N{\o}rgaard-Nielsen}, H.~U. and {Pagano}, L. and {Paoletti}, D. and {Partridge}, B. and {Patanchon}, G. and {Peiris}, H.~V. and {Perrotta}, F. and {Pettorino}, V. and {Piacentini}, F. and {Polastri}, L. and {Polenta}, G. and {Puget}, J.-L. and {Rachen}, J.~P. and {Reinecke}, M. and {Remazeilles}, M. and {Renzi}, A. and {Rocha}, G. and {Rosset}, C. and {Roudier}, G. and {Rubi{\~n}o-Mart{\'\i}n}, J.~A. and {Ruiz-Granados}, B. and {Salvati}, L. and {Sandri}, M. and {Savelainen}, M. and {Scott}, D. and {Shellard}, E.~P.~S. and {Sirignano}, C. and {Sirri}, G. and {Spencer}, L.~D. and {Sunyaev}, R. and {Suur-Uski}, A.-S. and {Tauber}, J.~A. and {Tavagnacco}, D. and {Tenti}, M. and {Toffolatti}, L. and {Tomasi}, M. and {Trombetti}, T. and {Valenziano}, L. and {Valiviita}, J. and {Van Tent}, B. and {Vibert}, L. and {Vielva}, P. and {Villa}, F. and {Vittorio}, N. and {Wandelt}, B.~D. and {Wehus}, I.~K. and {White}, M. and {White}, S.~D.~M. and {Zacchei}, A. and {Zonca}, A.},
        title = "{Planck 2018 results. VI. Cosmological parameters}",
      journal = {\aap},
         year = 2020,
        month = sep,
       volume = {641},
          eid = {A6},
        pages = {A6},
          doi = {10.1051/0004-6361/201833910},
archivePrefix = {arXiv},
       eprint = {1807.06209},
 primaryClass = {astro-ph.CO},
       adsurl = {https://ui.adsabs.harvard.edu/abs/2020A&A...641A...6P}
}

@ARTICLE{hirano14,
       author = {{Hirano}, Shingo and {Hosokawa}, Takashi and {Yoshida}, Naoki and {Umeda}, Hideyuki and {Omukai}, Kazuyuki and {Chiaki}, Gen and {Yorke}, Harold W.},
        title = "{One Hundred First Stars: Protostellar Evolution and the Final Masses}",
      journal = {\apj},
         year = 2014,
        month = feb,
       volume = {781},
       number = {2},
          eid = {60},
        pages = {60},
          doi = {10.1088/0004-637X/781/2/60},
archivePrefix = {arXiv},
       eprint = {1308.4456},
 primaryClass = {astro-ph.CO},
       adsurl = {https://ui.adsabs.harvard.edu/abs/2014ApJ...781...60H}
}

@ARTICLE{morishita25a,
       author = {{Morishita}, Takahiro and {Liu}, Zhaoran and {Stiavelli}, Massimo and {Treu}, Tommaso and {Trenti}, Michele and {Chartab}, Nima and {Roberts-Borsani}, Guido and {Vulcani}, Benedetta and {Bergamini}, Pietro and {Castellano}, Marco and {Grillo}, Claudio},
        title = "{Accelerated Emergence of Evolved Galaxies in Early Overdensities at z {\ensuremath{\sim}} 5.7}",
      journal = {\apj},
         year = 2025,
        month = apr,
       volume = {982},
       number = {2},
          eid = {153},
        pages = {153},
          doi = {10.3847/1538-4357/adb30f},
archivePrefix = {arXiv},
       eprint = {2408.10980},
 primaryClass = {astro-ph.GA},
       adsurl = {https://ui.adsabs.harvard.edu/abs/2025ApJ...982..153M}
}

@ARTICLE{schaerer02,
       author = {{Schaerer}, D.},
        title = "{On the properties of massive Population III stars and metal-free stellar populations}",
      journal = {\aap},
         year = 2002,
        month = jan,
       volume = {382},
        pages = {28-42},
          doi = {10.1051/0004-6361:20011619},
archivePrefix = {arXiv},
       eprint = {astro-ph/0110697},
 primaryClass = {astro-ph},
       adsurl = {https://ui.adsabs.harvard.edu/abs/2002A&A...382...28S}
}

@ARTICLE{visbal25,
       author = {{Visbal}, Eli and {Hazlett}, Ryan and {Bryan}, Greg L.},
        title = "{LAP1-B is the First Observed System Consistent with Theoretical Predictions for Population III Stars}",
      journal = {\apjl},
         year = 2025,
        month = nov,
       volume = {993},
       number = {1},
          eid = {L17},
        pages = {L17},
          doi = {10.3847/2041-8213/ae122f},
archivePrefix = {arXiv},
       eprint = {2508.03842},
 primaryClass = {astro-ph.GA},
       adsurl = {https://ui.adsabs.harvard.edu/abs/2025ApJ...993L..17V}
}

@ARTICLE{vanderwel22,
       author = {{van der Wel}, Arjen and {van Houdt}, Josha and {Bezanson}, Rachel and {Franx}, Marijn and {D'Eugenio}, Francesco and {Straatman}, Caroline and {Bell}, Eric F. and {Muzzin}, Adam and {Sobral}, David and {Maseda}, Michael V. and {de Graaff}, Anna and {Holden}, Bradford P.},
        title = "{The Mass Scale of High-redshift Galaxies: Virial Mass Estimates Calibrated with Stellar Dynamical Models from LEGA-C}",
      journal = {\apj},
         year = 2022,
        month = sep,
       volume = {936},
       number = {1},
          eid = {9},
        pages = {9},
          doi = {10.3847/1538-4357/ac83c5},
archivePrefix = {arXiv},
       eprint = {2208.12605},
 primaryClass = {astro-ph.GA},
       adsurl = {https://ui.adsabs.harvard.edu/abs/2022ApJ...936....9V}
}

@ARTICLE{kroupa01,
       author = {{Kroupa}, Pavel},
        title = "{On the variation of the initial mass function}",
      journal = {\mnras},
         year = 2001,
        month = apr,
       volume = {322},
       number = {2},
        pages = {231-246},
          doi = {10.1046/j.1365-8711.2001.04022.x},
archivePrefix = {arXiv},
       eprint = {astro-ph/0009005},
 primaryClass = {astro-ph},
       adsurl = {https://ui.adsabs.harvard.edu/abs/2001MNRAS.322..231K}
}

@ARTICLE{raiter10,
       author = {{Raiter}, A. and {Schaerer}, D. and {Fosbury}, R.~A.~E.},
        title = "{Predicted UV properties of very metal-poor starburst galaxies}",
      journal = {\aap},
         year = 2010,
        month = nov,
       volume = {523},
          eid = {A64},
        pages = {A64},
          doi = {10.1051/0004-6361/201015236},
archivePrefix = {arXiv},
       eprint = {1008.2114},
 primaryClass = {astro-ph.CO},
       adsurl = {https://ui.adsabs.harvard.edu/abs/2010A&A...523A..64R}
}

@ARTICLE{tumlinson06,
       author = {{Tumlinson}, Jason},
        title = "{Chemical Evolution in Hierarchical Models of Cosmic Structure. I. Constraints on the Early Stellar Initial Mass Function}",
      journal = {\apj},
         year = 2006,
        month = apr,
       volume = {641},
       number = {1},
        pages = {1-20},
          doi = {10.1086/500383},
archivePrefix = {arXiv},
       eprint = {astro-ph/0507442},
 primaryClass = {astro-ph},
       adsurl = {https://ui.adsabs.harvard.edu/abs/2006ApJ...641....1T}
}

@ARTICLE{zackrisson11,
       author = {{Zackrisson}, Erik and {Rydberg}, Claes-Erik and {Schaerer}, Daniel and {{\"O}stlin}, G{\"o}ran and {Tuli}, Manan},
        title = "{The Spectral Evolution of the First Galaxies. I. James Webb Space Telescope Detection Limits and Color Criteria for Population III Galaxies}",
      journal = {\apj},
         year = 2011,
        month = oct,
       volume = {740},
       number = {1},
          eid = {13},
        pages = {13},
          doi = {10.1088/0004-637X/740/1/13},
archivePrefix = {arXiv},
       eprint = {1105.0921},
 primaryClass = {astro-ph.CO},
       adsurl = {https://ui.adsabs.harvard.edu/abs/2011ApJ...740...13Z}
}

@ARTICLE{messa26,
       author = {{Messa}, M. and {Vanzella}, E. and {Morishita}, T. and {Stiavelli}, M. and {Treu}, T. and {Bergamini}, P. and {Liu}, Z. and {Zanella}, A. and {Bolamperti}, A. and {Verhamme}, A. and {Garel}, T. and {Grillo}, C. and {Rosati}, P.},
        title = "{A faint M$_{UV}$ = {\ensuremath{-}}14.5 Lyman-continuum leaker candidate in the epoch of reionization: Unprecedented Ly{\ensuremath{\alpha}} properties at z = 5.725}",
      journal = {\aap},
         year = 2026,
        month = mar,
       volume = {708},
          eid = {L2},
        pages = {L2},
          doi = {10.1051/0004-6361/202558660},
archivePrefix = {arXiv},
       eprint = {2512.16728},
 primaryClass = {astro-ph.GA},
       adsurl = {https://ui.adsabs.harvard.edu/abs/2026A&A...708L...2M}
}

@ARTICLE{hsiao25,
       author = {{Hsiao}, Tiger Yu-Yang and {Sun}, Fengwu and {Lin}, Xiaojing and {Coe}, Dan and {Egami}, Eiichi and {Eisenstein}, Daniel J. and {Fudamoto}, Yoshinobu and {Bunker}, Andrew J. and {Fan}, Xiaohui and {Harikane}, Yuichi and {Helton}, Jakob M. and {Kakiichi}, Koki and {Liu}, Yichen and {Liu}, Weizhe and {Maiolino}, Roberto and {Ouchi}, Masami and {Tee}, Wei Leong and {Wang}, Feige and {Wu}, Yunjing and {Xu}, Yi and {Yang}, Jinyi and {Zhu}, Yongda},
        title = "{SAPPHIRES: Extremely Metal-Poor Galaxy Candidates with $12+{\rm log(O/H)}<7.0$ at $z\sim5-7$ from Deep JWST/NIRCam Grism Observations}",
      journal = {arXiv e-prints},
         year = 2025,
        month = may,
          eid = {arXiv:2505.03873},
        pages = {arXiv:2505.03873},
          doi = {10.48550/arXiv.2505.03873},
archivePrefix = {arXiv},
       eprint = {2505.03873},
 primaryClass = {astro-ph.GA},
       adsurl = {https://ui.adsabs.harvard.edu/abs/2025arXiv250503873H}
}

@ARTICLE{eldridge17,
       author = {{Eldridge}, J.~J. and {Stanway}, E.~R. and {Xiao}, L. and {McClelland}, L.~A.~S. and {Taylor}, G. and {Ng}, M. and {Greis}, S.~M.~L. and {Bray}, J.~C.},
        title = "{Binary Population and Spectral Synthesis Version 2.1: Construction, Observational Verification, and New Results}",
      journal = {\pasa},
         year = 2017,
        month = nov,
       volume = {34},
          eid = {e058},
        pages = {e058},
          doi = {10.1017/pasa.2017.51},
archivePrefix = {arXiv},
       eprint = {1710.02154},
 primaryClass = {astro-ph.SR},
       adsurl = {https://ui.adsabs.harvard.edu/abs/2017PASA...34...58E}
}

@ARTICLE{nakajima25,
       author = {{Nakajima}, Kimihiko and {Ouchi}, Masami and {Harikane}, Yuichi and {Vanzella}, Eros and {Ono}, Yoshiaki and {Isobe}, Yuki and {Nishigaki}, Moka and {Tsujimoto}, Takuji and {Nakamura}, Fumitaka and {Xu}, Yi and {Umeda}, Hiroya and {Zhang}, Yechi},
        title = "{An Ultra-Faint, Chemically Primitive Galaxy Forming at the Epoch of Reionization}",
      journal = {arXiv e-prints},
         year = 2025,
        month = jun,
          eid = {arXiv:2506.11846},
        pages = {arXiv:2506.11846},
          doi = {10.48550/arXiv.2506.11846},
archivePrefix = {arXiv},
       eprint = {2506.11846},
 primaryClass = {astro-ph.GA},
       adsurl = {https://ui.adsabs.harvard.edu/abs/2025arXiv250611846N}
}

@ARTICLE{chemerynska24,
       author = {{Chemerynska}, Iryna and {Atek}, Hakim and {Dayal}, Pratika and {Furtak}, Lukas J. and {Feldmann}, Robert and {Greene}, Jenny E. and {Maseda}, Michael V. and {Nanayakkara}, Themiya and {Oesch}, Pascal A. and {Fujimoto}, Seiji and {Labb{\'e}}, Ivo and {Bezanson}, Rachel and {Brammer}, Gabriel and {Cutler}, Sam E. and {Leja}, Joel and {Pan}, Richard and {Price}, Sedona H. and {Wang}, Bingjie and {Weaver}, John R. and {Whitaker}, Katherine E.},
        title = "{The Extreme Low-mass End of the Mass{\textendash}Metallicity Relation at z {{\sim}} 7}",
      journal = {\apjl},
         year = 2024,
        month = nov,
       volume = {976},
       number = {1},
          eid = {L15},
        pages = {L15},
          doi = {10.3847/2041-8213/ad8dc9},
archivePrefix = {arXiv},
       eprint = {2407.17110},
 primaryClass = {astro-ph.GA},
       adsurl = {https://ui.adsabs.harvard.edu/abs/2024ApJ...976L..15C}
}

@ARTICLE{morishita24b,
       author = {{Morishita}, Takahiro and {Stiavelli}, Massimo and {Grillo}, Claudio and {Rosati}, Piero and {Schuldt}, Stefan and {Trenti}, Michele and {Bergamini}, Pietro and {Boyett}, Kit and {Chary}, Ranga-Ram and {Leethochawalit}, Nicha and {Roberts-Borsani}, Guido and {Treu}, Tommaso and {Vanzella}, Eros},
        title = "{Diverse Oxygen Abundance in Early Galaxies Unveiled by Auroral Line Analysis with JWST}",
      journal = {\apj},
         year = 2024,
        month = aug,
       volume = {971},
       number = {1},
          eid = {43},
        pages = {43},
          doi = {10.3847/1538-4357/ad5290},
archivePrefix = {arXiv},
       eprint = {2402.14084},
 primaryClass = {astro-ph.GA},
       adsurl = {https://ui.adsabs.harvard.edu/abs/2024ApJ...971...43M}
}

@ARTICLE{naidu24,
       author = {{Naidu}, Rohan P. and {Matthee}, Jorryt and {Kramarenko}, Ivan and {Weibel}, Andrea and {Brammer}, Gabriel and {Oesch}, Pascal A. and {Lechner}, Peter and {Furtak}, Lukas J. and {Di Cesare}, Claudia and {Torralba}, Alberto and {Kotiwale}, Gauri and {Bezanson}, Rachel and {Bouwens}, Rychard J. and {Chandra}, Vedant and {Claeyssens}, Ad{\'e}la{\"\i}de and {Danhaive}, A. Lola and {Frebel}, Anna and {de Graaff}, Anna and {Greene}, Jenny E. and {Heintz}, Kasper E. and {Ji}, Alexander P. and {Kashino}, Daichi and {Katz}, Harley and {Labbe}, Ivo and {Leja}, Joel and {Li}, Yijia and {Maseda}, Michael V. and {Richard}, Johan and {Shivaei}, Irene and {Simcoe}, Robert A. and {Sobral}, David and {Suess}, Katherine A. and {Tacchella}, Sandro and {Williams}, Christina C.},
        title = "{All the Little Things in Abell 2744: $>$1000 Gravitationally Lensed Dwarf Galaxies at $z=0-9$ from JWST NIRCam Grism Spectroscopy}",
      journal = {arXiv e-prints},
         year = 2024,
        month = oct,
          eid = {arXiv:2410.01874},
        pages = {arXiv:2410.01874},
          doi = {10.48550/arXiv.2410.01874},
archivePrefix = {arXiv},
       eprint = {2410.01874},
 primaryClass = {astro-ph.GA},
       adsurl = {https://ui.adsabs.harvard.edu/abs/2024arXiv241001874N}
}

@ARTICLE{sanders24,
       author = {{Sanders}, Ryan L. and {Shapley}, Alice E. and {Topping}, Michael W. and {Reddy}, Naveen A. and {Brammer}, Gabriel B.},
        title = "{Direct T $_{e}$-based Metallicities of z = 2{\textendash}9 Galaxies with JWST/NIRSpec: Empirical Metallicity Calibrations Applicable from Reionization to Cosmic Noon}",
      journal = {\apj},
         year = 2024,
        month = feb,
       volume = {962},
       number = {1},
          eid = {24},
        pages = {24},
          doi = {10.3847/1538-4357/ad15fc},
archivePrefix = {arXiv},
       eprint = {2303.08149},
 primaryClass = {astro-ph.GA},
       adsurl = {https://ui.adsabs.harvard.edu/abs/2024ApJ...962...24S}
}

@ARTICLE{norris14,
       author = {{Norris}, Mark A. and {Kannappan}, Sheila J. and {Forbes}, Duncan A. and {Romanowsky}, Aaron J. and {Brodie}, Jean P. and {Faifer}, Favio Ra{\'u}l and {Huxor}, Avon and {Maraston}, Claudia and {Moffett}, Amanda J. and {Penny}, Samantha J. and {Pota}, Vincenzo and {Smith-Castelli}, Anal{\'\i}a and {Strader}, Jay and {Bradley}, David and {Eckert}, Kathleen D. and {Fohring}, Dora and {McBride}, JoEllen and {Stark}, David V. and {Vaduvescu}, Ovidiu},
        title = "{The AIMSS Project - I. Bridging the star cluster-galaxy divide$^{★}${\textdagger}{\textdaggerdbl}{\textsection}{\textparagraph}}",
      journal = {\mnras},
         year = 2014,
        month = sep,
       volume = {443},
       number = {2},
        pages = {1151-1172},
          doi = {10.1093/mnras/stu1186},
archivePrefix = {arXiv},
       eprint = {1406.6065},
 primaryClass = {astro-ph.GA},
       adsurl = {https://ui.adsabs.harvard.edu/abs/2014MNRAS.443.1151N}
}

@ARTICLE{topping24,
       author = {{Topping}, Michael W. and {Stark}, Daniel P. and {Senchyna}, Peter and {Plat}, Adele and {Zitrin}, Adi and {Endsley}, Ryan and {Charlot}, St{\'e}phane and {Furtak}, Lukas J. and {Maseda}, Michael V. and {Smit}, Renske and {Mainali}, Ramesh and {Chevallard}, Jacopo and {Molyneux}, Stephen and {Rigby}, Jane R.},
        title = "{Metal-poor star formation at z > 6 with JWST: new insight into hard radiation fields and nitrogen enrichment on 20 pc scales}",
      journal = {\mnras},
         year = 2024,
        month = apr,
       volume = {529},
       number = {4},
        pages = {3301-3322},
          doi = {10.1093/mnras/stae682},
archivePrefix = {arXiv},
       eprint = {2401.08764},
 primaryClass = {astro-ph.GA},
       adsurl = {https://ui.adsabs.harvard.edu/abs/2024MNRAS.529.3301T}
}

@ARTICLE{cameron23,
       author = {{Cameron}, Alex J. and {Katz}, Harley and {Rey}, Martin P.},
        title = "{A novel approach to correcting T$_{e}$-based mass-metallicity relations}",
      journal = {\mnras},
         year = 2023,
        month = jun,
       volume = {522},
       number = {1},
        pages = {L89-L94},
          doi = {10.1093/mnrasl/slad046},
archivePrefix = {arXiv},
       eprint = {2210.14234},
 primaryClass = {astro-ph.GA},
       adsurl = {https://ui.adsabs.harvard.edu/abs/2023MNRAS.522L..89C}
}

@ARTICLE{vanzella23,
       author = {{Vanzella}, E. and {Loiacono}, F. and {Bergamini}, P. and {Me{\v{s}}tri{\'c}}, U. and {Castellano}, M. and {Rosati}, P. and {Meneghetti}, M. and {Grillo}, C. and {Calura}, F. and {Mignoli}, M. and {Brada{\v{c}}}, M. and {Adamo}, A. and {Rihtar{\v{s}}i{\v{c}}}, G. and {Dickinson}, M. and {Gronke}, M. and {Zanella}, A. and {Annibali}, F. and {Willott}, C. and {Messa}, M. and {Sani}, E. and {Acebron}, A. and {Bolamperti}, A. and {Comastri}, A. and {Gilli}, R. and {Caputi}, K.~I. and {Ricotti}, M. and {Gruppioni}, C. and {Ravindranath}, S. and {Mercurio}, A. and {Strait}, V. and {Martis}, N. and {Pascale}, R. and {Caminha}, G.~B. and {Annunziatella}, M. and {Nonino}, M.},
        title = "{An extremely metal-poor star complex in the reionization era: Approaching Population III stars with JWST}",
      journal = {\aap},
         year = 2023,
        month = oct,
       volume = {678},
          eid = {A173},
        pages = {A173},
          doi = {10.1051/0004-6361/202346981},
archivePrefix = {arXiv},
       eprint = {2305.14413},
 primaryClass = {astro-ph.GA},
       adsurl = {https://ui.adsabs.harvard.edu/abs/2023A&A...678A.173V}
}

@ARTICLE{laseter23,
       author = {{Laseter}, Isaac H. and {Maseda}, Michael V. and {Curti}, Mirko and {Maiolino}, Roberto and {D'Eugenio}, Francesco and {Cameron}, Alex J. and {Looser}, Tobias J. and {Arribas}, Santiago and {Baker}, William M. and {Bhatawdekar}, Rachana and {Boyett}, Kristan and {Bunker}, Andrew J. and {Carniani}, Stefano and {Charlot}, Stephane and {Chevallard}, Jacopo and {Curtis-lake}, Emma and {Egami}, Eiichi and {Eisenstein}, Daniel J. and {Hainline}, Kevin and {Hausen}, Ryan and {Ji}, Zhiyuan and {Kumari}, Nimisha and {Perna}, Michele and {Rawle}, Tim and {Rix}, Hans-Walter and {Robertson}, Brant and {Rodr{\'\i}guez Del Pino}, Bruno and {Sandles}, Lester and {Scholtz}, Jan and {Smit}, Renske and {Tacchella}, Sandro and {{\"U}bler}, Hannah and {Williams}, Christina C. and {Willott}, Chris and {Witstok}, Joris},
        title = "{JADES: Detecting [OIII]{\ensuremath{\lambda}}4363 emitters and testing strong line calibrations in the high-z Universe with ultra-deep JWST/NIRSpec spectroscopy up to z {\ensuremath{\sim}} 9.5}",
      journal = {\aap},
         year = 2024,
        month = jan,
       volume = {681},
          eid = {A70},
        pages = {A70},
          doi = {10.1051/0004-6361/202347133},
archivePrefix = {arXiv},
       eprint = {2306.03120},
 primaryClass = {astro-ph.GA},
       adsurl = {https://ui.adsabs.harvard.edu/abs/2024A&A...681A..70L}
}

@ARTICLE{astropy18,
       author = {{Astropy Collaboration} and {Price-Whelan}, A.~M. and {Sip{\H{o}}cz}, B.~M. and {G{\"u}nther}, H.~M. and {Lim}, P.~L. and {Crawford}, S.~M. and {Conseil}, S. and {Shupe}, D.~L. and {Craig}, M.~W. and {Dencheva}, N. and {Ginsburg}, A. and {VanderPlas}, J.~T. and {Bradley}, L.~D. and {P{\'e}rez-Su{\'a}rez}, D. and {de Val-Borro}, M. and {Aldcroft}, T.~L. and {Cruz}, K.~L. and {Robitaille}, T.~P. and {Tollerud}, E.~J. and {Ardelean}, C. and {Babej}, T. and {Bach}, Y.~P. and {Bachetti}, M. and {Bakanov}, A.~V. and {Bamford}, S.~P. and {Barentsen}, G. and {Barmby}, P. and {Baumbach}, A. and {Berry}, K.~L. and {Biscani}, F. and {Boquien}, M. and {Bostroem}, K.~A. and {Bouma}, L.~G. and {Brammer}, G.~B. and {Bray}, E.~M. and {Breytenbach}, H. and {Buddelmeijer}, H. and {Burke}, D.~J. and {Calderone}, G. and {Cano Rodr{\'\i}guez}, J.~L. and {Cara}, M. and {Cardoso}, J.~V.~M. and {Cheedella}, S. and {Copin}, Y. and {Corrales}, L. and {Crichton}, D. and {D'Avella}, D. and {Deil}, C. and {Depagne}, {\'E}. and {Dietrich}, J.~P. and {Donath}, A. and {Droettboom}, M. and {Earl}, N. and {Erben}, T. and {Fabbro}, S. and {Ferreira}, L.~A. and {Finethy}, T. and {Fox}, R.~T. and {Garrison}, L.~H. and {Gibbons}, S.~L.~J. and {Goldstein}, D.~A. and {Gommers}, R. and {Greco}, J.~P. and {Greenfield}, P. and {Groener}, A.~M. and {Grollier}, F. and {Hagen}, A. and {Hirst}, P. and {Homeier}, D. and {Horton}, A.~J. and {Hosseinzadeh}, G. and {Hu}, L. and {Hunkeler}, J.~S. and {Ivezi{\'c}}, {\v{Z}}. and {Jain}, A. and {Jenness}, T. and {Kanarek}, G. and {Kendrew}, S. and {Kern}, N.~S. and {Kerzendorf}, W.~E. and {Khvalko}, A. and {King}, J. and {Kirkby}, D. and {Kulkarni}, A.~M. and {Kumar}, A. and {Lee}, A. and {Lenz}, D. and {Littlefair}, S.~P. and {Ma}, Z. and {Macleod}, D.~M. and {Mastropietro}, M. and {McCully}, C. and {Montagnac}, S. and {Morris}, B.~M. and {Mueller}, M. and {Mumford}, S.~J. and {Muna}, D. and {Murphy}, N.~A. and {Nelson}, S. and {Nguyen}, G.~H. and {Ninan}, J.~P. and {N{\"o}the}, M. and {Ogaz}, S. and {Oh}, S. and {Parejko}, J.~K. and {Parley}, N. and {Pascual}, S. and {Patil}, R. and {Patil}, A.~A. and {Plunkett}, A.~L. and {Prochaska}, J.~X. and {Rastogi}, T. and {Reddy Janga}, V. and {Sabater}, J. and {Sakurikar}, P. and {Seifert}, M. and {Sherbert}, L.~E. and {Sherwood-Taylor}, H. and {Shih}, A.~Y. and {Sick}, J. and {Silbiger}, M.~T. and {Singanamalla}, S. and {Singer}, L.~P. and {Sladen}, P.~H. and {Sooley}, K.~A. and {Sornarajah}, S. and {Streicher}, O. and {Teuben}, P. and {Thomas}, S.~W. and {Tremblay}, G.~R. and {Turner}, J.~E.~H. and {Terr{\'o}n}, V. and {van Kerkwijk}, M.~H. and {de la Vega}, A. and {Watkins}, L.~L. and {Weaver}, B.~A. and {Whitmore}, J.~B. and {Woillez}, J. and {Zabalza}, V. and {Astropy Contributors}},
        title = "{The Astropy Project: Building an Open-science Project and Status of the v2.0 Core Package}",
      journal = {\aj},
         year = 2018,
        month = sep,
       volume = {156},
       number = {3},
          eid = {123},
        pages = {123},
          doi = {10.3847/1538-3881/aabc4f},
archivePrefix = {arXiv},
       eprint = {1801.02634},
 primaryClass = {astro-ph.IM},
       adsurl = {https://ui.adsabs.harvard.edu/abs/2018AJ....156..123A}
}

@ARTICLE{astropy13,
       author = {{Astropy Collaboration} and {Robitaille}, Thomas P. and {Tollerud}, Erik J. and {Greenfield}, Perry and {Droettboom}, Michael and {Bray}, Erik and {Aldcroft}, Tom and {Davis}, Matt and {Ginsburg}, Adam and {Price-Whelan}, Adrian M. and {Kerzendorf}, Wolfgang E. and {Conley}, Alexander and {Crighton}, Neil and {Barbary}, Kyle and {Muna}, Demitri and {Ferguson}, Henry and {Grollier}, Fr{\'e}d{\'e}ric and {Parikh}, Madhura M. and {Nair}, Prasanth H. and {Unther}, Hans M. and {Deil}, Christoph and {Woillez}, Julien and {Conseil}, Simon and {Kramer}, Roban and {Turner}, James E.~H. and {Singer}, Leo and {Fox}, Ryan and {Weaver}, Benjamin A. and {Zabalza}, Victor and {Edwards}, Zachary I. and {Azalee Bostroem}, K. and {Burke}, D.~J. and {Casey}, Andrew R. and {Crawford}, Steven M. and {Dencheva}, Nadia and {Ely}, Justin and {Jenness}, Tim and {Labrie}, Kathleen and {Lim}, Pey Lian and {Pierfederici}, Francesco and {Pontzen}, Andrew and {Ptak}, Andy and {Refsdal}, Brian and {Servillat}, Mathieu and {Streicher}, Ole},
        title = "{Astropy: A community Python package for astronomy}",
      journal = {\aap},
         year = 2013,
        month = oct,
       volume = {558},
          eid = {A33},
        pages = {A33},
          doi = {10.1051/0004-6361/201322068},
archivePrefix = {arXiv},
       eprint = {1307.6212},
 primaryClass = {astro-ph.IM},
       adsurl = {https://ui.adsabs.harvard.edu/abs/2013A&A...558A..33A}
}

@ARTICLE{bergamini23b,
       author = {{Bergamini}, Pietro and {Acebron}, Ana and {Grillo}, Claudio and {Rosati}, Piero and {Caminha}, Gabriel Bartosch and {Mercurio}, Amata and {Vanzella}, Eros and {Mason}, Charlotte and {Treu}, Tommaso and {Angora}, Giuseppe and {Brammer}, Gabriel B. and {Meneghetti}, Massimo and {Nonino}, Mario and {Boyett}, Kristan and {Brada{\v{c}}}, Maru{\v{s}}a and {Castellano}, Marco and {Fontana}, Adriano and {Morishita}, Takahiro and {Paris}, Diego and {Prieto-Lyon}, Gonzalo and {Roberts-Borsani}, Guido and {Roy}, Namrata and {Santini}, Paola and {Vulcani}, Benedetta and {Wang}, Xin and {Yang}, Lilan},
        title = "{The GLASS-JWST Early Release Science Program. III. Strong-lensing Model of Abell 2744 and Its Infalling Regions}",
      journal = {\apj},
         year = 2023,
        month = jul,
       volume = {952},
       number = {1},
          eid = {84},
        pages = {84},
          doi = {10.3847/1538-4357/acd643},
archivePrefix = {arXiv},
       eprint = {2303.10210},
 primaryClass = {astro-ph.GA},
       adsurl = {https://ui.adsabs.harvard.edu/abs/2023ApJ...952...84B}
}

@Article{numpy,
 title         = {Array programming with {NumPy}},
 author        = {Charles R. Harris and K. Jarrod Millman and St{\'{e}}fan J.
                 van der Walt and Ralf Gommers and Pauli Virtanen and David
                 Cournapeau and Eric Wieser and Julian Taylor and Sebastian
                 Berg and Nathaniel J. Smith and Robert Kern and Matti Picus
                 and Stephan Hoyer and Marten H. van Kerkwijk and Matthew
                 Brett and Allan Haldane and Jaime Fern{\'{a}}ndez del
                 R{\'{i}}o and Mark Wiebe and Pearu Peterson and Pierre
                 G{\'{e}}rard-Marchant and Kevin Sheppard and Tyler Reddy and
                 Warren Weckesser and Hameer Abbasi and Christoph Gohlke and
                 Travis E. Oliphant},
 year          = {2020},
 month         = sep,
 journal       = {Nature},
 volume        = {585},
 number        = {7825},
 pages         = {357--362},
 doi           = {10.1038/s41586-020-2649-2},
 publisher     = {Springer Science and Business Media {LLC}},
 url           = {https://doi.org/10.1038/s41586-020-2649-2}
}

@ARTICLE{richard21,
       author = {{Richard}, Johan and {Claeyssens}, Ad{\'e}la{\"\i}de and {Lagattuta}, David and {Guaita}, Lucia and {Bauer}, Franz Erik and {Pello}, Roser and {Carton}, David and {Bacon}, Roland and {Soucail}, Genevi{\`e}ve and {Lyon}, Gonzalo Prieto and {Kneib}, Jean-Paul and {Mahler}, Guillaume and {Cl{\'e}ment}, Benjamin and {Mercier}, Wilfried and {Variu}, Andrei and {Tamone}, Am{\'e}lie and {Ebeling}, Harald and {Schmidt}, Kasper B. and {Nanayakkara}, Themiya and {Maseda}, Michael and {Weilbacher}, Peter M. and {Bouch{\'e}}, Nicolas and {Bouwens}, Rychard J. and {Wisotzki}, Lutz and {de la Vieuville}, Geoffroy and {Martinez}, Johany and {Patr{\'\i}cio}, Vera},
        title = "{An atlas of MUSE observations towards twelve massive lensing clusters}",
      journal = {\aap},
         year = 2021,
        month = feb,
       volume = {646},
          eid = {A83},
        pages = {A83},
          doi = {10.1051/0004-6361/202039462},
archivePrefix = {arXiv},
       eprint = {2009.09784},
 primaryClass = {astro-ph.GA},
       adsurl = {https://ui.adsabs.harvard.edu/abs/2021A&A...646A..83R}
}

@ARTICLE{chabrier03,
   author = {{Chabrier}, G.},
    title = "{Galactic Stellar and Substellar Initial Mass Function}",
  journal = {\pasp},
   eprint = {astro-ph/0304382},
     year = 2003,
    month = jul,
   volume = 115,
    pages = {763-795},
      doi = {10.1086/376392},
   adsurl = {http://adsabs.harvard.edu/abs/2003PASP..115..763C}
}

@ARTICLE{morishita19,
       author = {{Morishita}, T. and {Abramson}, L.~E. and {Treu}, T. and
         {Brammer}, G.~B. and {Jones}, T. and {Kelly}, P. and {Stiavelli}, M. and
         {Trenti}, M. and {Vulcani}, B. and {Wang}, X.},
        title = "{Massive Dead Galaxies at z ̃ 2 with HST Grism Spectroscopy. I. Star Formation Histories and Metallicity Enrichment}",
      journal = {\apj},
         year = 2019,
        month = jun,
       volume = {877},
       number = {2},
          eid = {141},
        pages = {141},
          doi = {10.3847/1538-4357/ab1d53},
archivePrefix = {arXiv},
       eprint = {1812.06980},
 primaryClass = {astro-ph.GA},
       adsurl = {https://ui.adsabs.harvard.edu/abs/2019ApJ...877..141M}
}
